\documentclass[traditabstract]{aa} 
\usepackage{graphicx}
\usepackage{txfonts}
\usepackage{wasysym}
\newcommand{\hevelius}{{\em Hevelius}}
\newcommand{\keplere}{{\em KeplerE}}

\newcommand{\kepler}{{\em Kepler}}

\newcommand{\gtap}{\mathrel{\hbox{\rlap{\lower.55ex \hbox {$\sim$}}
                   \kern-.3em \raise.4ex \hbox{$>$}}}}
\newcommand{\ltap}{\mathrel{\hbox{\rlap{\lower.55ex \hbox {$\sim$}}
                   \kern-.3em \raise.4ex \hbox{$<$}}}}

\begin{document}
  \title{The Star Catalogue of Hevelius}
  \subtitle{Machine-readable version and comparison with the modern Hipparcos Catalogue}

  \author{Frank Verbunt\inst{1} \and Robert H. van Gent\inst{2,3}}

  \institute{Astronomical Institute, Utrecht University, PO Box 80\,000,
    3508 TA Utrecht, The Netherlands; \email{f.w.m.verbunt@uu.nl}
  \and URU-Explokart, Faculty of Geosciences, Utrecht University, PO Box 80\,115,
    3508 TC Utrecht, The Netherlands
  \and Institute for the History and Foundations of Science,  PO Box 80\,000,
    3508 TA Utrecht, The Netherlands }

  \date{Received \today / Accepted}

  \abstract{The catalogue by Johannes Hevelius with the positions and
    magnitudes of 1564 entries was published by his wife Elisabeth
    Koopman in 1690. We provide a machine-readable version of the
    catalogue, and briefly discuss its accuracy on the basis of
    comparison with data from the modern Hipparcos 
    Catalogue\thanks{The full Table \hevelius\  (Table\,\ref{t:hevelius}) 
    is available in electronic from only at the CDS via
    anonymous ftp to cdsarc.u-strasbg.fr (130.79.128.5)
    or via http://cdsweb.u-strasbg.fr/cgi-bin/qcat?J/A+A/} .  We
    compare our results with an earlier analysis by Rybka (1984),
    finding good overall agreement. The magnitudes given by Hevelius
    correlate well with modern values. The accuracy of his position
    measurements is similar to that of Brahe, with $\sigma=2$\arcmin\ for
    longitudes and latitudes, but with more errors $>5$\arcmin\ than
    expected for a Gaussian distribution. The position accuracy decreases
    slowly with magnitude. The fraction of stars with position errors 
    larger than a degree is 1.5\%, rather smaller than the fraction of 5\%\
    in the star catalogue of Brahe.
    }

  \maketitle

  \section{Introduction}

  Even though a major improvement on earlier work, the star catalogue
  produced by Tycho Brahe (1598, 1602), and re-edited by Kepler
  (1627), contains occasional large errors. Johannes Hevelius decided
  to produce a better and larger catalogue, which was printed after
  his death by his wife and collaborator Elisabeth Koopman (Hevelius
  1690). The title page has 1687, the year in which the catalogue was
  printed, but publication followed only in 1690. The extent of the
  contribution by Koopman to measuring the stars and producing the
  catalogue is not known; her presence on several images of
  instruments used by Hevelius suggests that it was significant.  A
  brief but informative and well-illustrated description of the life
  of Hevelius and of his star catalogue is given by Volkoff et al.\
  (1971) in a book celebrating the acquisition by the Brigham Young
  University of Hevelius' manuscript for the catalogue.

In 1679 Halley visited Hevelius and his observatory, and verified that
measurements with the instruments of Hevelius, fitted with naked-eye
sights, were more accurate than measurements with contemporary
instruments with telescopic sights (Volkoff et al.\ 1971, p.41-45).
Hevelius' star catalogue was studied among others by Baily (1843), and a
modern comprehensive analysis was made by Rybka (1984), who
confirmed that the measurements by Hevelius were superior to those by
his contemporaries.

Our study of the star catalogue by Brahe (Verbunt \&\ Van Gent 2010,
hereafter Paper\,I), showed the superiority of the modern {\em
  Hipparcos Catalogue} (ESA 1997) in the analysis of old star
positions, due to its better completeness, accuracy and homogeneity as
compared to earlier catalogues.  In this paper we present a
machine-readable version of the star catalogue of Hevelius, as printed
in Hevelius (1690). In addition to the numbers given by Hevelius this
version provides a cross-correlation with the catalogue of Brahe;
identifications with stars from the {\em Hipparcos Catalogue} (ESA
1997) and on the basis of these the accuracy of the positions and
magnitudes tabulated by Hevelius; and a comparison of our
identifications with those of Rybka (1984).

\begin{table*}
\caption{Constellations in the star catalogue of Hevelius.
\label{t:const}}
\begin{tabular}{rc@{}rrr@{ }r@{ }rrl|rc@{}rrr@{ }r@{ }rrl}
C &      & $N$ & $N_{OT}$ & $N_K$ & $N_{SC}$ & $N_e$ & H  & &C &  & $N$ & $N_{OT}$ & $N_K$ & $N_{SC}$ & $N_e$ & H\\
\hline
 1 & And & 47 & 23 &  0 &  0 &  0&     1 &Andromeda        &
31 & Leo & 50 & 40 &  0 &  0 &  1&   852 &Leo              \\
 2 & Atn & 19 &  7 &  0 &  0 &  0&    48 &Antinous         &
32 & LMi & 18 &  0 &  7 &  0 &  0&   902 &{\bf Leo Minor}        \\
 3 & Aqr & 48 & 41 &  0 &  0 &  1&    67 &Aquarius         &
33 & Lep & 16 & 13 &  0 &  0 &  0&   920 &Lepus            \\
 4 & Aql & 23 & 12 &  0 &  0 &  0&   115 &Aquila           &
34 & Lib & 21 & 15 &  0 &  0 &  1&   936 &Libra            \\
 5 & Ari & 27 & 21 &  0 &  0 &  0&   138 &Aries            &
35 & Lyn & 19 &  0 &  4 &  0 &  0&   957 &{\bf Lynx sive Tigris}             \\
 6 & Aur & 40 & 27 &  2 &  0 &  0&   165 &Auriga           &
36 & Lyr & 17 & 11 &  2 &  0 &  0&   976 &Lyra             \\
 7 & Boo & 52 & 26 &  0 &  0 &  0&   205 &Bootes           &
37 & Mon & 19 & 12 &  0 &  0 &  0&   993 &{\bf Monoceros}        \\
 8 & Cnc & 29 & 17 &  1 &  0 &  0&   257 &Cancer           &
38 & Arg &  5 &  4 &  0 &  0 &  1&  1012 &Navis            \\
 9 & CMa & 22 & 13 &  0 &  4 &  1&   286 &Canis Maior      &
39 & Ori & 62 & 52 &  0 &  0 &  0&  1017 &Orion            \\
10 & CMi & 13 &  3 &  0 &  0 &  0&   308 &Canis Minor      &
40 & Peg & 38 & 23 &  0 &  0 &  0&  1079 &Pegasus          \\
11 & CVn & 23 &  0 &  2 &  0 &  0&   321 &{\bf Canes Venatici}   &
41 & Per & 48 & 28 &  0 &  0 &  0&  1117 &Perseus          \\
12 & Cam & 32 &  0 & 14 &  0 &  0&   344 &{\bf Camelopardalis}   &
42 & Psc & 39 & 37 &  0 &  0 &  0&  1165 &Pisces           \\
13 & Cap & 30 & 28 &  0 &  0 &  1&   376 &Capricornus      &
43 & Sge &  5 &  5 &  0 &  0 &  0&  1204 &Sagitta          \\
14 & Cas & 38 & 27 &  2 &  0 &  1&   406 &Cassiopeia       &
44 & Sgr & 26 & 14 &  0 &  8 &  4&  1209 &Sagittarius      \\
15 & Cep & 51 & 11 &  4 &  0 &  0&   444 &Cepheus          &
45 & Sco & 20 & 13 &  0 &  0 &  0&  1235 &Scorpius         \\
16 & Cer &  4 &  0 &  0 &  0 &  0&   495 &{\bf Cerberus}         &
46 & Sct &  7 &  0 &  0 &  0 &  0&  1255 &{\bf Scutum Sobiesc.}  \\
17 & Cet & 46 & 22 &  1 &  0 &  1&   499 &Cetus            &
47 & Oph & 44 & 23 &  1 &  4 &  4&  1262 &Serpentarius     \\
18 & Com & 21 & 14 &  0 &  0 &  0&   545 &Coma Berenices   &
48 & Ser & 22 & 18 &  0 &  0 &  0&  1306 &Serpens          \\
19 & CrB &  8 &  8 &  0 &  0 &  0&   566 &Corona Borealis  &
49 & Sex & 12 &  0 &  1 &  0 &  0&  1328 &{\bf Sextans Uraniae}  \\
20 & Crv &  8 &  7 &  0 &  0 &  0&   574 &Corvus           &
50 & Tau & 51 & 43 &  0 &  0 &  0&  1340 &Taurus           \\
21 & Crt & 10 &  8 &  0 &  0 &  0&   582 &Crater           &
51 & Tri &  9 &  4 &  0 &  0 &  0&  1391 &Triangulum       \\
22 & Cyg & 47 & 20 &  0 &  0 &  0&   592 &Cygnus           &
52 & TrM &  3 &  0 &  0 &  0 &  0&  1400 &{\bf Triangulum Minus} \\
23 & Del & 14 & 10 &  0 &  0 &  0&   639 &Delphinus        &
53 & Vir & 50 & 39 &  0 &  0 &  0&  1403 &Virgo            \\
24 & Dra & 40 & 32 &  2 &  0 &  0&   653 &Draco            &
54 & UMa & 73 & 34 &  1 &  0 &  0&  1453 &Ursa Maior       \\
25 & Equ &  6 &  4 &  0 &  0 &  0&   693 &Equuleus         &
55 & UMi & 12 &  9 &  0 &  0 &  0&  1526 &Ursa Minor       \\
26 & Eri & 29 & 17 &  0 & 11 &  2&   699 &Eridanus         &
56 & Vul & 27 &  0 &  3 &  0 &  0&  1538 &{\bf Vulpecula}        \\
27 & Gem & 38 & 29 &  0 &  0 &  0&   728 &Gemini           \\
28 & Her & 45 & 28 &  0 &  0 &  0&  766 &Hercules         \\
29 & Hya & \phantom{15}31 & 23 &  0 &  1 &  0&   811 &Hydra            \\
30 & Lac & 10 &  0 &  0 &  0 &  0&   842 &{\bf Lacerta}          &
all & &1564 &915 & 47 & 28 & 18
\end{tabular}
\tablefoot{For each constellation the table gives
its abbreviation, the number of stars in it, and the numbers among these 
of stars indicated to be in \keplere\ $N_{OT}$, in \keplere\ without
indication $N_K$ and in {\em Secunda Classis} $N_{SC}$, the number
of empty entries $N_e$, and the H sequence number of the first star
in each constellation. New constallations by Hevelius -- partially
based on Plancius -- are indicated with bold face.}
\end{table*}

In the following we refer to (our machine-readable version of) the
catalogue of Hevelius (1690) as \hevelius, to Kepler's 1627 edition of
Brahe's catalogue as \kepler, and to our emended version of this
edition as \keplere. As we will see, Hevelius also refers to the {\em
  Secunda Classis}, the star list that immediately follows Brahe's
catalogue in Kepler (1627), and gives positions and magnitudes of those
stars from the catalogue of Hipparchos/Ptolemaios that Brahe omitted
from his own catalogue.  Individual entries in \hevelius\ are numbered
according to the order in which they appear, i.e.\, H\,350 is the
350th entry. A K-number refers to an entry in \keplere\ (K$\leq$1004)
or in {\em Secunda Classis} (K$\geq$1005).  The sequence number within
a constellation is indicated by a number following the abbreviated
name of the constellation: Vul\,3 is the third entry in the
constellation Vulpecula in \hevelius.

  \section{Description of the Catalogue \label{s:description}}

The catalogue by Hevelius is organized by constellation, and the
constellations are ordered alphabetically. There are 56 constellations
in the catalogue, of which 11 are new with respect to the star
catalogue of Brahe (Table\,\ref{t:const}). Cerberus, Lacerta sive
Stellio, Scutum Sobiescianum, Sextans Uraniae and Triangulum Minus are
new constellations introduced by Hevelius.  Camelopardalis and
Monoceros are constellations introduced by the Dutch astronomer and
cartographer Plancius for a globe by Van den Keere in 1612. Hevelius
divided other constellations from Plancius, Jordanis Fluvius (River
Jordan) and Tigris Fluv.\ / Euphrates Fluv.\ (Rivers Tigris and
Euphrates) into Canes Venatici, Leo Minor, Lynx (sive Tigris), and
Vulpecula cum Anser (see Van der Krogt 1993, p.190-196).

The catalogue of Hevelius contains 1564 entries, including 18 empty
ones for which Hevelius gives no own measurements, but only positions
from other catalogues. 13 entries are repeat entries, (almost)
identical to entries elsewhere in the catalogue
(Table\,\ref{t:doubles}). Thus Hevelius gives his magnitudes and
positions for 1533 independent entries.

For each entry a brief description is followed first by the sequence
number of this entry in the corresponding constellation in Brahe's
catalogue ({\em Ordo Tychonis}), by the magnitude given to this star
by Brahe ({\em Magnitudo Tychonis}), and by the magnitude as
determined by Hevelius ({\em Magnitudo Hevelii}). The magnitudes are
given in integers, but are sometimes qualified by an second number 1
higher or lower than the first number (e.g.\ 6.7.), by the word {\em
  fere} (approximately), or otherwise (Table\,\ref{t:magqual}).

The magnitude as measured by Hevelius is followed by the 
position of the entry in ecliptic coordinates, given in degrees $D$,
minutes $M$, seconds $S$ and sign ({\em Gr., Min., Sec., Sig.}). All
numbers are integers. For the longitude, the sign is the zodiacal
sign, indicated with its symbol in Hevelius' catalogue, replaced by us
with an integer number $Z$ from 1 to 12 as shown in Table\,2 of
Paper\,I (from Aries = 1 to Pisces = 12).  For the latitude the sign
is an A or B indicating Australis (south) or Borealis (north). The
longitude and latitude in decimal degrees follow as
$$ \lambda = (Z-1)*30 + G + {M\over60} + {S\over3600} $$
and
$$ \beta = \pm \left(G + {M\over60} + {S\over3600}\right)
 \qquad +/- \quad\mathrm{for}\quad \mathrm{B/A} $$
The equinox of the coordinates is given as Annum Christi Completum MDCLX,
i.e.\ 1661.0.

These data are followed for each entry by the positions, where
available, in ecliptic coordinates as given by Tycho, Wilhelm Landgrave of
Hesse, Riccioli, Ulugh Beg, and Ptolemaios. The final information for
each entry is the equatorial position as determined by Hevelius.

Of all this information our machine-readable version of Hevelius' catalogue only
retains the sequence number in Brahe, and the magnitude and ecliptic position
according to Hevelius.

Hevelius gives {\em Ordo Tychonis} numbers for 939 entries: 911 of these
indeed are entries in \keplere, usually but not always in the same
constellation as in Hevelius; the 28 others are in the {\em Secunda
  Classis} of Kepler (1627).  Magnitudes of stars first catalogued by
Hevelius are labelled IH (i.e.\ Johannes Hevelius). In most cases
entries which do not have a Brahe sequence number have magnitudes
labelled IH, but exceptions occur when the star is in another earlier
catalogue, in particular the {\em Secunda Classis} of Kepler (1627) or
in Riccioli (1651). Occasionally a star with a magnitude not
labelled IH, thus presumably considered a previously catalogued star
by Hevelius, nonetheless is without position in any of the catalogues
tabulated by Hevelius. Details are given in Sect.\,\ref{s:remarks}.

\section{Identification procedure \label{s:identification}}

The procedure that we follow for the identification of each star from
the catalogue of Hevelius is {\em mutatis mutandis} identical to the
procedure that we followed for the catalogue of Brahe, and we refer to
Paper\,I for details.  Briefly, we select all stars from the {\em
  Hipparcos Catalogue} with a Johnson visual magnitude brighter than
6.0, we correct their equatorial positions for proper motion between
the Hipparcos epoch 1991.25 and 1661.0, then precess the resulting
equatorial coordinates from the Hipparcos equinox 2000.0 to 1661.0,
and finally convert the coordinates from equatorial to ecliptic, using
the obliquity appropriate for 1661.

For each entry in the Hevelius catalogue we find the nearest -- in
terms of angular separation -- counterpart in the Hipparcos
Catalogue. In general, this counterpart is selected by us as a secure
identification, and given an identification flag 1. If a
much brighter star is at a marginally larger angular distance, we
select that star as the secure counterpart, and give it flag 2.
Especially for larger angular distances we may decide that the
identification is uncertain (flag 3); and occasionally several
Hipparcos stars appear to be comparably plausible as counterparts for
the same entry (flag 4). An entry for which we do not find a plausible
identification is flagged 5; and an entry which is identified with an
Hipparcos star that already is the identification of another entry --
i.e.\ a repeat entry -- is flagged 6.

It is indicative of the high accuracy of the Hevelius catalogue that
the number of problematic identifications (flags 3-5) is much smaller
than in our analysis of Brahe's catalogue, notwithstanding the rather
larger number of entries.  In 5 cases we accept an identification with
Hipparcos magnitude $V$=6.1, and in one case each with $V$=6.3 and $V$=6.5.

In thirteen cases two entries are identified with the same
Hipparcos star. These are listed in Table\,\ref{t:doubles}
In seven pairs both stars have exactly identical coordinates,
in four their coordinates differ by $\le1$\arcmin. In some cases
there are nearby unidentified stars, and it would be tempting
to assign one of the pair to such a star, if it weren't for the
too large offset required.

\begin{table}
\caption{\small \it Repeated entries in \hevelius. \label{t:doubles}}
\begin{tabular}{rrrrcl}
\phantom{H}H & \phantom{H}H & HIP\phantom{HP} & $d_{1,2}$(\arcmin) &
$\Delta M$ & Figure: location\\
  16 & 1202 &   4463 &  1.0 & 0 & \ref{f:andromeda}: $-$4.4,$-$16.7 \\
 321 & 1460 &  63125 &  0.0 & 0 & \ref{f:canesven}: $-$4.1,$-$2.0 \\
 322 & 1488 &  61317 &  0.0 & 0 & \ref{f:canesven}: $-9.2$,$-$1.0 \\
 349 &  689 &  59504 &  0.2 & 0 & \ref{f:camelopardalis}: 11.4,17.3 \\
 692 & 1536 &  73199 &  9.1 & 1 & \ref{f:ursaminor}: 10.4,3.4 \\
 887 &  919 &  52457 &  0.4 & 1 & \ref{f:leo}: $-$1.6,6.7 \\
 889 &  916 &  53417 &  0.0 & 0 & \ref{f:leo}: 0.3,9.2 \\
 890 &  918 &  54951 &  0.0 & 0 & \ref{f:leo}: 5.0,9.5 \\
1091 & 1557 & 106140 &  0.0 & 0 & \ref{f:pegasus}: $-$14.5,11.4 \\
1145 & 1151 &  17313 & 15.7 & 1 & \ref{f:perseus}: 0.7,$-$11.0 \\
1150 & 1164 &  17886 &  0.0 & 0 & \ref{f:perseus}: 1.9,$-$12.1 \\
1154 & 1163 &  14382 &  0.0 & 0 & \ref{f:perseus}: 0.9,12.5 \\
1299 & 1326 &  80179 &  0.3 & 1 & \ref{f:serpens}: $-$12.7,5.7 
\end{tabular}
\end{table}

\subsection{Identifications by Rybka}

To compare the identifications from Rybka (1984) with those by us, we
convert his identifications to an Hipparcos number. In most cases
Rybka gives a Bayer or Flamsteed identification (e.g.\ $\alpha$\,And,
7\,And), or an identification from the FK4 catalogue or from the SAO
catalogue. We convert these to HR numbers through the {\em Bright
  Stars Catalogue} (Hoffleit \&\ Warren 1991) and from HR via HD
number to Hipparcos numbers. If this fails, we use the {\em SIMBAD}
database. For H\,319 (=CMi\,12), Rybka gives 12\,CMi as Flamsteed
counterpart. We do not find a star from the {\em Hipparcos catalogue}
with $V$$<$6.5 within 45\arcmin\ from 12\,CMi, however; our
identification for H\,319, HIP\,39311, corresponds to the 13th star in
Navis in Flamsteed's catalogue.

\begin{table}
\caption{Magnitude qualifiers used by Hevelius. \label{t:magqual}}
\begin{tabular}{lcc|lcc}
.$n$+1 or .$n$+1. & + & 22 &   3.0.6  & V & H\,593\\
.$n$$-$1 or .$n$$-$1. & - & 8 & $n$ imo major & M &  H\,965\\
fere & f & 9 & non nisi tub.\ vis. & T & H\,956 \\
imo $n$+1 & I &  H\,88 & non nisi tubo visibilis & T & H\,1250 \\
imo $n$$-$1  & i & H\,530
\end{tabular}
\tablefoot{For each qualifier for magnitude $n$ we give
  our indication in the machine-readable \hevelius\ and the
  number of occurrences.  If it occurs only once, the H number is
  given. }
\end{table}

In some cases Rybka (1984) only provides an identification from
Flamsteed's (1725) edition of Hevelius' catalogue: these look like
Flamsteed numbers but have an H attached: e.g. 32H\,Cam. The stars in
Flamsteed's edition are in a different order than in \hevelius, but
otherwise have identical numbers (i.e.\ magnitudes and both equatorial
and ecliptic position for 1661.0). Thus 32H\,Cam is a self-reference,
and does not provide an identification. In some cases we have found
Hipparcos identifications for Flamsteeds Hevelius numbers via {\em
  SIMBAD}, but is 6 cases we have not, and we classify these as
unidentified in Rybka (1984).

In 32 cases our identification suggests an emendation to Rybka's
identification. These are discussed in Sect.\,\ref{s:rybkae}.  In 21
cases this emendation leads to a different corresponding Hipparcos
number, in 11 cases the correspondence is not affected.

\section{The machine-readable catalogue\label{s:machine}}

\begin{table*}
\caption{First lines of the machine-readable table \hevelius.\label{t:hevelius}}
\begin{tabular}{rrrrrrr@{}crrrrrrrrrrrrrrrr}
H & K\phantom{2} & C     &&  & OT &\multicolumn{2}{l}{$V_H$}&Z&G& M& S&  G& M& S& &  HIP\phantom{e}&I&B&R&$V$&
$\Delta\lambda$ & $\Delta\beta$ & $\Delta$(\arcmin)\\
\hline 
 1& 442& 1&=And& 1&  1& 2&&  1&09&36&36& 25&43&12&B&   677& 1&1&1& 2.1& $-$1.1& $-$2.1& 2.3\\
 2& 454& 1&=And& 2& 13& 2&&  1&25&43&00& 25&56&54&B&  5447& 1&1&1& 2.1& $-$2.2& $-$0.8& 2.2\\
 3& 457& 1&=And& 3& 16& 2&&  2&09&30&57& 27&47&13&B&  9640& 1&1&1& 2.1& $-$0.5& $-$0.5& 0.7\\
\multicolumn{2}{c}{\ldots} \\
23& 458& 1&=And&23& 17& 4&&  2&09&54&07& 36&48&27&B&  8068& 1&1&1& 4.0& $-$1.2& 0.7& 1.2\\
24&   0& 1&=And&24&  0& 6&&  1&06&55&45& 17&01&22&B&  1168& 1&0&1& 4.8& $-$3.5& 0.9& 3.4\\
25&   0& 1&=And&25&  0& 6&&  1&05&21&30& 23&06&26&B&118131& 1&0&1& 4.6& 0.6& 2.6& 2.7\\
29&   0& 1&=And&29&  0& 6&&  1&11&02&15& 48&37&00&B&114570& 1&0&1& 4.5& $-$0.8& $-$3.2& 3.2\\
30&   0& 1&=And&30&  0& 6&&  1&11&54&10& 47&19&00&B&115152& 1&0&3& 5.4& $-$10.0& $-$3.2& 7.5\\
31&   0& 1&=And&31&  0& 6&&  1&14&54&53& 17&55&57&B&  3269& 3&0&1& 6.1& $-$9.5& 32.5& 33.7\\
32&   0& 1&=And&32&  0& 9&&  1&23&10&10& 33&23&04&B&     0& 1&0&1& 3.4& $-$1.6& $-$3.5& 3.8\\
44&   0& 1&=And&44&  0& 6&+& 2&12&48&01& 25&31&33&B& 11090& 1&0&1& 5.8& $-$0.4& 4.4& 4.4\\
\end{tabular}
\tablefoot{For  explanation of the columns see Sect.\,\ref{s:machine}.}
\end{table*}

The machine-readable table \hevelius\ contains the following
information (see Table\,\ref{t:hevelius}).  The first column gives the
sequence number H.  The second column the sequence number K of the
corresponding star in \keplere\ (K$\leq$1004) or in {\em Secunda
  Classis} (1005$\leq$K$<$2000): these numbers are used only when
Hevelius gives a Brahe sequence number (OT) or an ecliptic position
from Kepler, that provides an unambiguous correspondence. We give the
first star in {\em Secunda Classis} the number K\,1005, and continue
the numbering for the following stars in order of appearance 
(see Verbunt \&\ Van Gent 2010, Paper III, in preparation; also for the exception in
Sagittarius). In some cases an entry in \hevelius\ is identified
with the same {\em Hipparcos} number as an entry in \keplere, even
though Hevelius does not indicate a Brahe sequence number or
position. We consider the correspondence in these cases probable but
not secure, and indicate them with 2000+K in column two. For example,
H\,201 is identified with HIP\,23522 as is K\,11 in \keplere; the
second column in the machine-readable table \hevelius\ gives 2011.

The third and fourth columns indicate the constellation: indicated
with its sequence number in the catalogue and with the modern
abbreviation, as listed in Table\,\ref{t:const}. For some
constellations no longer in use (Antinous, (Argo) Navis, and
Triangulum Minus) we introduce an abbreviation.  Column 5 gives the
sequence number of the entry within the constellation in
\hevelius.

Columns 6--16 copy information from the original catalogue.  Column 6
gives the {\em Ordo Tychonis} OT (see Sect.\,\ref{s:description} and
Sect.\,\ref{s:remarks} for details).  Column 7 gives the magnitude:
when the entry is indicated as {\em non nisi tubo visibilis} (not
visible unless with a tube [i.e.\ telescope]) by Hevelius, we give it
magnitude 8; when indicated nebulous by Hevelius, we give it magnitude
9.  Column 8 give magnitude qualifiers, as detailed in
Table\,\ref{t:magqual}. Columns 9-12 give the ecliptic longitude,
($Z$, $G$, $M$ and $S$) and columns 13-16 the ecliptic latitutde ($G$,
$M$, $S$ and A/B). For this notation, see Sect.\,\ref{s:description}.

Columns 17--24 provide additional information from our analysis.
Column 17 gives the Hipparcos number of our identification, and column
18 a flag indicating the quality of the identification, as explained in
Sect.\,\ref{s:identification} (see also Table\,5 in Paper\,I).

Column 19 flags the identification of the corresponding entry in
\keplere: a 0 if that entry is not identified, a 1 if its
identification is the same as the one here in column\,17, a 2 if its
identification is to the other of a pair of possible identifications,
and a 3 for a different identification. Column 20 flags the
identifications by Rybka (1984), with the same notation.  Column
21 gives the visual (Johnson) magnitude of the {\em Hipparcos} object
given in column 17. Columns 22, 23 give the difference in longitude
and latitude between the correct position (based on information from
the {\em Hipparcos Catalogue}) and the catalogue entry. Note that the
tabulation in Hevelius gives minutes and seconds, which we convert to
decimal minutes $M_\mathrm{H}$ to compute columns 22 and 23.  If the
catalogue entry for minutes as computed from the position and proper
motion in {\em Hipparcos Catalogue} is $M_\mathrm{HIP}$, and
$M_\mathrm{H}$ is the value from the Hevelius Catalogue, then columns
22 and 23 give $M_\mathrm{HIP}-M_\mathrm{H}$. Column 24 gives the
difference between correct and tabulated position in \arcmin.

\section{Analysis and discussion}

\begin{table}
\caption{Frequency of flags B of identifications of corresponding
 entries in \keplere\  as a function of our flags I. \label{t:kepler}}
\begin{tabular}{l|rrrrr}
I\verb+\+B & 0 & 1 & 2 & 3 & all \\
\hline 
1 &   2 &  873 &   2 &  15 &  892 \\
2 &   0 &    5 &   0 &   0 &    5 \\
3 &   0 &    0 &   0 &   0 &    0 \\
4 &   0 &    3 &   1 &   1 &    5 \\
5 &   2 &    0 &   0 &   0 &    2 \\
6 &   0 &    1 &   0 &   0 &    1 \\
all & 4 &  882 &   3 &  16 &  905
\end{tabular}
\end{table}

\subsection{Comparison with Brahe\label{s:brahe}}

915 entries in \hevelius\ can be matched unambigously with an entry in
\keplere: 911 (including the repeat entry H\,793) through their {\em
  Ordo Tychonis}, and 4 without OT through the position according to
Tycho as given by Hevelius. 10 of these 914 independent entries have
no position by Hevelius, among them SN\,1572.  Table\,\ref{t:kepler}
compares our identifications of the entries common to \hevelius\ and
\keplere: in most cases we found the same identification for both
catalogues. The 16 exceptions arise because Hevelius gives a position
very different from the position in \keplere: striking examples are
H\,14/K\,463, H\,1005/K\,961 and H\,1099/K\,441, as detailed in
Sect.\,\ref{s:notes}.

90 entries in \keplere, among which 12 which we were unable to
identify with an Hipparcos star, have no explicit counterpart in
\hevelius.  The 12 unidentified stars probably have a wrong position
in \keplere, which would explain why Hevelius found no star at that
position. For the others we checked whether their Hipparcos
identifications occur also in \hevelius. This is the case for 57. It
appears likely that Hevelius and Brahe observed the same star in these
cases, but we cannot exclude an occasional chance coincidence.  Many
of the remaining 21 entries in \keplere\ that are not matched with an
entry in \hevelius\ have a very large position error $\Delta$ in \keplere.
The unmatched entries also include 4 stars in Argo and all 4 stars
in Centaurus.

Some remarkable features of entries in \keplere\ are present in
\hevelius\ as well. We mention in particular the three stars
in Capricornus which are denoted `nebulous' both in Hevelius and Brahe,
and H\,1188/K\,801 which both in \kepler\ and \hevelius\ are given a B
for northern latitude, whereas the correct latitude is S for southern
(see Fig.\,C.44 in Paper\,I).

\begin{table}
\caption{Frequency of flags R of identifications by Rybka 
as a function of our flags I. \label{t:rybka}}
\begin{tabular}{l|rrrrr|rrrrr}
        & \multicolumn{5}{c}{in KeplerE} & \multicolumn{5}{c}{full catalogue} \\
I\verb+\+R & 0 & 1 & 2 & 3 & all  & 0 & 1 & 2 & 3 & all  \\
\hline
1 &   5 &  880 &   3 &   4 &  892 & 20 & 1426 &   4 &  29 & 1479 \\
2 &   0 &    5 &   0 &   0 &    5 &  0 &   10 &   0 &   1 &   11 \\
3 &   0 &    0 &   0 &   0 &    0 &  2 &    9 &   0 &   4 &   15 \\
4 &   0 &    2 &   1 &   2 &    5 &  1 &    5 &   2 &   4 &   12 \\
5 &   0 &    0 &   0 &   2 &    2 &  7 &    0 &   0 &   9 &   16 \\
6 &   0 &    1 &   0 &   0 &    1 &  1 &   11 &   0 &   1 &   13 \\
all & 5 &  888 &   4 &   8 &  905 & 31 & 1461 &   6 &  48 & 1546 
\end{tabular}
\tablefoot{Left: for stars in \keplere, right: for all entries in \hevelius.}
\end{table}

\subsection{Comparison with Rybka (1984)}

In Table\,\ref{t:rybka} we compare the identifications as found by us
with those given by Rybka (1984), separately for the stars in
\keplere\ and for all stars in \hevelius. In most cases the
identifications are identical, but there are differences. We have
identified 24 stars (among which 1 repeated entry) that Rybka could
not identify. In 6 cases where two stars are plausible
counterparts we choose the stars that Rybka did not choose. In 48
cases (among which 1 repeated entry) we do not agree with the
identification in Rybka (1984); this includes 9 stars which we cannot
identify. This number does {\em not} include the 21 cases where our
emendation to Rybka leads to a different Hipparcos identification (see
Sect.\,\ref{s:rybkae}). In some cases the identification given by
Rybka (1984) has such a large positional offset, or is so faint,
that we consider our rejection secure; in other cases we choose
another closer and/or brighter star as a more plausible counterpart.
Details may be found in Sect.\,\ref{s:notes}. 

\subsection{Accuracy}

\begin{figure}
\includegraphics[angle=270,width=\columnwidth]{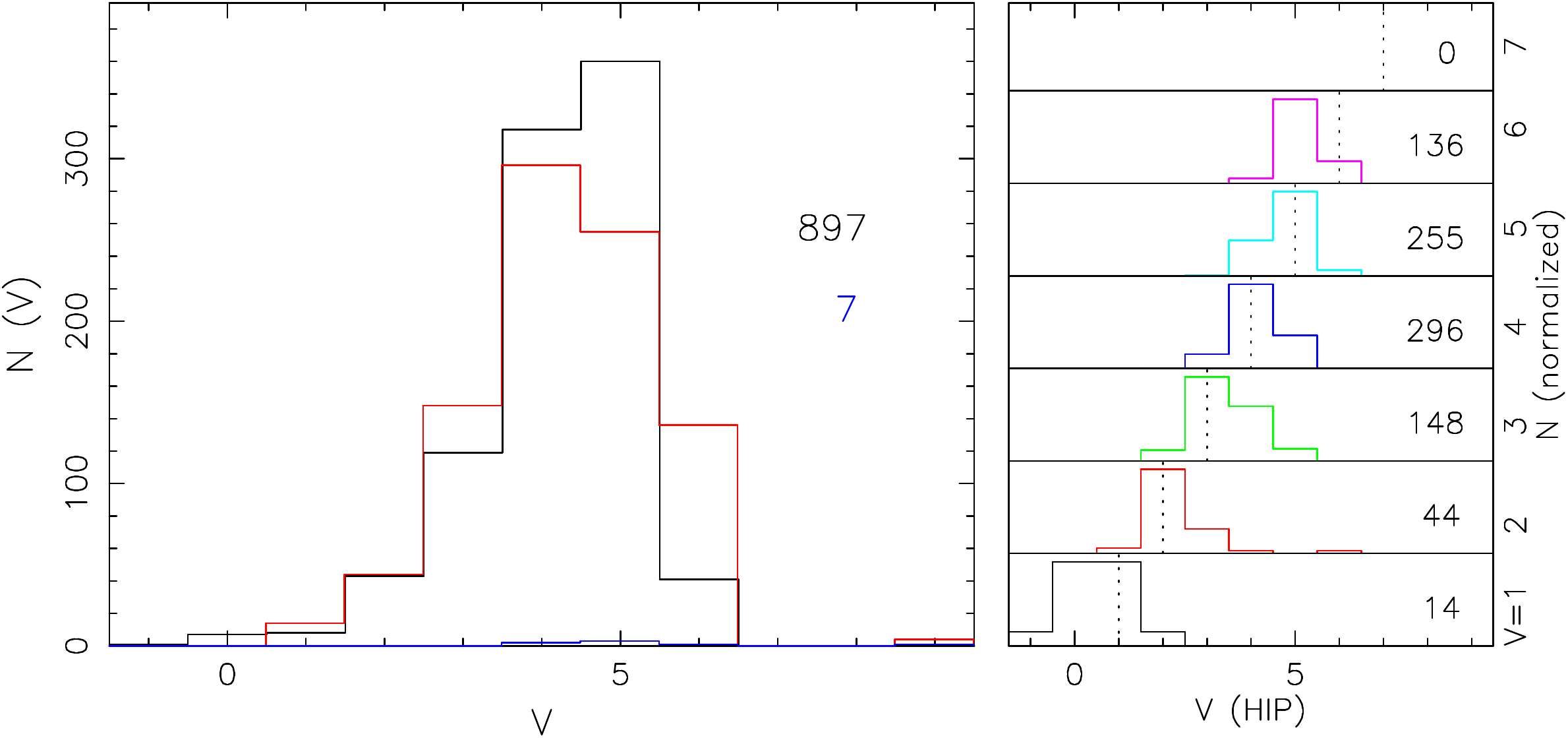}

\includegraphics[angle=270,width=\columnwidth]{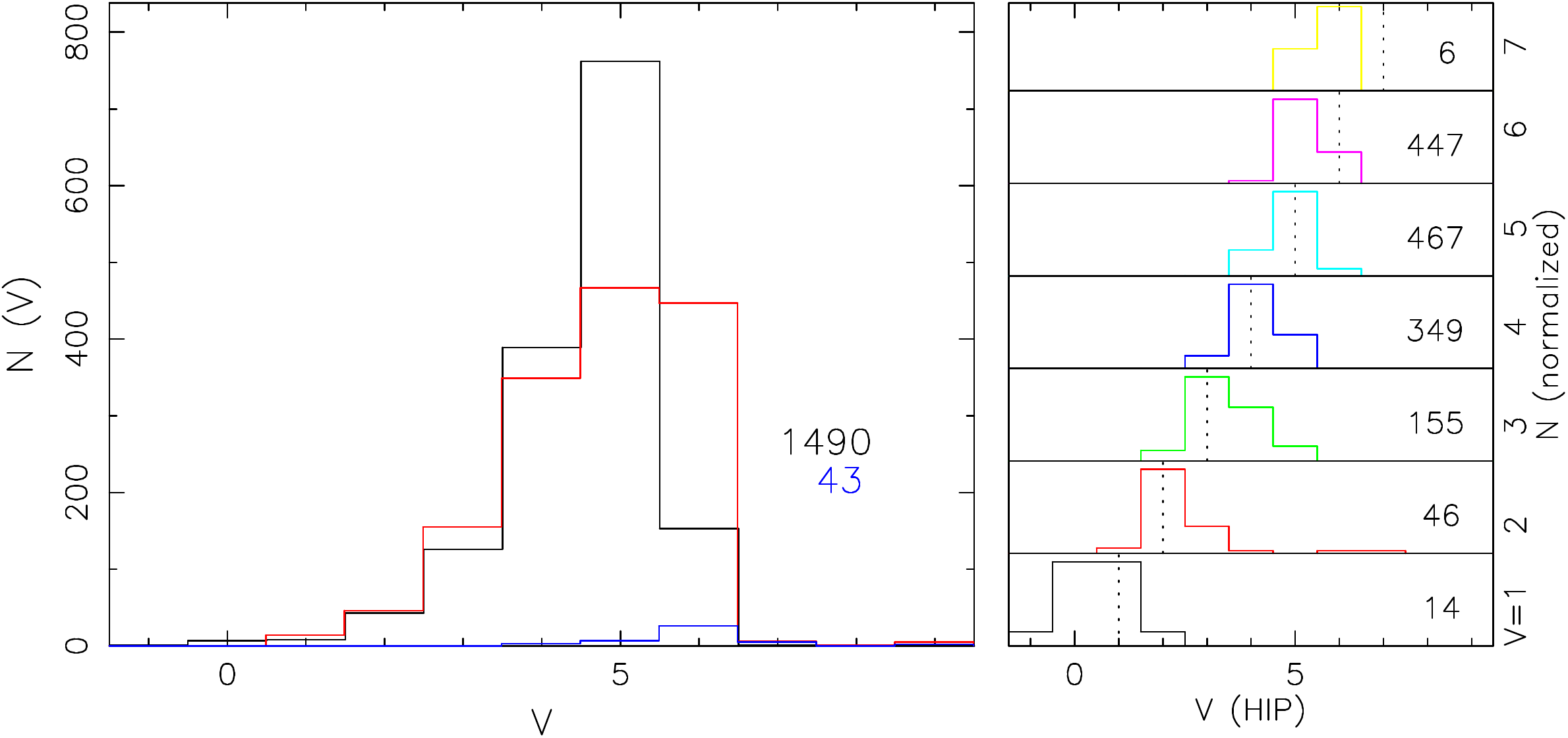}

\caption{Distribution of the magnitudes for all stars in \hevelius\
 (below), and for only those stars that have a counterpart in 
  \keplere\ (above). In the large frames the histograms indicate the
  magnitudes according to Hevelius for stars which we have securely
  identified (red; flags 1-2 ) or not
  securely identified (blue, flags 3-5), and the magnitudes from the
  {\em Hipparcos} catalogue for securely identified stars (black). The
  numbers of securely and not-securely identified stars are indicated.
  The small frames give the {\em Hipparcos} magnitude distributions
  for securely identified stars for each magnitude according to Hevelius
  separately.  The number of securely identified stars at each (Hevelius)
  magnitude is indicated.
 \label{f:magnitudes}}
\end{figure}

Table\,\ref{t:rybka} shows that there are 16 stars, i.e.\ one percent
of the total, in \hevelius\ that we are not able to identify. We do
not count in these the empty entries.  As in the case of the Brahe
catalogue, we would have to accept fainter counterparts or larger
position errors to identify these; in both cases the probability of
chance coincidences would increase. Other entries in \hevelius\ which
do not have an identification in the {\em Hipparcos Catalogue} are
H\,32 (= M\,31), H\,259 (= Praesepe) and H\,1540 (= Nova Vul 1670; see
Sect.\,\ref{s:notes}).

\begin{figure}
\includegraphics[angle=270,width=0.48\columnwidth]{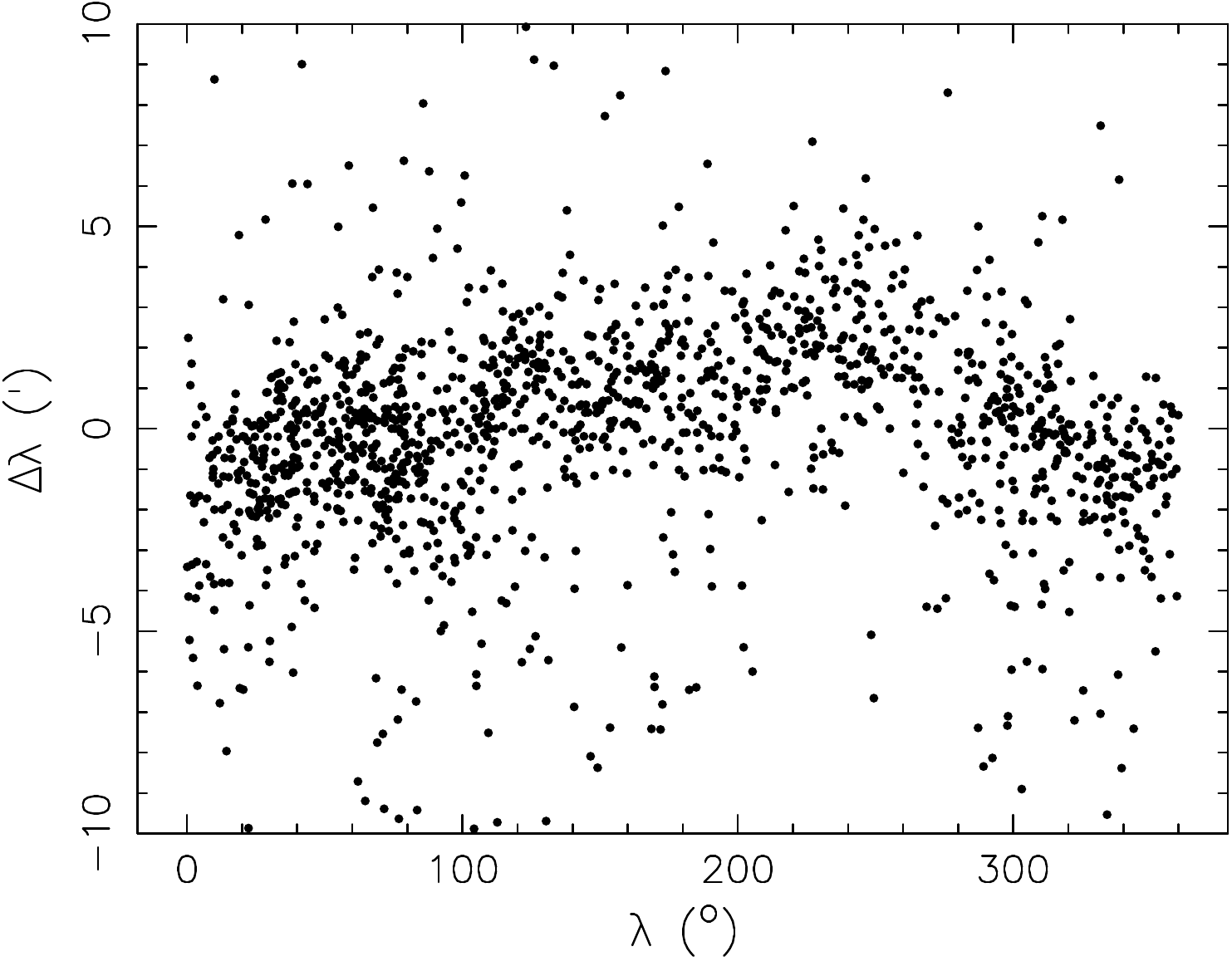}
\includegraphics[angle=270,width=0.48\columnwidth]{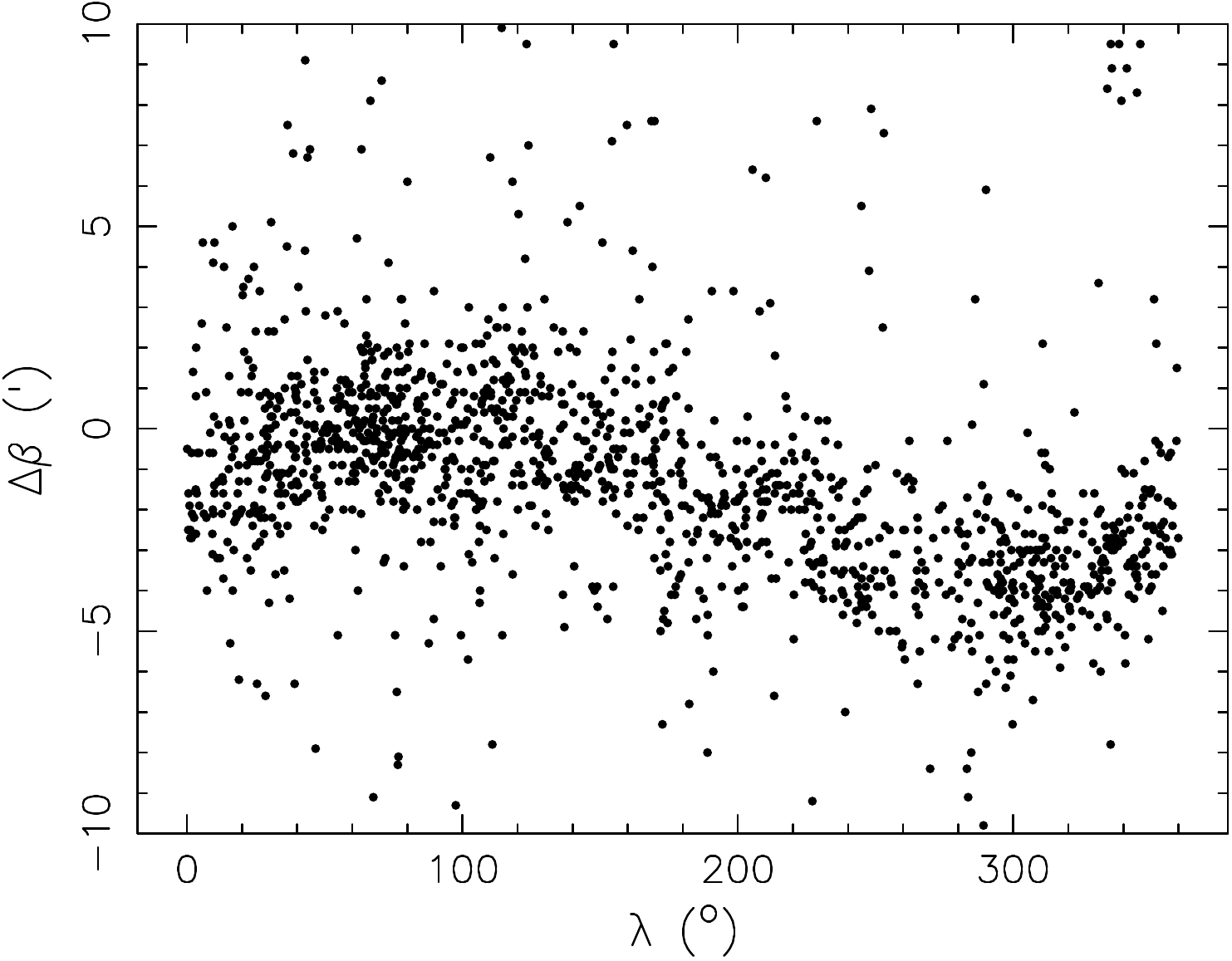}

\includegraphics[angle=270,width=\columnwidth]{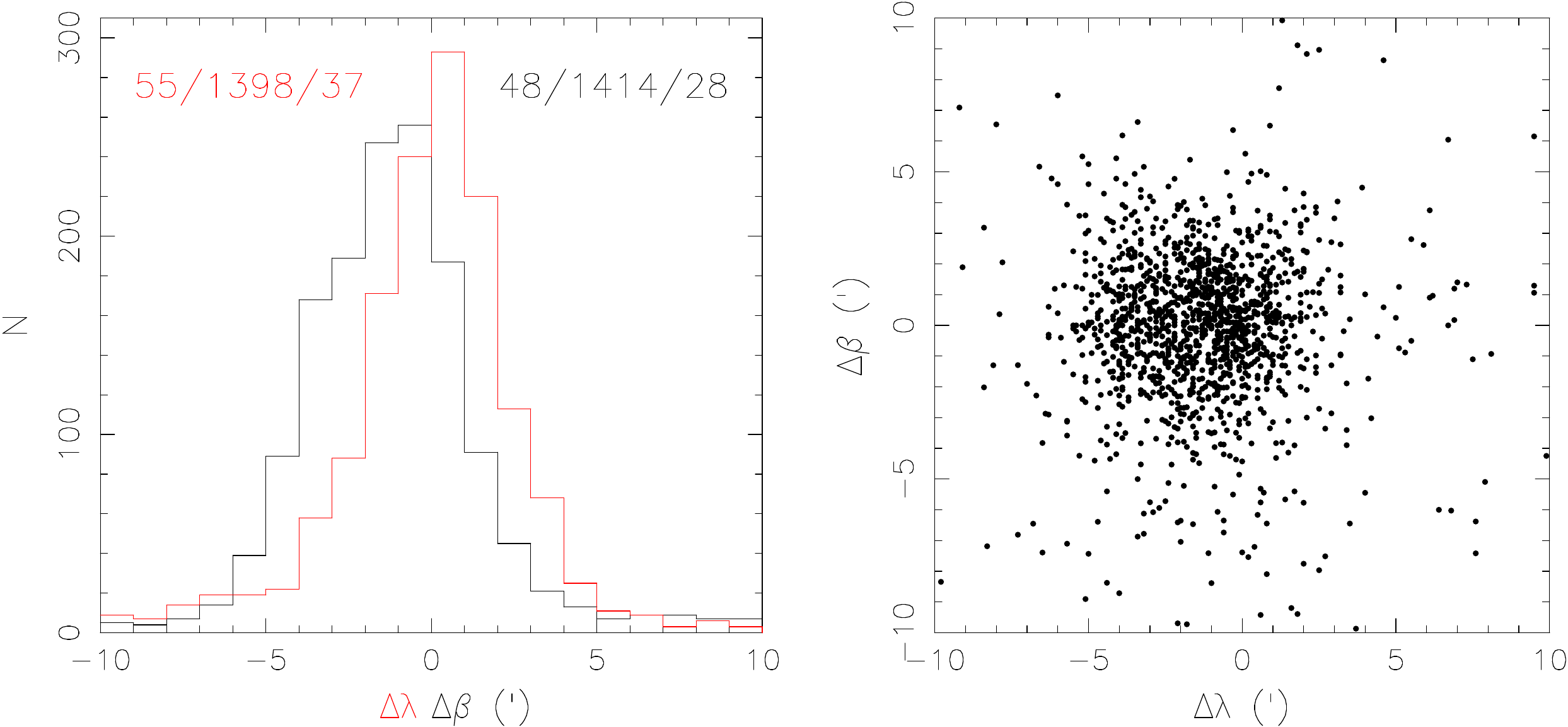}

\caption{Above: Correlations of the differences in longitude
  $\Delta\lambda\equiv(\lambda_\mathrm{HIP}-\lambda)\cos\beta$ and
  latitude $\Delta\beta\equiv\beta_\mathrm{HIP}-\beta$ of the entries
  in \hevelius\ and their secure {\em Hipparcos} counterparts
  (converted to 1661) with longitude. Below left: Distributions of
  $\Delta\lambda$ and $\Delta\beta$. The numbers of sources with
  $\Delta\lambda,\Delta\beta<-10\arcmin$, of sources included in the
  histogram ($-10\arcmin<\Delta\lambda,\Delta\beta<10\arcmin$), and of
  sources with $\Delta\lambda,\Delta\beta>10\arcmin$ are
  indicated. Below right: correlation between $\Delta\lambda$ and
  $\Delta\beta$. \label{f:dlongdlat}}
\end{figure}

Figure\,\ref{f:magnitudes} illustrates that the magnitudes assigned by
Hevelius correlate well with those of their counterparts in the {\em
  Hipparcos Catalogue}. Only the higest magnitudes assigned by
Hevelius, 6 and 7, tend to be too high. 

\begin{figure}
\includegraphics[angle=270,width=\columnwidth]{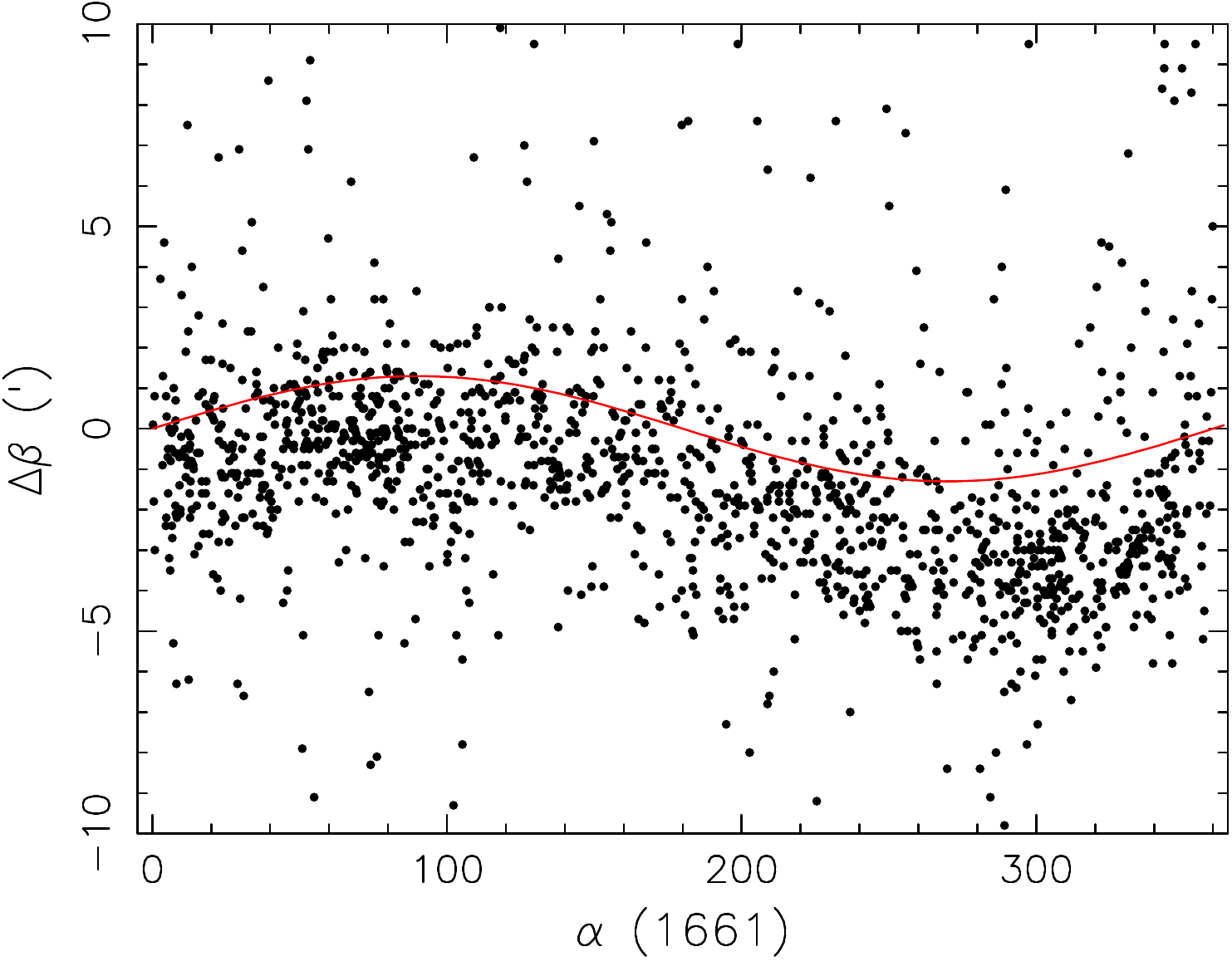}

\caption{Distribution of the errors in latitude in \hevelius\ as a function of
right ascension, together with the r.h.s.\ of Eq.\,\ref{e:alphadb}.
 \label{f:alphahdb}}
\end{figure}

In Figure\,\ref{f:dlongdlat} we show the error distributions
separately for the longitudes and latitudes, as well as their
correlation. The correlation distribution is roughly spherical,
i.e.\ the errors in longitude and latitude are mostly independent.
Gaussians that fit the central regions ($-$5\arcmin,+5\arcmin) of the
distributions of $\Delta\lambda$ and $\Delta\beta$ both have
widths $\sigma\simeq2$\arcmin; both predict fewer points at errors
larger than 5\arcmin\ than observed. 
The numbers of errors with absolute values larger than 10\arcmin\ 
correspond to less than 10\%\ of the number of identified entries,
a similar percentage as in \keplere.
The widths of the peak of the error distributions (near 2\arcmin) and
the fraction of larger errors are thus similar in \hevelius\ to those
in \keplere, which is an impressive achievement since the number of
stars has increased by more than 50\%\, mostly at the fainter magnitudes
5 and 6.

Figure\,\ref{f:dlongdlat} further shows that the errors in longitude
$\Delta\lambda$ increase with the distance to the zero point \aries;
and that the errors in latitude $\Delta\beta$ have a roughly
sinusoidal dependence on longitude. The latter dependence may be
explained by an error in the value of the obliquity. Hevelius measured
the obliquity in several years, and found values around
$\epsilon_H$=23\fdg506 (Rybka 1984, p.37); according to modern theory
the obliquity in 1661 was $\epsilon$=23\fdg483. For small declinations
$\delta$, the resulting error in latitude
$\Delta\beta\equiv\beta-\beta_H$ due to the error
$\Delta\epsilon\equiv\epsilon-\epsilon_H$ after converting equatorial
to ecliptic coordinates may be written
\begin{equation}
\cos\beta\Delta\beta \simeq -\sin\alpha\cos\epsilon\Delta\epsilon\simeq
1\farcm3\sin\alpha
\label{e:alphadb}\end{equation}
The observed relation between $\Delta\beta$ and $\alpha$ is shown in
Fig.\,\ref{f:alphahdb} together with the curve 1\farcm3$\sin\alpha$
which roughly matches the phase and amplitude of the
$\alpha$-dependence of $\Delta\beta$.

\begin{figure}
\includegraphics[angle=270,width=\columnwidth]{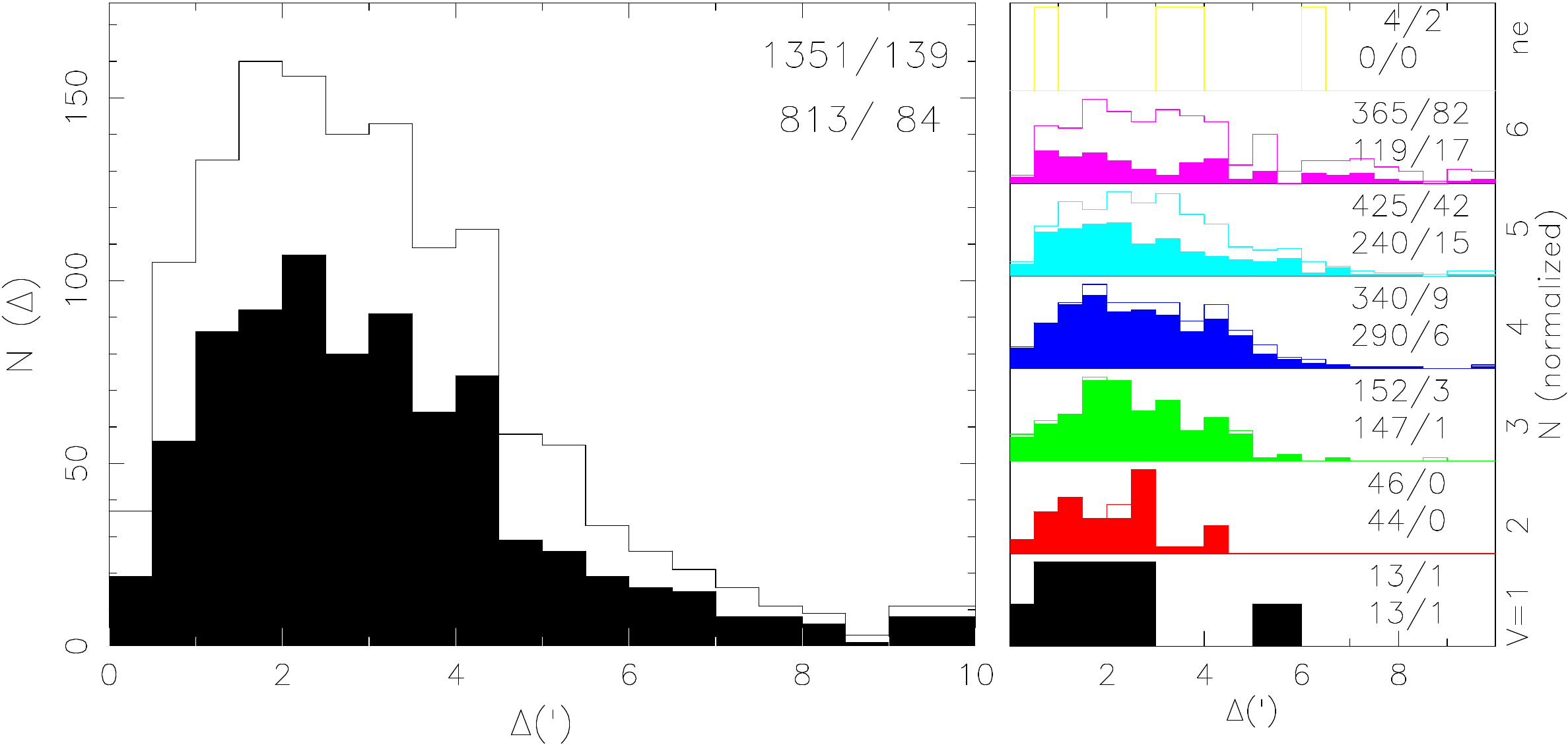}

\caption{Distribution of the position errors $\Delta$ for all stars
in \hevelius\ (open histogram) and for only those stars that have
a counterpart in \keplere\ (solid histograms) for all securely identified stars
(left) and for the securely identified stars at each Hevelius magnitude
separately (right). The numbers indicate the number of stars included
in the plot (i.e.\ with $\Delta<10\arcmin$) and those excluded 
($\Delta\ge10\arcmin$) \label{f:delta}}
\end{figure}

\begin{figure}
\includegraphics[angle=270,width=\columnwidth]{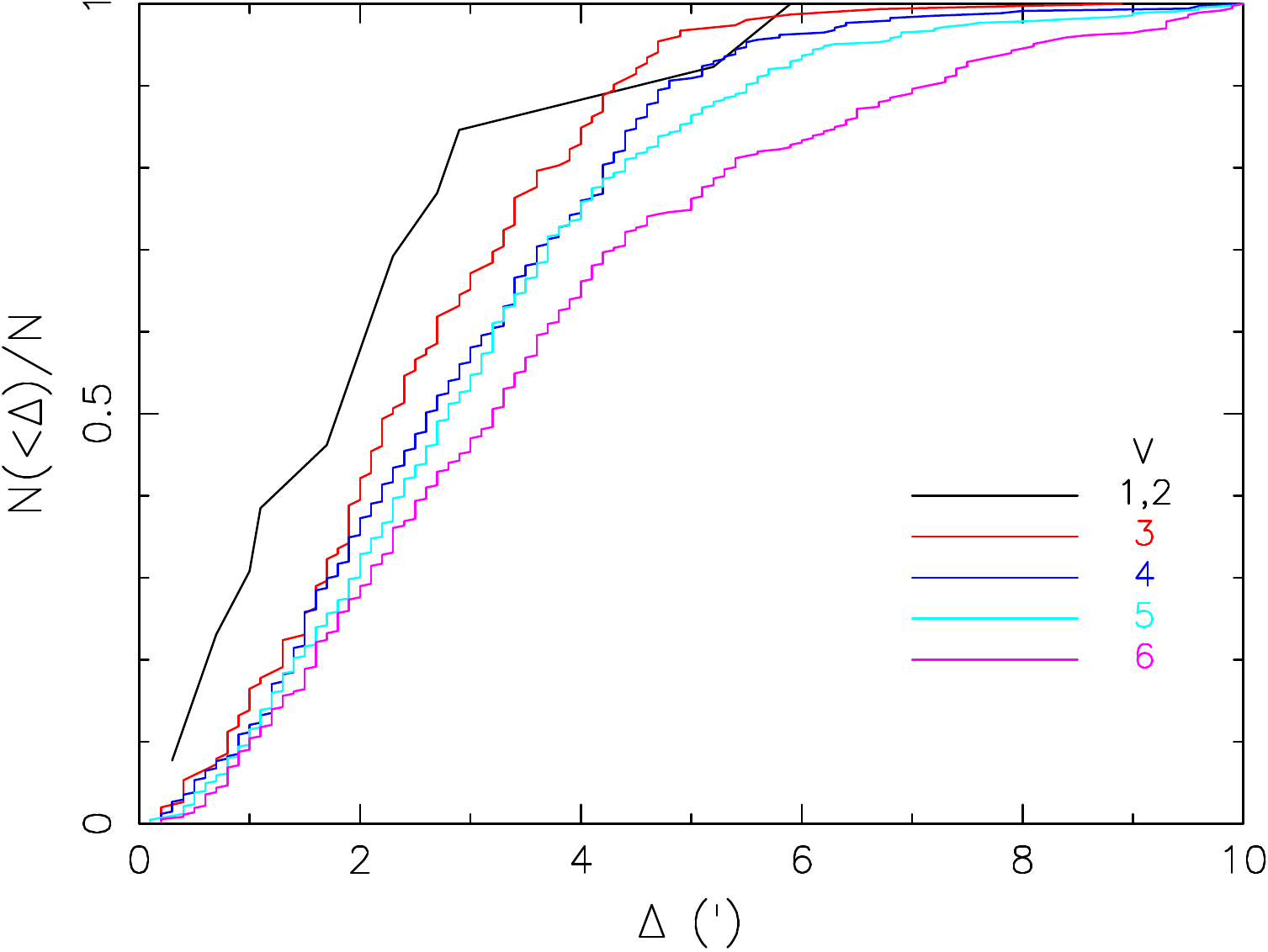}

\caption{Cumulative distribution of the position errors $\Delta$ in
  \hevelius\ as a function of Hevelius magnitude, showing a systematic
  increase in median error with magnitude.\label{f:ksplots}}
\end{figure}

The average offset of longitude is virtually zero; the latitudes have
an average offset of -1\farcm4. This average offset may be due to an
underestimate by Hevelius of refraction. The distribution of the total
position errors $\Delta$ in \hevelius\ is shown in
Fig.\,\ref{f:delta}.  This distribution peaks roughly at the value of
the width 2\arcmin\ of the separate distributions in $\Delta\lambda$,
$\Delta\beta$, as expected (see explanation in Paper\,I).  The number
of stars with large position errors is markedly smaller in
\hevelius\ than in \keplere. In particular, the number of stars with
position errors larger than a degree is 21 (on a total of 1517
identified entries) in \hevelius\ as compared to 47 (on a total of 977
identified entries) in \keplere. Similarly, the number of unidentified
stars is 16 (of 1533 independent entries) in \hevelius\ and 14 (of 992
independent entries) in \keplere.  It may be concluded that the
overall accuracy of the star catalogue of Hevelius is better than that
of the star catalogue of Brahe/Kepler.

\begin{figure}
\includegraphics[angle=270,width=\columnwidth]{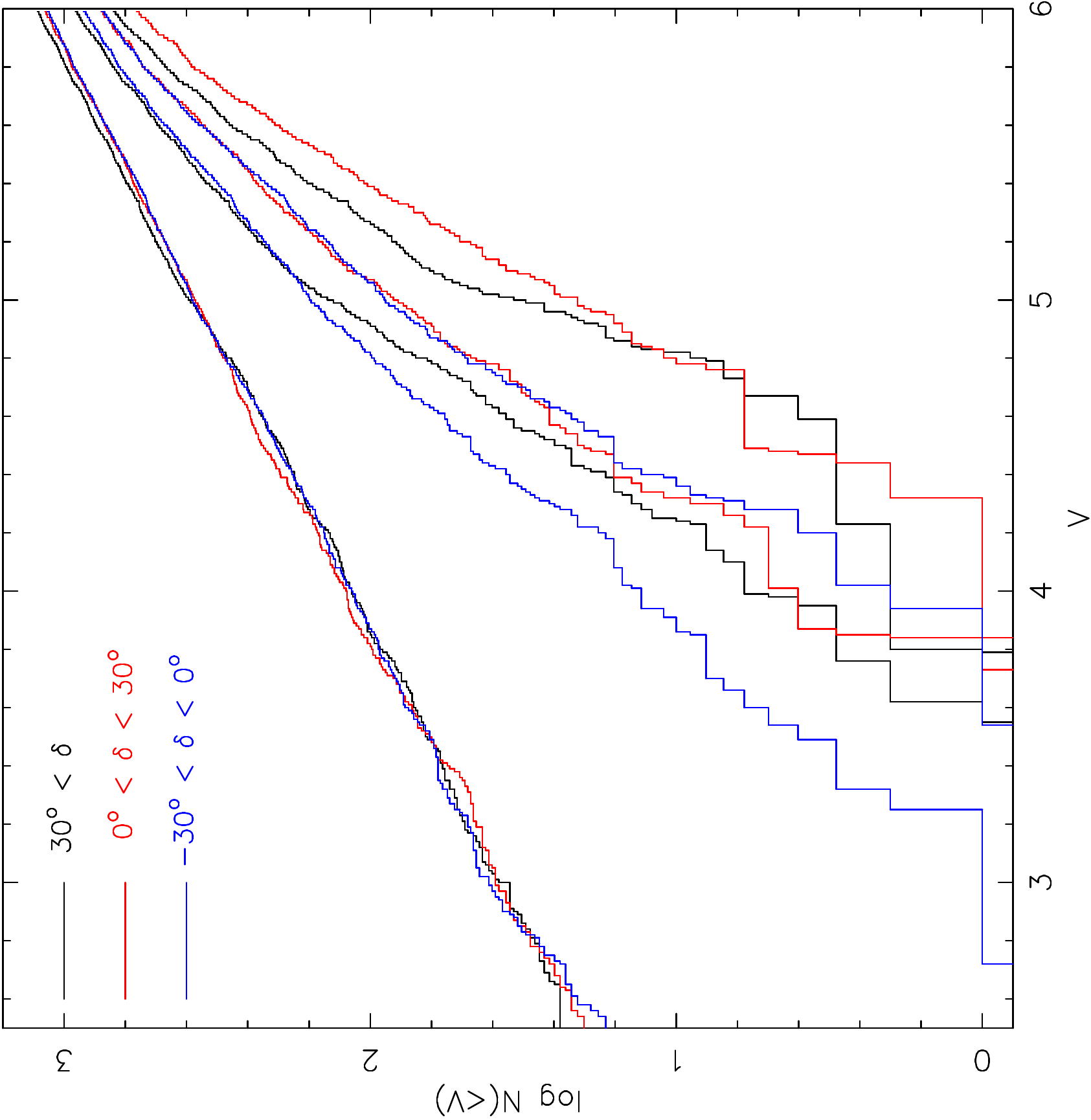}

\caption{Completeness of \hevelius\ and \keplere\ as a function of
  magnitude and declination, as illustrated by cumulative magnitude
  distributions. For each range of declination (equinox 1631) the top
  curve shows the magnitudes for all stars in the {\em Hipparcos
  Catalogue}, the middle curve the Hipparcos stars {\em not} in \keplere,
  and the lower curve the Hipparcos stars {\em not} in \hevelius.
  \label{f:complet}}
\end{figure}

In Fig.\,\ref{f:ksplots} we show the cumulative error distributions
for each Hevelius magnitude separately, taking magnitudes 1 and 2
together, and limiting the distributions to $\Delta<10$\arcmin. It is
seen that the median error increases slowly but systematically with
magnitude.

\subsection{New and old stars: completeness}

In Fig.\,\ref{f:complet} we investigate the completeness of
\hevelius\ and \keplere\ as function of magnitude, for three
declination ranges. For this purpose we select from \hevelius\ and
\keplere\ only those entries which we have identified, and which are
not repeat entries, i.e.\ entries with I=1-4.  For selecting the
Hipparcos stars in the latitude ranges we convert their positions to
an equinox halfway between Brahe and Hevelius, viz.\ 1631.0. At
magnitudes $V$$<$4 there are 348 stars from the {\em Hipparcos
  Catalogue} with $\delta>-30^\circ$, of which 5 are absent from
\hevelius\, and 23 from \keplere\ (of which 13 with
$\delta<0^\circ$). At magnitudes $V$$<$5 there are 1138 Hipparcos
stars with $\delta>-30^\circ$, of which \hevelius\ misses 141 stars
(89 with $\delta<0^\circ$) and \keplere\ 389 (156 with
$\delta<0^\circ$). Finally, about 3500 Hipparcos stars with $V$$<$6
have $\delta>-30^\circ$, and of these some 2000 are absent from
\hevelius\ and 2500 from \keplere, which is just another way of saying
that \hevelius\ and \keplere\ contain about 1500 and 1000 stars
visible to the naked eye, respectively. It may be noted here that the
latitude of Gdansk is about 1\fdg5 further south than that of Hven.
Nonetheless, as shown in Fig.\,\ref{f:complet} \keplere\ is more
incomplete already at brighter magnitudes, also in the northern parts
of the sky.

How many new stars did Hevelius observe?
In the manuscript of the catalogue, a note dated 1681 March 31 states
that 946 stars of Tycho and 617 new stars were observed (Volkoff et
al.\ 1971, p.72).  This gives a total of 1563, very close to the total
of 1564 entries given in Table\,\ref{t:const}, but spuriously so since
Hevelius gives no own measurements for 18 of the 1564 entries.
Hevelius indicates, through an OT number or a position from Tycho, for
915 entries that they are from \kepler\ (Table\,\ref{t:const}).  For
905 of these \hevelius\ gives his own measurements (see
Sect.\,\ref{s:brahe}).  To obtain the higher number of 946 stars from
the note, we have two options. One is to add the 28 stars from the
{\em Secunda Classis}. (These were also measured by Brahe, according to
Kepler, albeit with less accuracy.) This option leaves us with too
small a number. The other option is to add the 47 entries for which
our identification corresponds to an identification in \keplere.  This
would imply that Hevelius was aware that more stars from his catalogue
corresponded to stars in \kepler\ than the 915 entries marked by him
as such through OT or Tycho position.

If we subtract from the total number of 1564 entries in \hevelius\ all
962 that have a counterpart in \keplere\ and further subtract the 28
entries that have a counterpart in {\em Secunda Classis}, we find a number
of 574 entries first measured by Hevelius. 573 of these are stars, the
other one is M\,31.

Cerberus, Lacerta, Scutum, Sextans and Triangulum Minus, the truly new
constellations by Hevelius, contain a total of 36 stars, of which only
one possibly corresponds to a star in \keplere\ (H\,1328 / K959: the
position error of K\,959 is 2\fdg5, so a chance coincidence is
possible). Monoceros and Camelopardalis, two constellations retained
by Hevelius from Plancius, contain twelve and up to fourteen stars
from \keplere, respectively (Table\,\ref{t:const}). The four
constellations fashioned by Hevelius from two constellations by
Plancius contain up to sixteen stars from \keplere. We use the
qualification `up to' for the stars listed under $N_K$ in
Table\,\ref{t:const} because some of the correspondences between
\hevelius\ and \keplere\ may be chance coincidences. The six
constellations retained of refashioned from Plancius by Hevelius
contain 138 stars, so even accepting all 42 correspondences as real,
we still find that a large majority of stars in these constellations
was first observed by Hevelius.

\begin{acknowledgements}
This research has made use of the SIMBAD database, operated at CDS,
Strasbourg, France.  We thank Dr. Jaros\l aw W\l odarczyk for
providing us with a copy of Rybka (1984), and Oliwia Madej for help
in reading parts of it.  This research is supported
by the Netherlands Organisation for Scientific Research under grant
614.000.425.
\end{acknowledgements}


\begin{appendix}
\section{Annotations and Emendations \label{s:emendations}}

\subsection{Annotations to Hevelius \label{s:remarks}}

In 28 cases the {\em Ordo Tychonis} given in \hevelius\ is larger than
the number of stars in the corresponding constellation in \keplere,
and the OT actually refers to a star in {\em Secundis Classis}.  Stars
in {\em Secunda Classis} are usually given OTs that follow by
continuation of the numbering from the last star in the corresponding
constellation in \keplere.  We identify entries from \hevelius\ with
stars from \keplere\ or {\em Secunda Classis} when the positions given
by Hevelius as due to Tycho have a longitude higher by 51\arcmin\ (the
correction applied by Hevelius for precession from 1601 to 1661) and
the same latitude as in Brahe's catalogue, accepting round-off errors
of 0.5 in M (minute) for each coordinate.
Such is the case for: H\,296, H\,298, H\,299, H\,302 (CMa); H\,716 --
H\,726 (Eri); H\,834 (Hya); H\,1218 -- H\,1221, H\,1229 -- H\,1232
(Sgr); H\,1275 -- H\,1278 (Oph). For two stars in the Pleiades, see
below.

In 21 cases the {\em Ordo Tychonis} given in \hevelius\ refers to a 
different constellation in \keplere. These are
H\,258 (Cnc, OT refers to Hya);
H\,273 (Cnc, OT refers to CMi);
H\,993 -- H\,1002 (Mon, OTs refer to Hya);
H\,1202 (Psc, OT refers to And);
H\,1241 -- H\,1243 (Sco, OTs refer to Lib);
H\,1314, H\,1317 -- H\,1320 (Ser, OTs refer to Oph).

In 9 cases a position from Tycho is given by Hevelius, but no OT
number. In 4 cases the star is in \keplere, viz.\ H\,415 (Cas,
SN\,1572) = K\,267; H\,521 (Cet) = K\,919 (Eri); H\,1004, H\,1005
(Mon) = K\,958, K\,961 (Arg; with an emendation, see
Sect.\,\ref{s:notes}) \\ 
\noindent 
and in 5 cases the star is in {\em Secunda Classis}: H\,777 (Her, SC:
Oph); H\,1003 (Mon, SC:Hya); H\,1213, H\,1216 (Sgr, SC: Sgr); H\,1289
(Oph, SC: Oph).  In one case, H\,1016 (in Arg), a Tycho position by
Hevelius ($\lambda$ = \leo\ 01\,51, 49\,00\,A) does not match a
position either in \keplere, or in {\em Secunda Classis}, although it
is close to two stars in Argo in {\em Secunda Classis} (viz.\ K\,1167
and K\,1168).

In 5 cases a magnitude from Tycho is given by Hevelius, but
no OT number or Kepler position:
H\,321, H\,322 (in RS CVn, see Table\,\ref{t:doubles};
H\,906, H\,909, H\,910 (in LMi)

In 15 cases Hevelius does not mark the magnitude IH, thereby
indicating that the star has been previously catalogued, but does not
give a Tycho magnitude or position. These are H\,108 (Aqr), H\,520
(Cet; position from Riccioli), H\,611 and H\,613 -- H\,617 (Cyg),
H\,835 (Hya), H\,911, H\,913, H\,914, H\,916, H\,918, H\,919 (in LMi),
H\,1467, H\,1468 (UMa).

\begin{figure}
\centerline{\includegraphics[angle=270,width=\columnwidth]{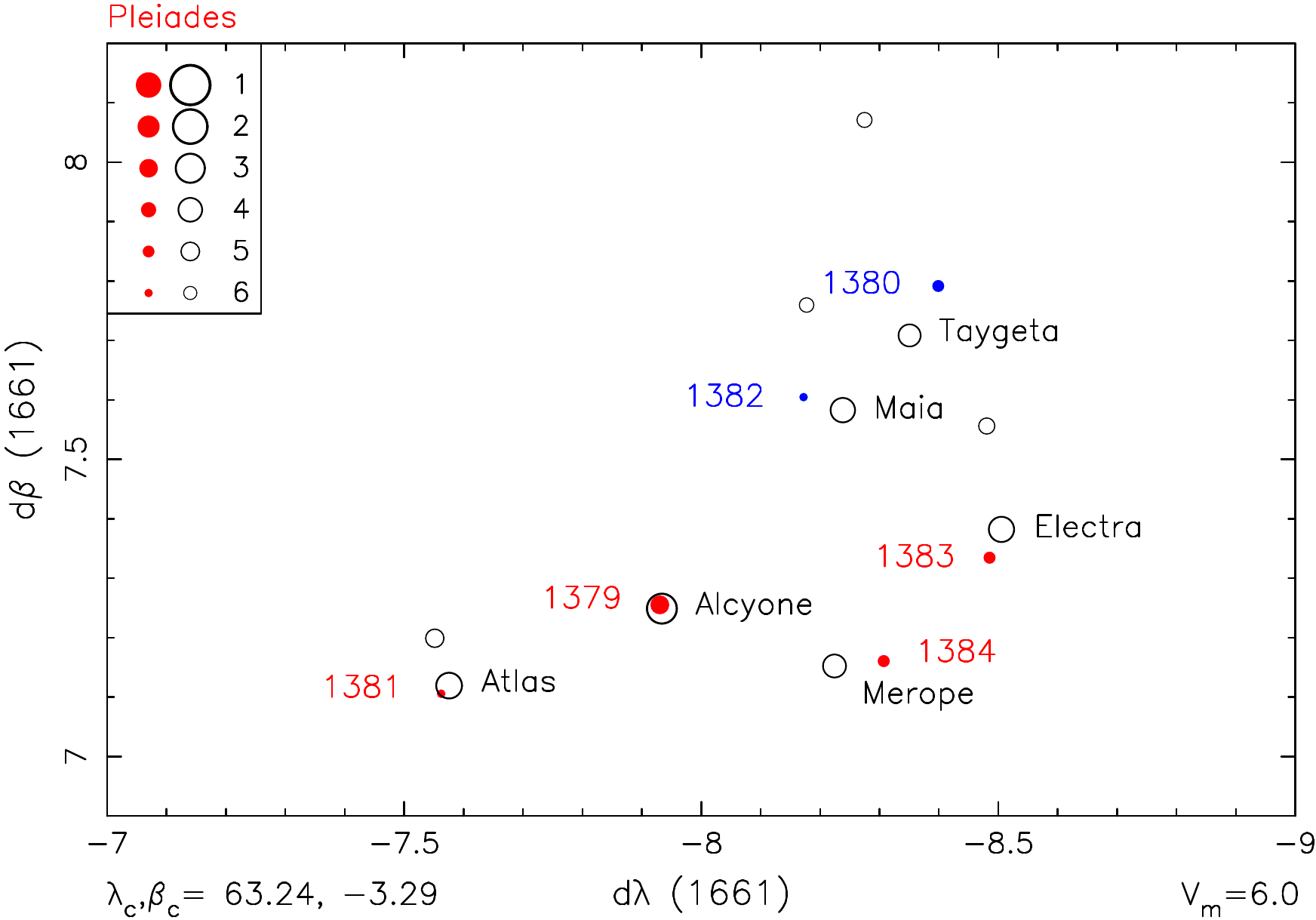}}
\caption{Pleiades. In red H numbers of stars also present in \keplere, in blue
stars added by Hevelius. 
\label{f:pleiades}}
\end{figure}

\hevelius\ lists six stars in the Pleiades (see
Fig.\,\ref{f:pleiades}). These correspond to entries in \keplere\ for
those four stars for which OTs are given, and to entries in {\em
  Secunda Classis} for the two other stars. If we analyse the
positions as measured by Hevelius, we find that the entries from
\keplere\ that we identified with Electra and Merope are coupled in
\hevelius\ to positions that match Maia and Electra. We have emended
these, as listed below where the first Taurus sequence number is our
emended version and the second the OT as given by Hevelius.

\begin{tabular}{lllll}
Alcyone & H\,1379 & HIP\,17702 & Tau\,32 & 32 \\
Taygeta & H\,1380 & HIP\,17531 &  -- & 30 \\
Atlas   & H\,1381 & HIP\,17847 & Tau\,33 & 33 \\
Maia    & H\,1382 & HIP\,17573 &  -- &  0 \\
Electra & H\,1383 & HIP\,17499 & Tau\,30 & 31 \\
Merope  & H\,1384 & HIP\,17608 & Tau\,31 & 0 \\
\end{tabular}

\subsection{Emendations to Rybka \label{s:rybkae}}

For each emendation we list the H sequence number, the original text in
Rybka (1984), our emendation, and a 1 (one) if the emendation changes
the identification, or is required to obtain a unique identification.
When Rybka gives two names for one identification, one of which 
corresponds to our identification, and the other must be emended
to do so, we do not give a 1.

\begin{tabular}{rll@{}l|rll@{}l}
H\phantom{5} & Rybka & emend. & &H\phantom{5} & Rybka & emend. \\
\hline
55 & $\upsilon$\,Aql & $\nu$\,Aql & 1 &602 &  $\iota$\,Cyg & $\iota^2$\,Cyg & 1 \\
64 & 1\,Psc & 1\,Aqr & 1 &671 & 63\,Dra & 64\,Dra \\
66 & 5\,Psc & 5\,Aqr & 1 &678 & $\psi$\,Dra & $\psi^1$\,Dra & 1 \\
86 & $\tau$\,Aqr & $\tau^2$\,Aqr & &733 & 6\,Gem & $\sigma$\,Gem & 1 \\
148 & $\rho$\,Ari &  $\rho^3$\,Ari & 1 &828 & $\phi$\,Hya & $\phi^3$\,Hya \\
212 & $\kappa$\,Boo & $\kappa^2$\,Boo& &926 & $\upsilon$\,Lep & $\nu$\,Lep & 1 \\
228 & $\xi$\,CrB & $\zeta^1$\,CrB & 1 &942 & $\zeta$\,Lib & $\zeta^4$\,Lib & 1 \\
265 & $\mu$\,Cnc & $\mu^2$\,Cnc &&1003 & 30\,Hya & 30\,Mon \\
284 & $\pi$\,Cnc & $\pi^2$\,Cnc & 1 &1004 & 31\,Hya & 31\,Mon & 1 \\
348 & UZ\,Cam & VZ\,Cam && 1047 & $\psi$\,Ori & $\psi^2$\,Ori & 1 \\
403 & $\xi$\,Cap & $\xi^2$\,Cap & 1 & 1084 & $\chi$\,Peg & $\gamma$\,Peg  & 1 \\
460 & 16\,Cep & 24\,Cep & 1 &1094 & $\pi$\,Peg & $\pi^2$\,Peg \\
461 & 24\,Cep & 16\,Cep & 1 &1196 & $\nu$\,Psc & $\upsilon$\,Psc \\
466 & 7\,Cas & 1\,Cas & 1 &1227 & 81\,Sgr & 61\,Sgr \\
538 & 32\,Eri & 22\,Eri & 1 & 1367 & $\omega$\,Tau & $\omega^2$\,Tau & 1 \\
559 & 2\,Com & 2\,Boo & 1 &1372 & $\sigma$\,Tau & $\sigma^2$\,Tau 
\end{tabular}

\end{appendix}
\begin{appendix}
\section{Notes on individual identifications \label{s:notes}}

\begin{figure}
\centerline{\includegraphics[angle=270,width=\columnwidth]{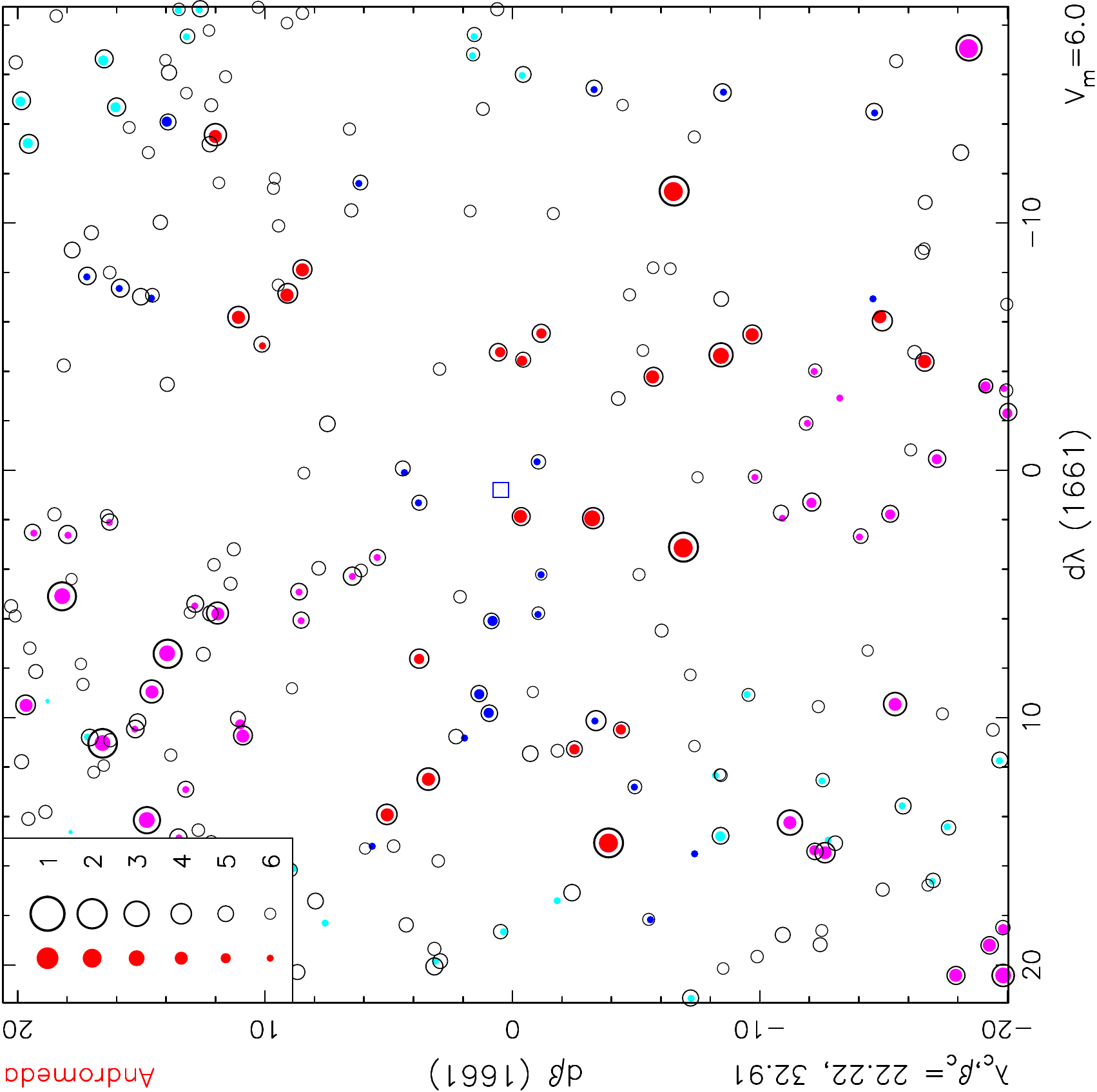}}
\caption{Andromeda. The nebulous object near the center is M\,31.
 \label{f:andromeda}}
\end{figure}

\noindent H\,14, near $-$14,12 in Fig.\,\ref{f:andromeda}, corresponds
to K\,463, but in \hevelius\ is much closer to the other stars in Andromeda,
i.e.\ much further south, than in \keplere.

\noindent H\,16 is identified with 65\,Psc (=HIP\,3885) by Rybka, but this star
is at 4\fdg5.

\noindent H\,30 is identified by Rybka with HIP\,115022, brighter but
further ($V$=4.8, $d$=17\farcm2).

\noindent H\,32 is the Andromeda Nebula M\,31.

\begin{figure}
\centerline{\includegraphics[angle=270,width=0.9\columnwidth]{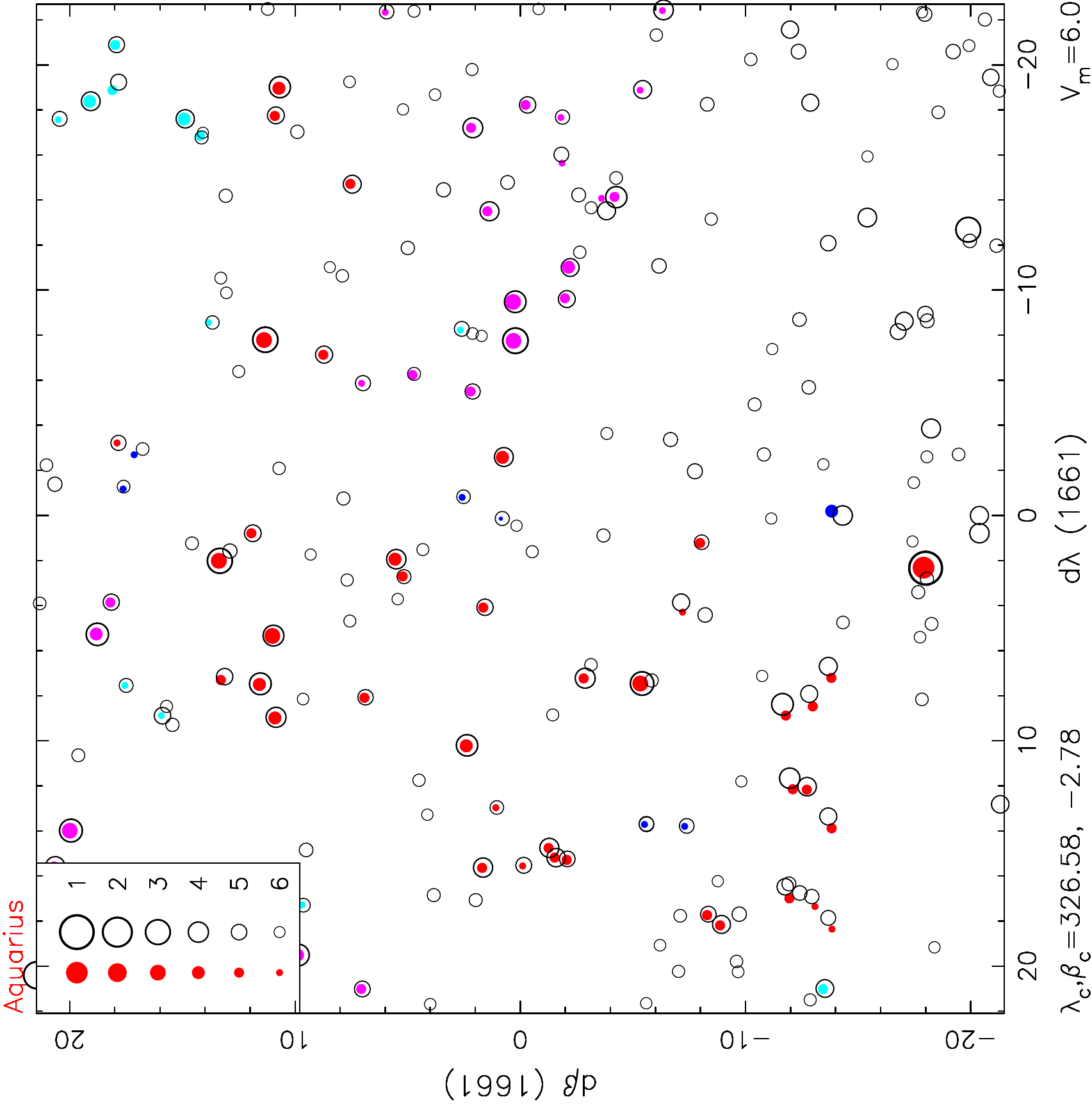}}
\caption{Aquarius
 \label{f:aquarius}}
\end{figure}

\begin{figure}
\centerline{\includegraphics[angle=270,width=0.9\columnwidth]{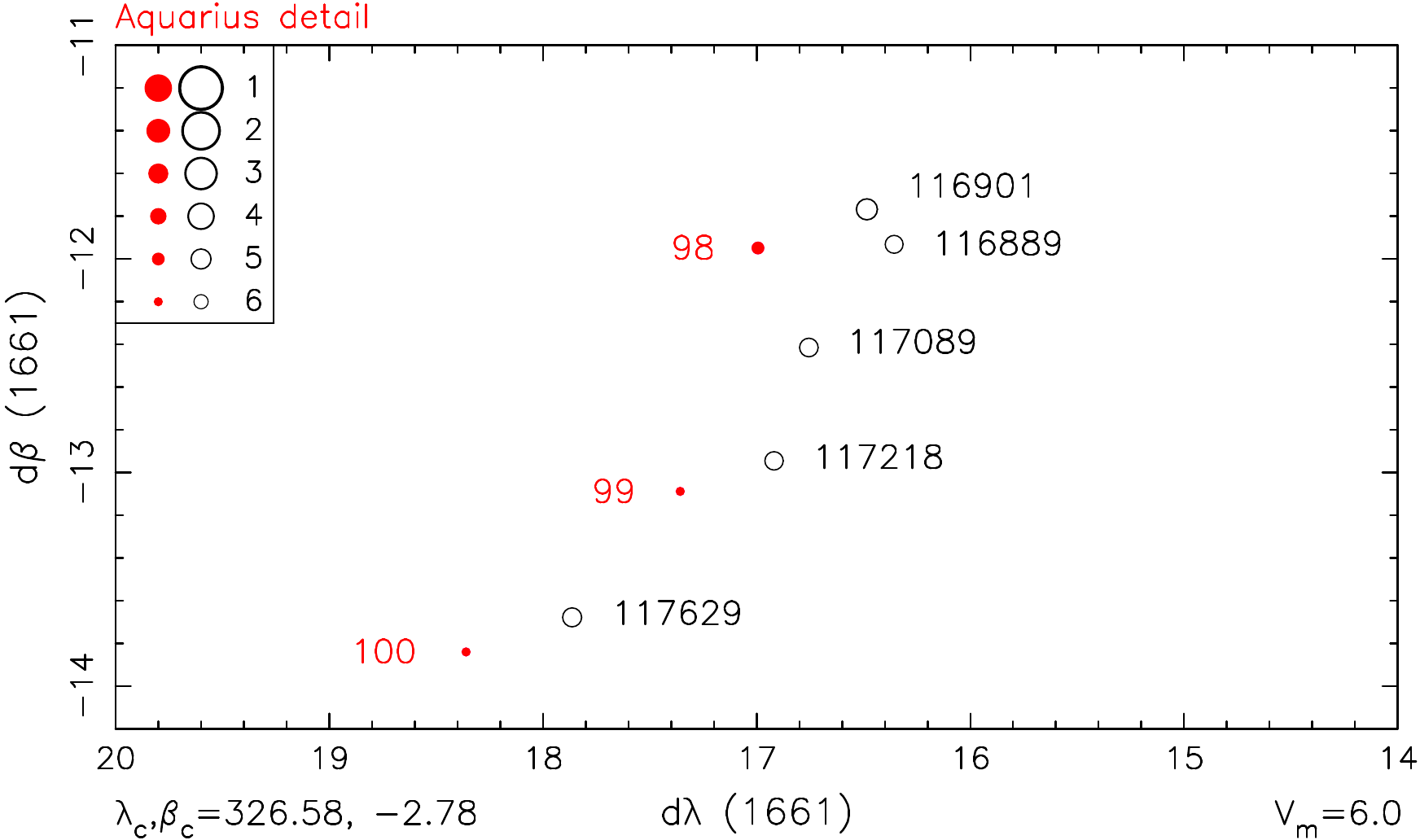}}
\caption{Aquarius detail
 \label{f:aqrdetail}}
\end{figure}

\noindent H\,98 is marginally closer to HIP\,117089 ($V$=5.2, $d$-31\farcm5), 
but we agree with Rybka that the brighter HIP\,116901 is the counterpart,
given the relative offsets of other stars in the area to their counterparts
(Figs.\,\ref{f:aquarius},\ref{f:aqrdetail}). 

\noindent H\,99, about equally distant from H\,98 and H\,100 in
\hevelius, is identified with HIP\,117218 (Fig.\,\ref{f:aqrdetail});
its counterpart in \keplere\ lies closer to the counterpart of H\,98
and is therefore identified with HIP\,117089.

\begin{figure}
\centerline{\includegraphics[angle=270,width=\columnwidth]{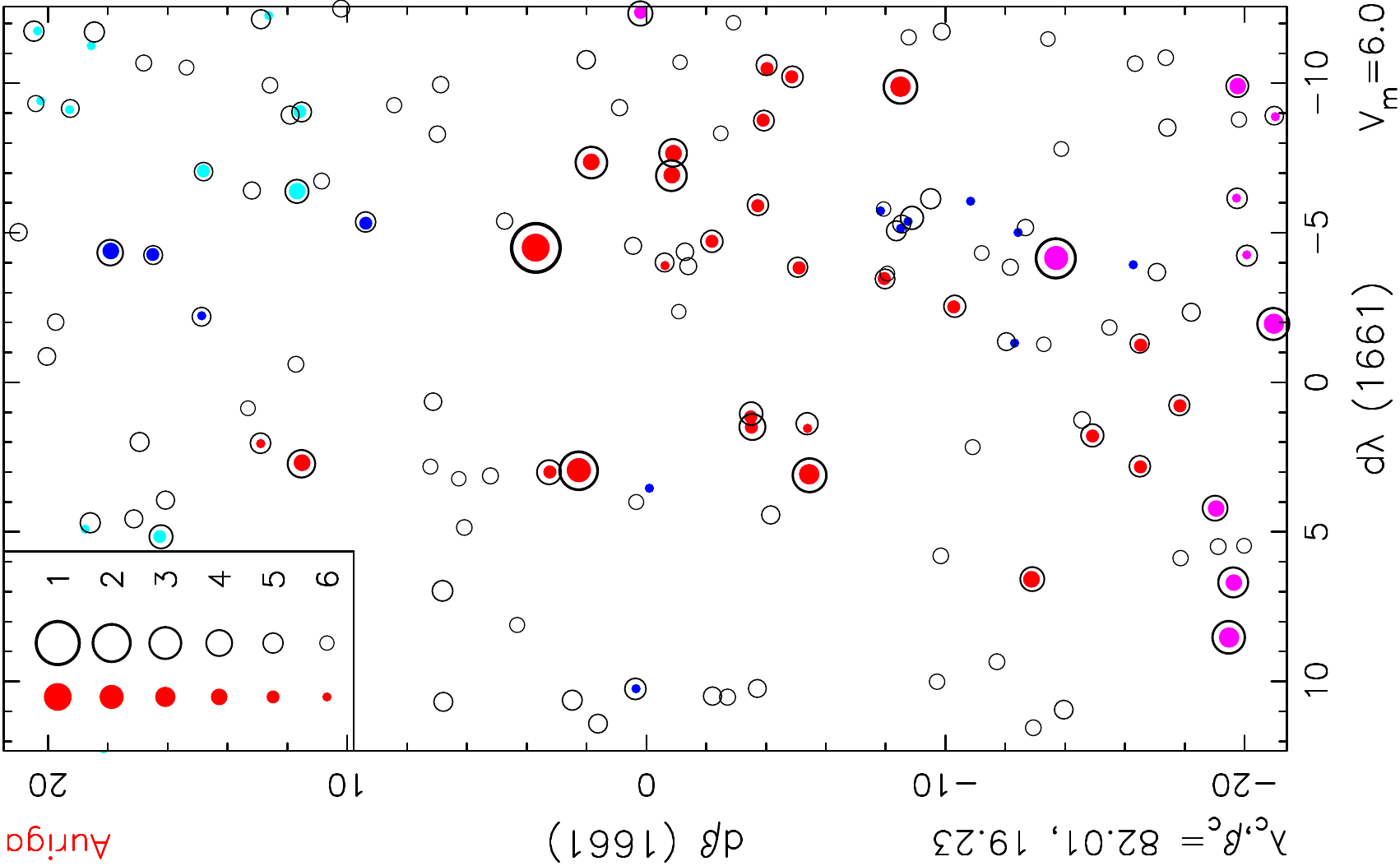}}
\caption{Auriga
 \label{f:auriga}}
\end{figure}

\noindent H\,174 near +3,+3 in Fig.\,\ref{f:auriga}, is
close to $\beta$\,Aur (= HIP\,28360) in \hevelius,
i.e.\ much further south than its counterpart K\,310 in \keplere.

\begin{figure}
\centerline{\includegraphics[angle=270,width=\columnwidth]{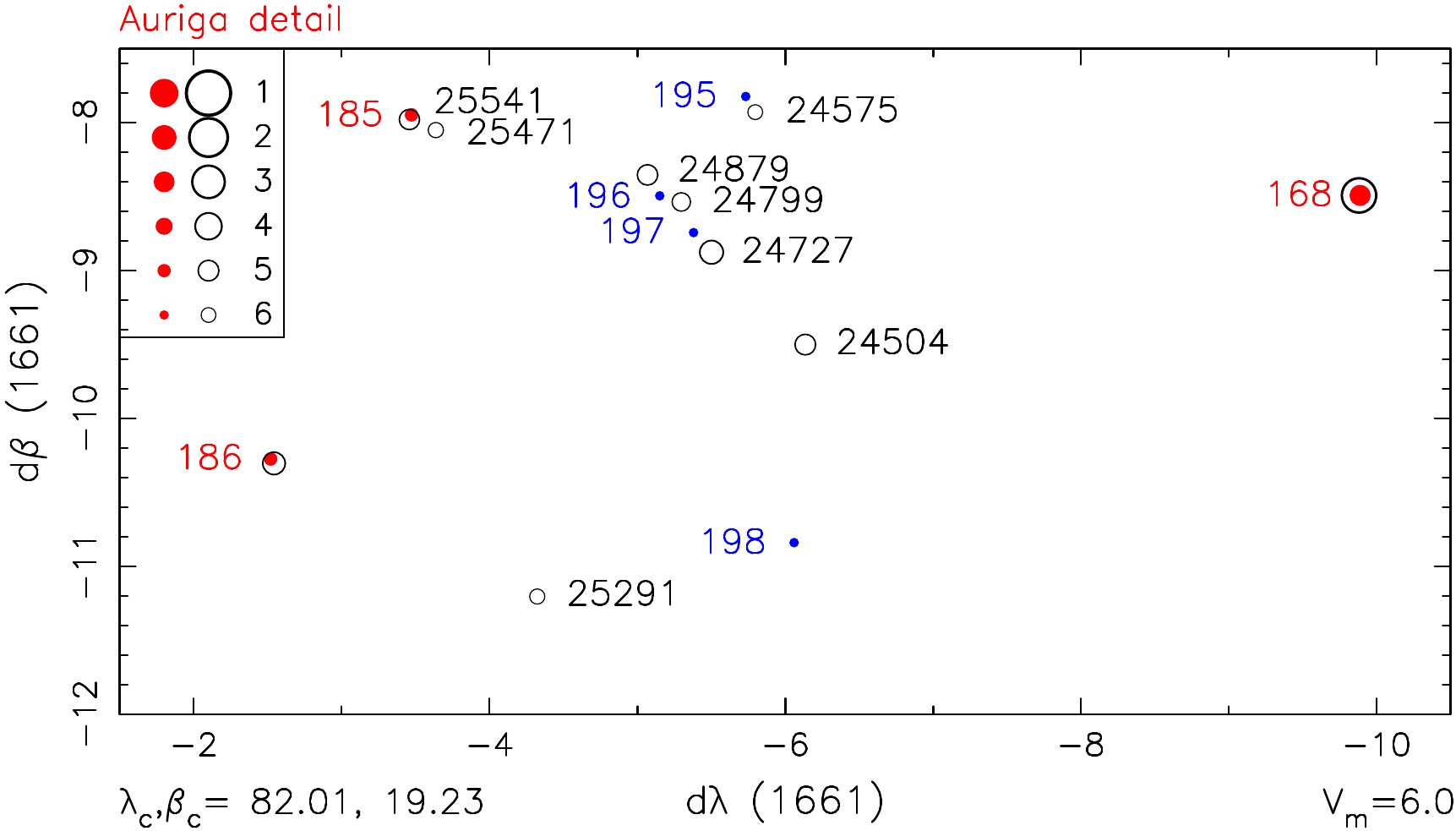}}
\caption{Auriga detail
 \label{f:aurdetail}}
\end{figure}

\noindent H\,185. Its counterpart in \keplere\ lies between HIP\,25541
and HIP\,24879 and is identified with the latter (Fig.\,\ref{f:aurdetail}).

\noindent H\,196, H\,197 and H\,198 are identified by Rybka with
HIP\,24727, HIP\,24504 and unidentified respectively. 
Fig.\,\ref{f:aurdetail} shows that our identifications HIP\,24799,
HIP\,24727 and HIP\,24504 are better matches. Alternatives to our
identifications for H\,196 and H\,198 are HIP\,24879 and HIP\,25291,
respectively.

\noindent H\,200, near $-$5,$-$13 in Fig.\,\ref{f:auriga}, is identified
by Rybka with HIP\,25492 (near $-$4,$-$12). Our identification is a better
positional match.

\begin{figure}
\centerline{\includegraphics[angle=270,width=\columnwidth]{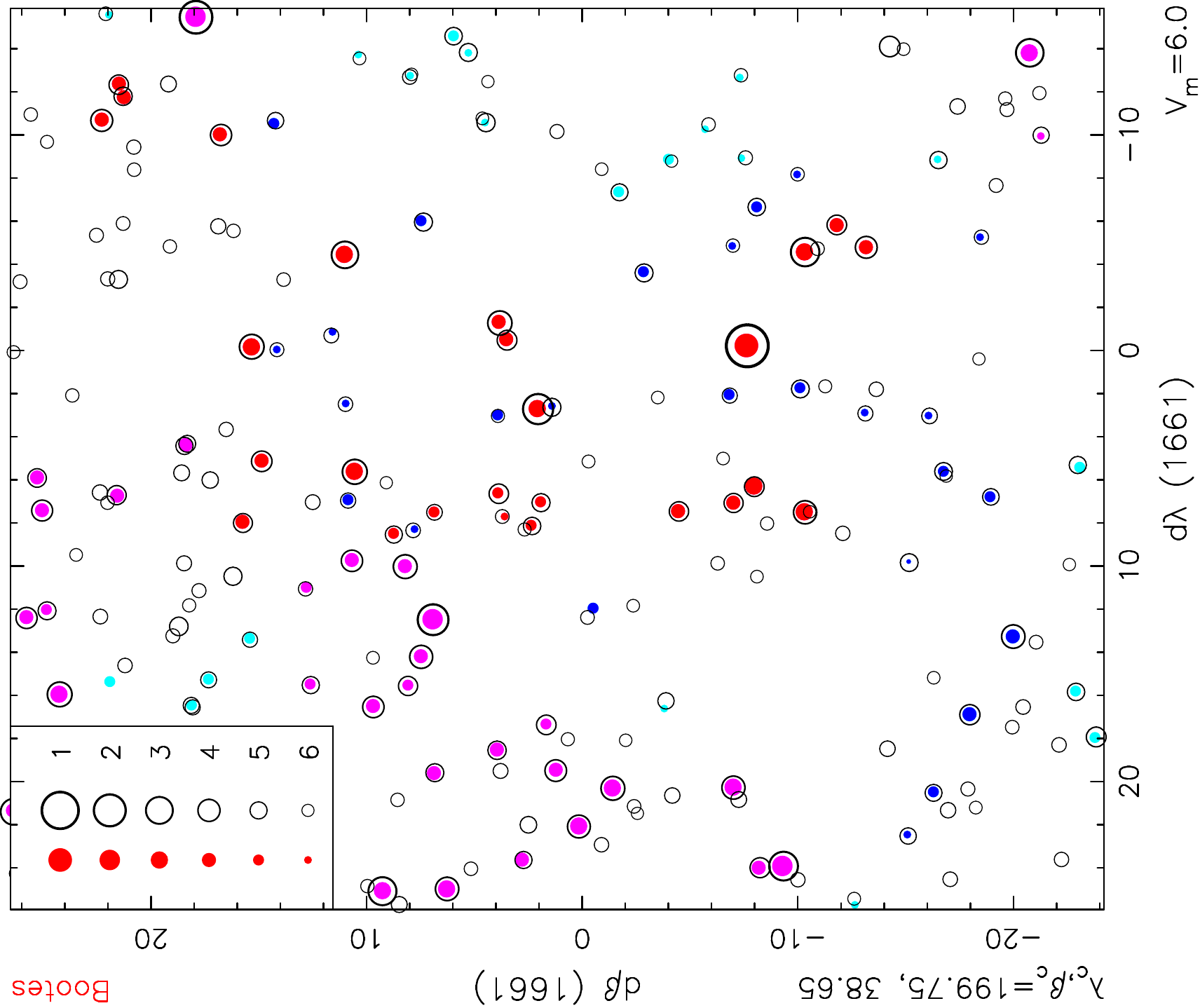}}
\caption{Bootes
 \label{f:bootes}}
\end{figure}

\noindent H\,234, near 5.6,$-$16.8 in Fig.\,\ref{f:bootes}, is identified
by Rybka with HIP70738 ($V$=5.9, $d$=16\farcm2); our identification is
brighter and nearer.

\begin{figure}
\centerline{\includegraphics[angle=270,width=\columnwidth]{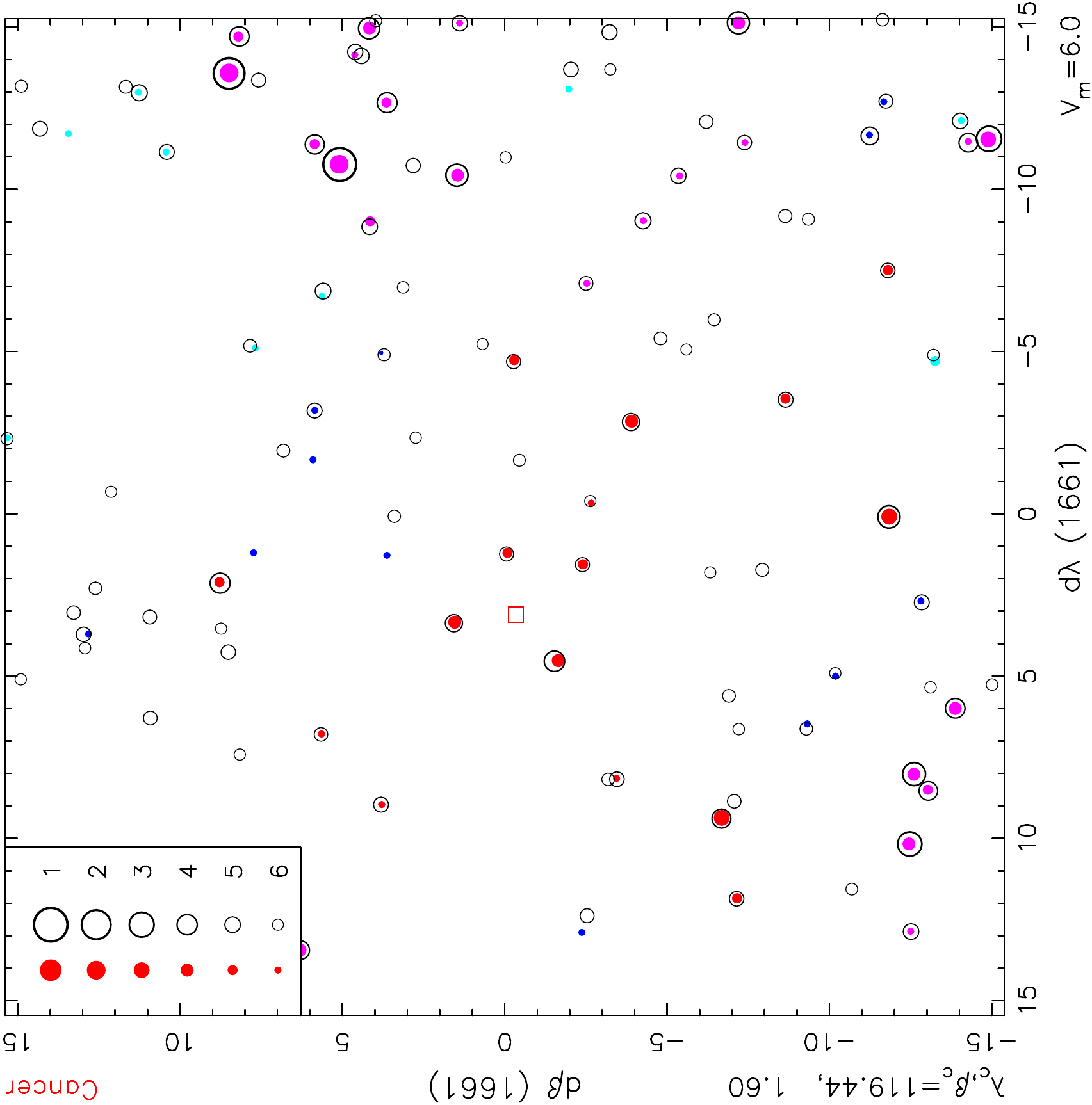}}
\caption{Cancer
 \label{f:cancer}}
\end{figure}

\noindent H\,282-283, near $-2$,+6 and +1.3,+3.6 in Fig.\,\ref{f:cancer},
only have distant counterparts in the Hipparcos Catalogue with $V\le$6.0. Rybka
gives fainter counterparts for each, viz.\ HIP\,41404 ($V$=6.3, $d$=1\arcmin)
and HIP\,41940 ($V$=6.4, $d$=53\arcmin).

\begin{figure}
\centerline{\includegraphics[angle=270,width=\columnwidth]{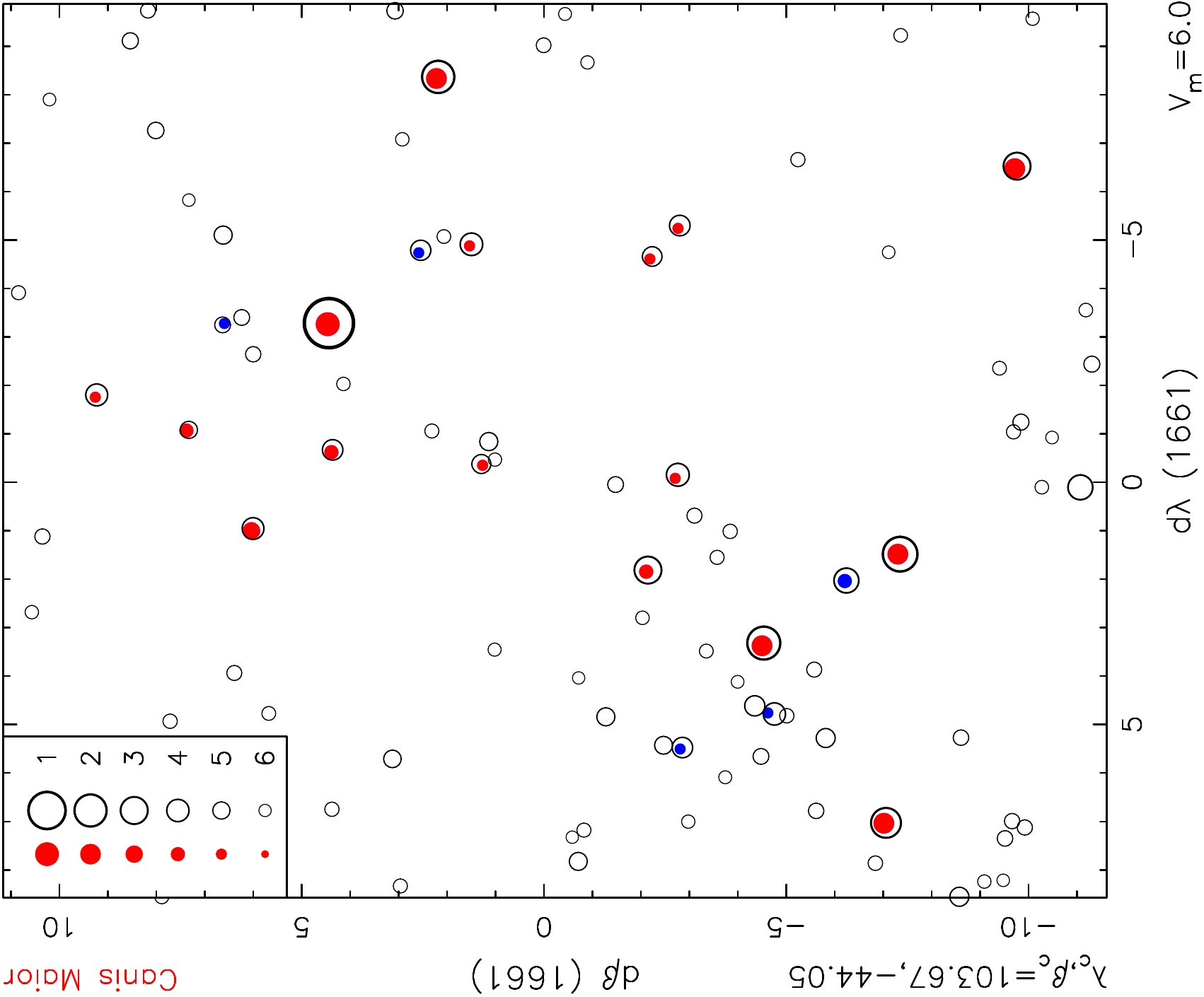}}
\caption{Canis Maior
 \label{f:canismaior}}
\end{figure}

\noindent H\,296, near $-$0.4,1.3 in Fig.\,\ref{f:canismaior}, is `difficult 
to identify' according to Rybka, but clear with Hipparcos data.

\noindent H\,306, near 5.5,$-$2.8 in Fig.\,\ref{f:canismaior}, is identified by Rybka
with HIP\,34981 near 4.6,$-$4.3.

\begin{figure}
\centerline{\includegraphics[angle=270,width=\columnwidth]{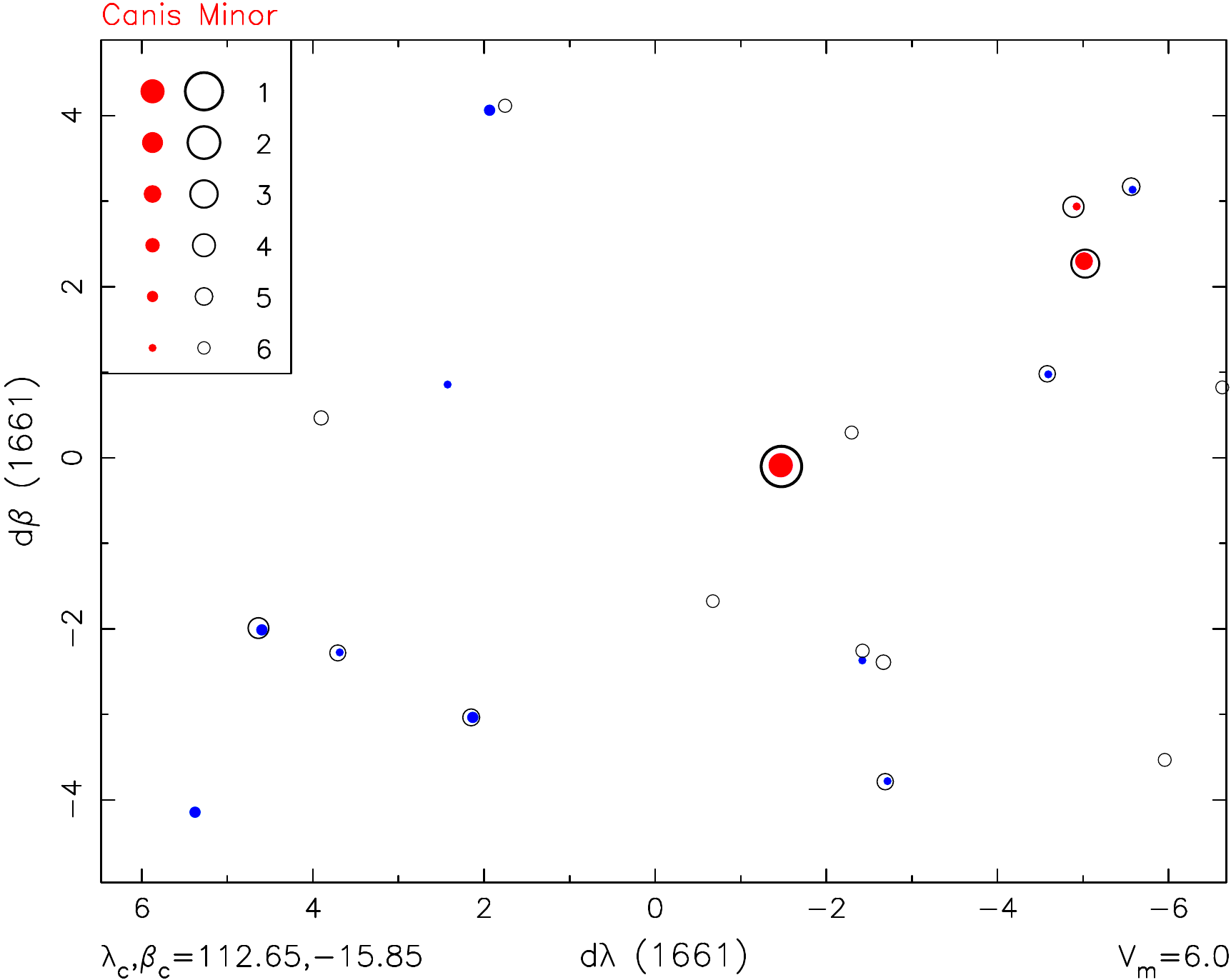}}
\caption{Canis Minor
 \label{f:canisminor}}
\end{figure}

\noindent H\,313, near $-$2.4,$-$2.4 in Fig.\,\ref{f:canisminor}, is identified 
by Rybka with HIP\,36723, slightly brighter but further ($V$=5.6, $d$=14\farcm9)
than our identification.

\noindent H\,314, near 2.4,0.9 in Fig.\,\ref{f:canisminor}, is identified with
Rybka with HIP\,39213, indeed the nearest star (3.9,0.5) but in our view too far,
at 1\fdg5.

\noindent H\,321 is the combined light of $\alpha^1$\,CVn (HIP\,63121, $V$=5.6) and 
 $\alpha^2$\,CVn (HIP\,63125 $V$=2.9), separated by 19\arcsec.

\noindent H\,326, is identified by us with the first of two possibilities given by
Rybka.

\begin{figure}
\centerline{\includegraphics[angle=270,width=\columnwidth]{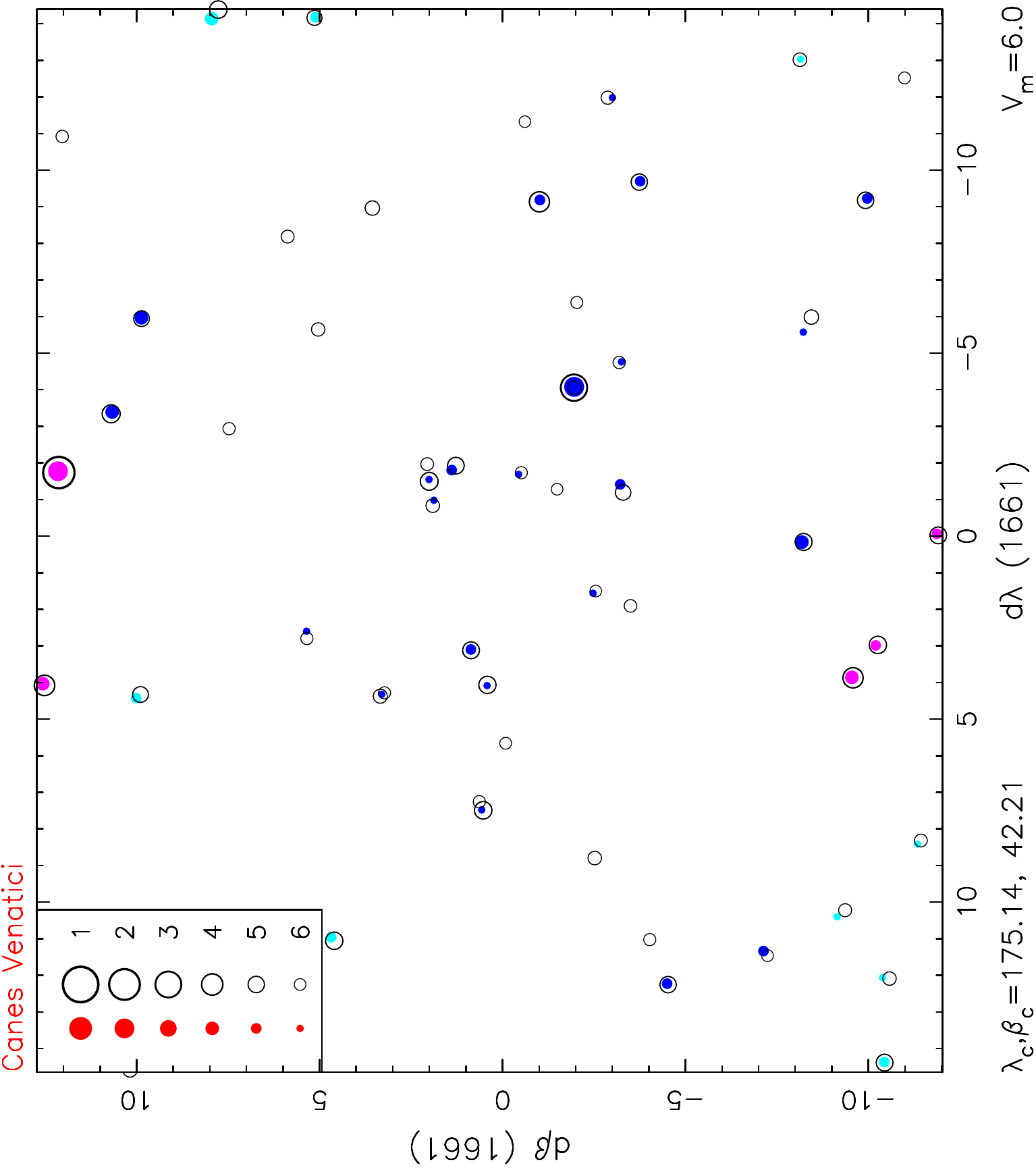}}
\caption{Canes Venatici
 \label{f:canesven}}
\end{figure}

\noindent H\,329, near 1.2,$-$2.5 in  Fig.\,\ref{f:canesven}, is identified by Rybka
with HIP\,64927 (near 1.9,$-$3.5), slightly brighter but much further than our
identification.

\noindent H\,331, near $-$1.8,1.4 in  Fig.\,\ref{f:canesven}, is identified by Rybka
with HIP\,64692 (near $-$2.0,2.1), fainter and further than our identification.

\noindent H\,338, is identified by us with the second of two possibilities given by
Rybka.

\begin{figure}
\centerline{\includegraphics[angle=270,width=\columnwidth]{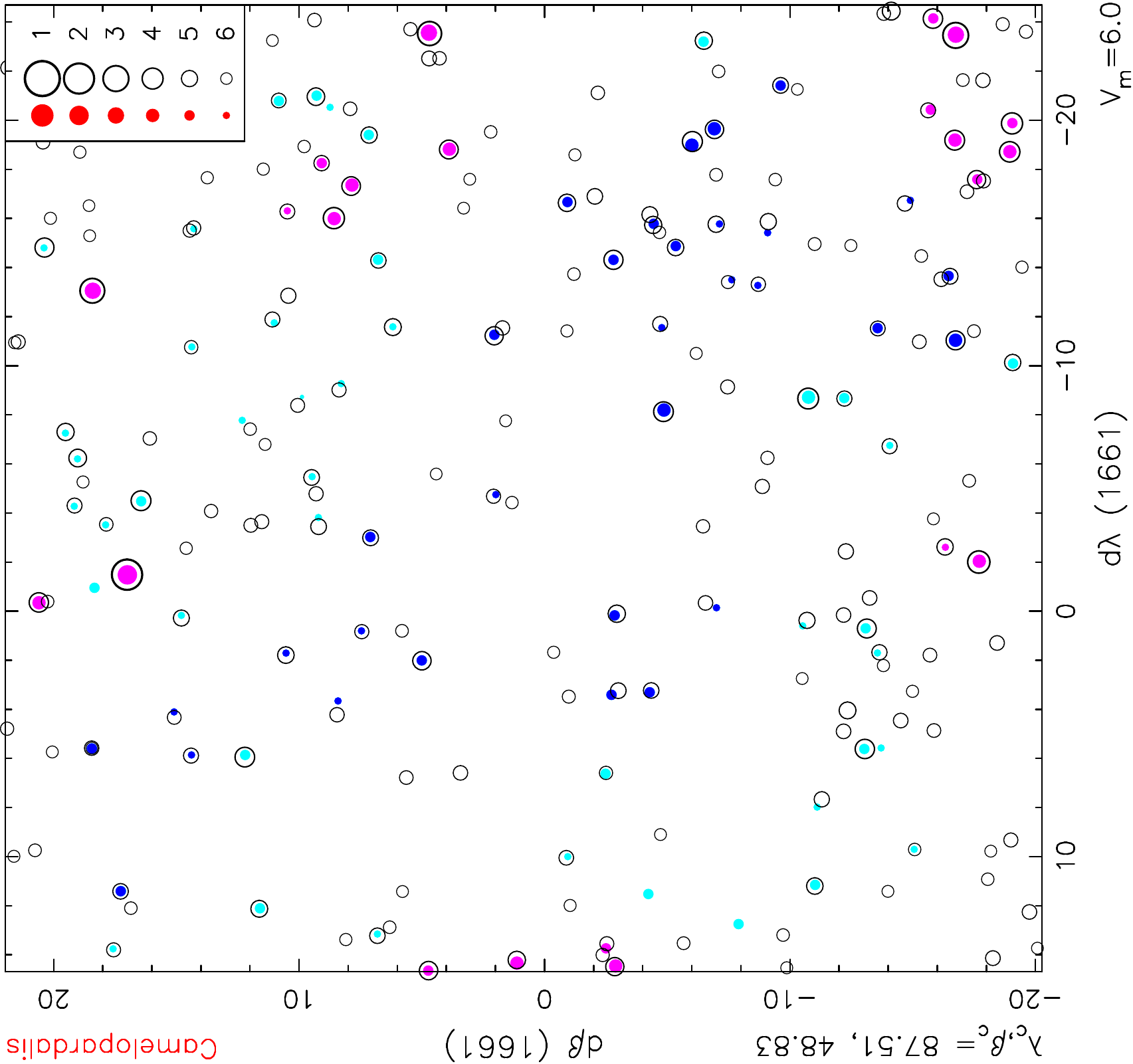}}
\caption{Camelopardalis
 \label{f:camelopardalis}}
\end{figure}

\noindent H\,346, near 4.1,15.1 in Fig.\,\ref{f:camelopardalis}, is identified by
Rybka with HIP\,49688, both fainter and further ($V$=6.3, $d$=42\arcmin) than our
identification

\begin{figure}
\centerline{\includegraphics[angle=270,width=\columnwidth]{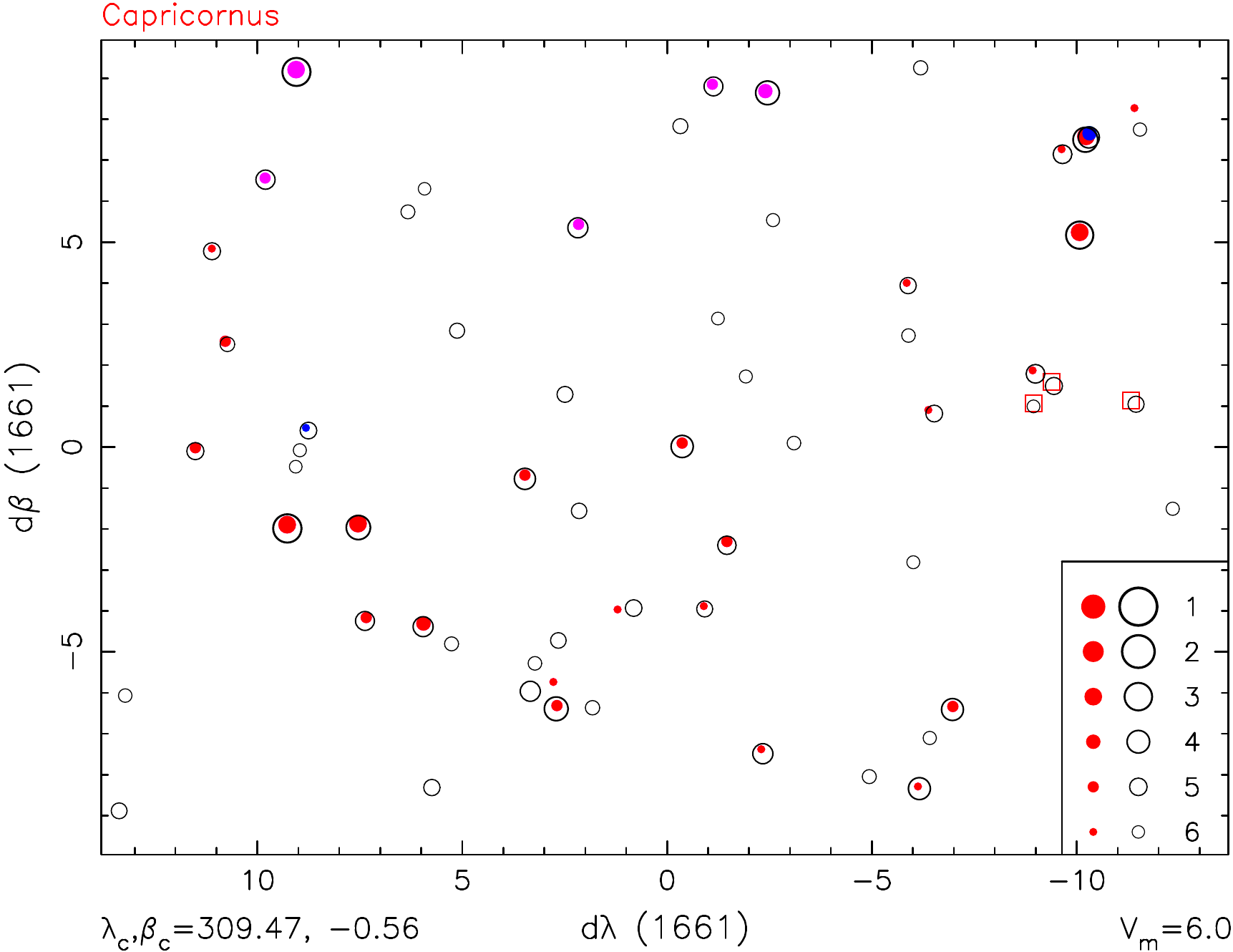}}
\caption{Capricornus
 \label{f:capricornus}}
\end{figure}

\noindent H\,376, near $-$10,+7.6 in Fig.\ref{f:capricornus}, is
closer to HIP\,100027 ($\alpha^1$\,Cap, $d$=4\farcm6) than to HIP\,100064
($\alpha^2$\,Cap, $d$=5\farcm1) but should still be identified with the
latter. H\,404 is the counterpart of HIP\,100027, described by
Hevelius as {\em In cornu borealis duarum contiguarum occidentalis}
(In the northern horn of the two adjacent ones the western one).

\noindent H\,400 {\em Non videtur amplius} (Is no longer seen) according to Hevelius,
which is understandable as Brahe's position was erroneous (Rawlins 1993, see also
Paper\,I).

\noindent H\,405, near 8.8,0.5 in Fig.\ref{f:capricornus}, is identified by
Rybka with HIP\,107517 fainter and much further ($V$=5.6, $d=3^\circ$) than
our identification. Rybka gives HIP\,107517 also as the identification of
H\,399, where we agree.

\begin{figure}
\centerline{\includegraphics[angle=270,width=\columnwidth]{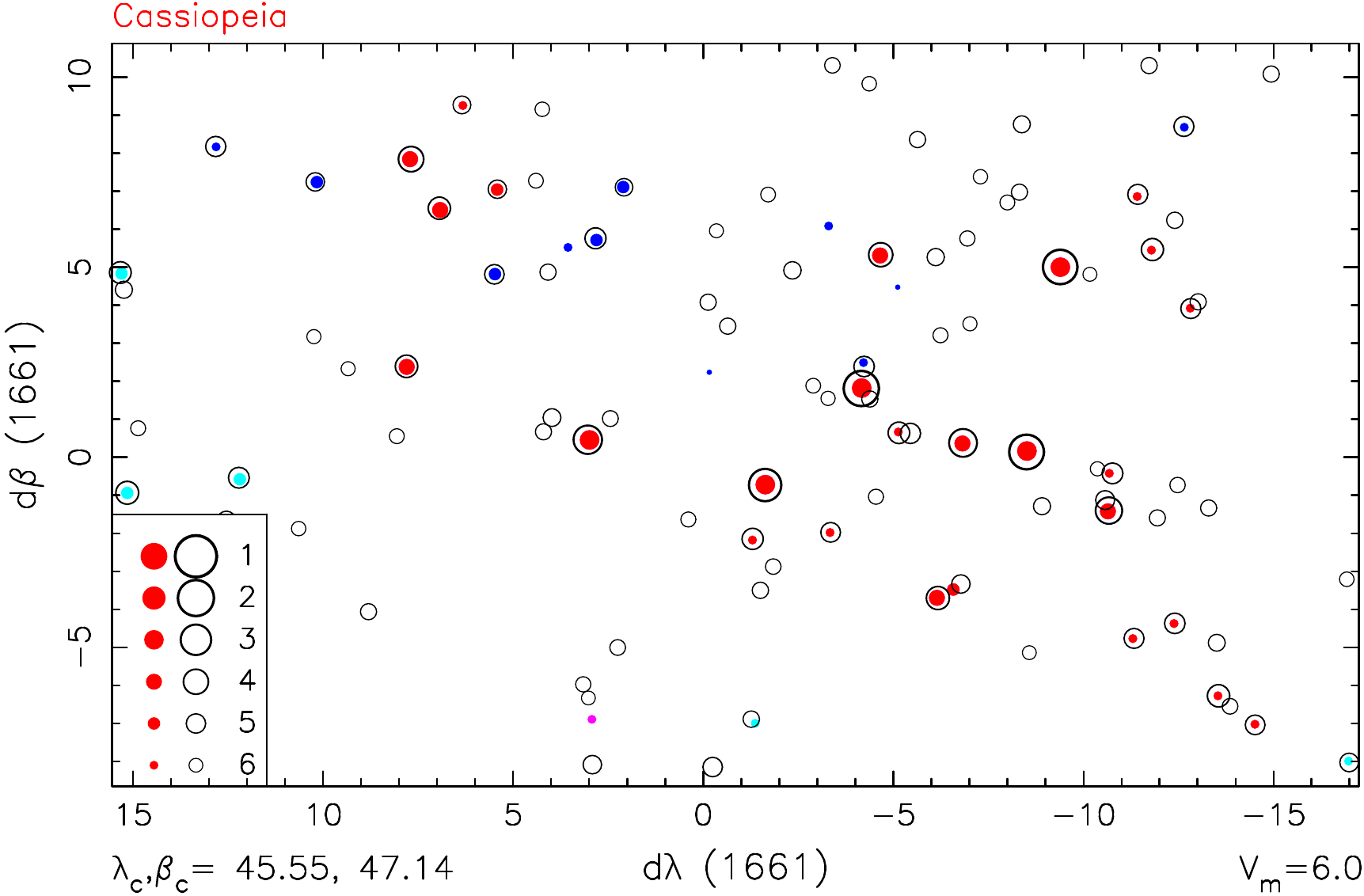}}
\caption{Cassiopeia
 \label{f:cassiopeia}}
\end{figure}

\noindent H\,429-432, four stars near 6,7 in Fig.\,\ref{f:cassiopeia},
had wrong positions in Brahe's catalogue and were assigned other
identifications, even though this left relatively bright stars in
Cassiopeia unidentified (see notes with K\,236-238 and Fig.\,C.14 in
Paper\,I; and Rawlins 1993, D\,585-588). Hevelius decided to identify
these entries in Brahe with the brighter stars.

\noindent H\,436, near $-$5.1,4.5 in Fig.\,\ref{f:cassiopeia} is identified
by Rybka with HIP\,2876 near $-$6.1,3.2, but this star is too far to be 
a reliable counterpart; we leave this star unidentified.

\noindent H\,438, near $-$0.2,2.2 in Fig.\,\ref{f:cassiopeia} is
identified by Rybka with the the Mira variable RV Cas, but this star
is almost 17$^\circ$ from the position of H\,438.

\begin{figure}
\centerline{\includegraphics[angle=270,width=\columnwidth]{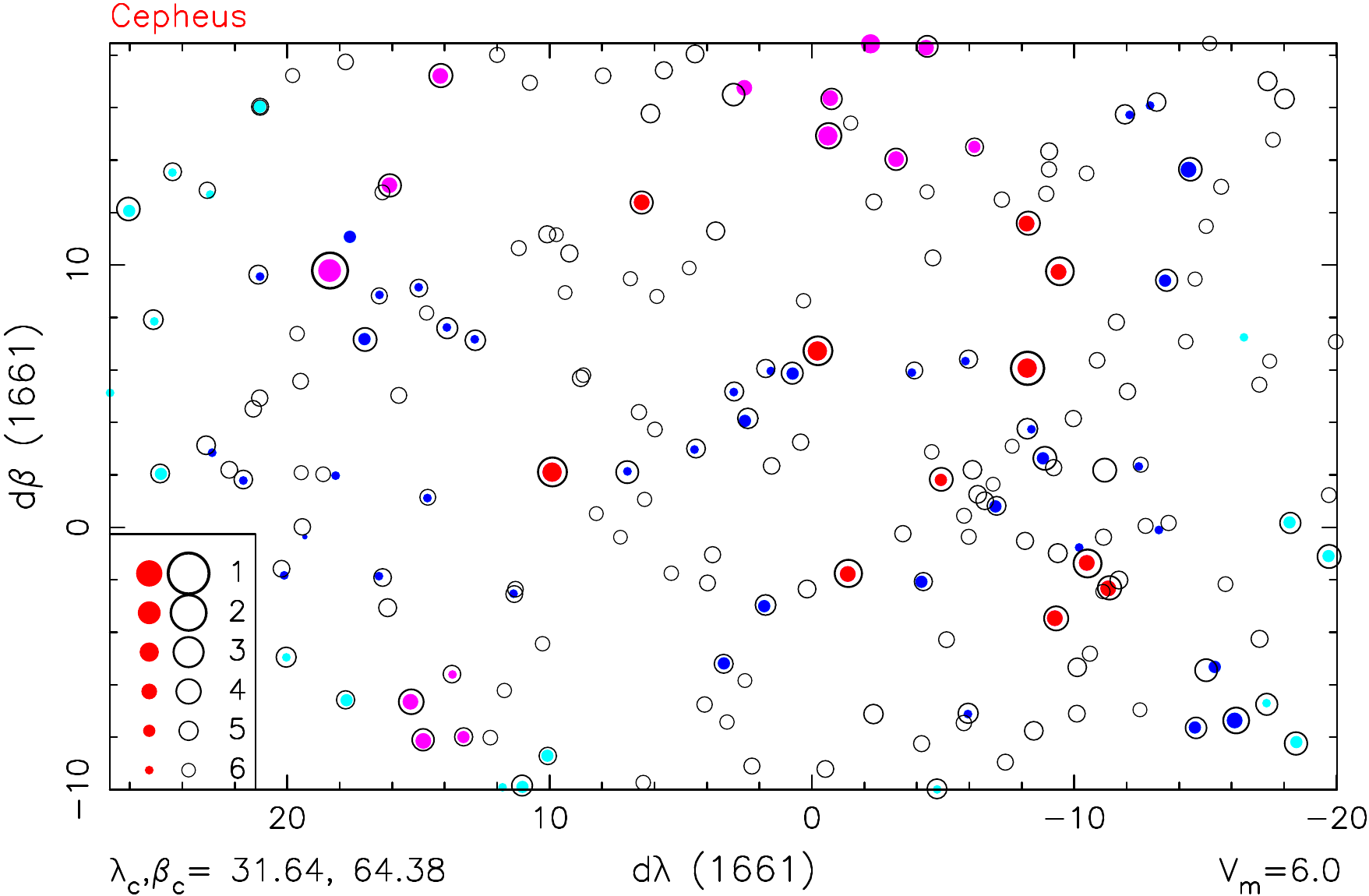}}
\caption{Cepheus
 \label{f:cepheus}}
\end{figure}

\noindent H\,462, near 11.4,$-$2.5 in Fig.\,\ref{f:cepheus}, is
identified by Rybka with HIP\,1296, almost 3$^\circ$ away.

\noindent H\,463. We list the same counterpart as Rybka; an alternative is 
HIP\,108165, $V$=5.7, $d$=32\farcm7.

\noindent H\,467-468, near $-$12.9,16.1 and $-$12.1,15.7 in
Fig.\,\ref{f:cepheus} are identified by Rybka with HIP\,100097 and HIP\,101084,
near $-$15.6,13.0 and $-$15.0,11.5.

\noindent H\,492, near 17.6,11.8 in Fig\,\ref{f:cepheus}, is
identified by Rybka with HIP\,84535, but this star is faint and rather
far ($V$=6.3, $d$=16\farcm5).

\begin{figure}
\centerline{\includegraphics[angle=270,width=\columnwidth]{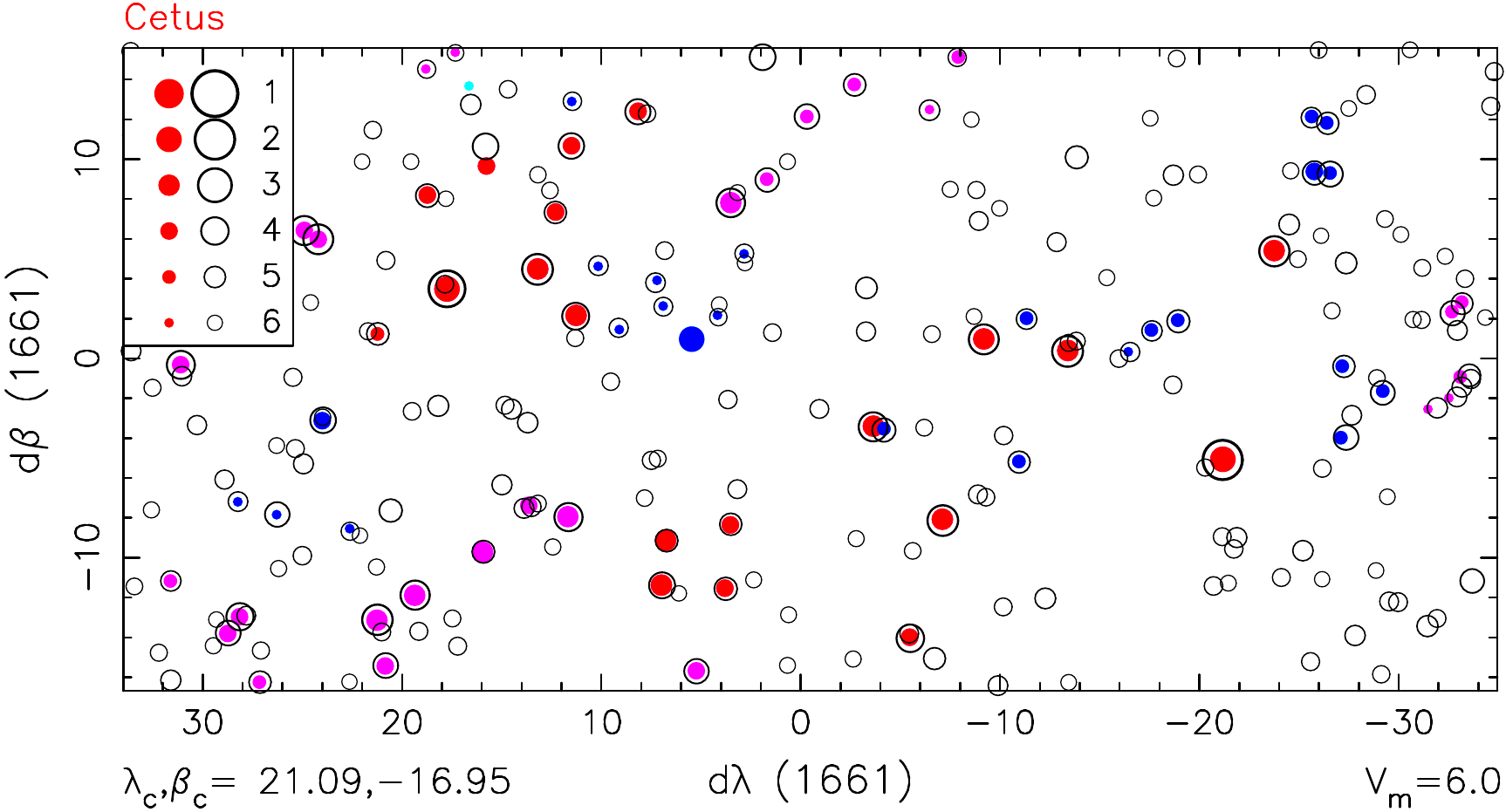}}
\caption{Cetus
 \label{f:cetus}}
\end{figure}

\noindent H\,523, near 5.5,1.0 in Fig.\,\ref{f:cetus}, is described as
{\em Nova in collo Ceti} (New star in the neck of the Whale) by Hevelius,
and is Mira Ceti ($o$\,Cet). It is listed in the Hipparcos catalogue at $V$=6.5
(sc.\ average) but its magnitude at maximum is close to 2.

\noindent H\,525 is erroneously given the same identification by Rybka as H\,524.

\begin{figure}
\centerline{\includegraphics[angle=270,width=\columnwidth]{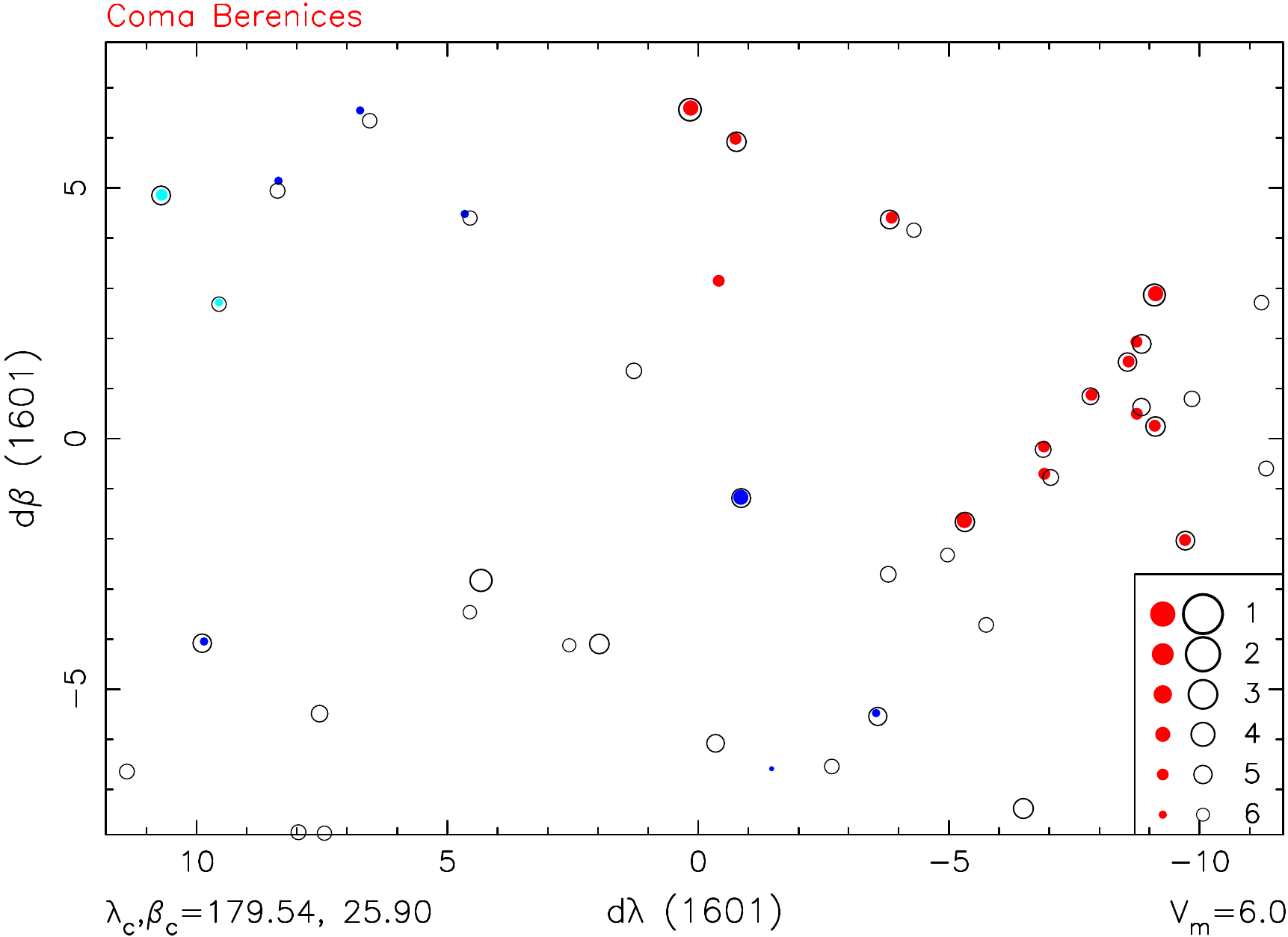}}
\caption{Coma Berenices
 \label{f:comaber}}
\end{figure}

\noindent H\,557, near $-$0.4,3.2 in Fig.\,\ref{f:comaber}, is
identified by Rybka with HIP\,62886, near $-$0.9,$-$1.2, but we
identify H\,562 with this star.  We consider H\,557 unidentified, just
as its corresponding entry K\,483 in Brahe's catalogue.

\noindent H\,562, near $-$0.9,$-$1.2 in Fig.\,\ref{f:comaber}, is
identified by Rybka with HIP\,6424 =$\alpha$\,Com, near 4.3,$-$2.8,
clearly not compatible with its position.

\begin{figure}
\centerline{\includegraphics[angle=270,width=\columnwidth]{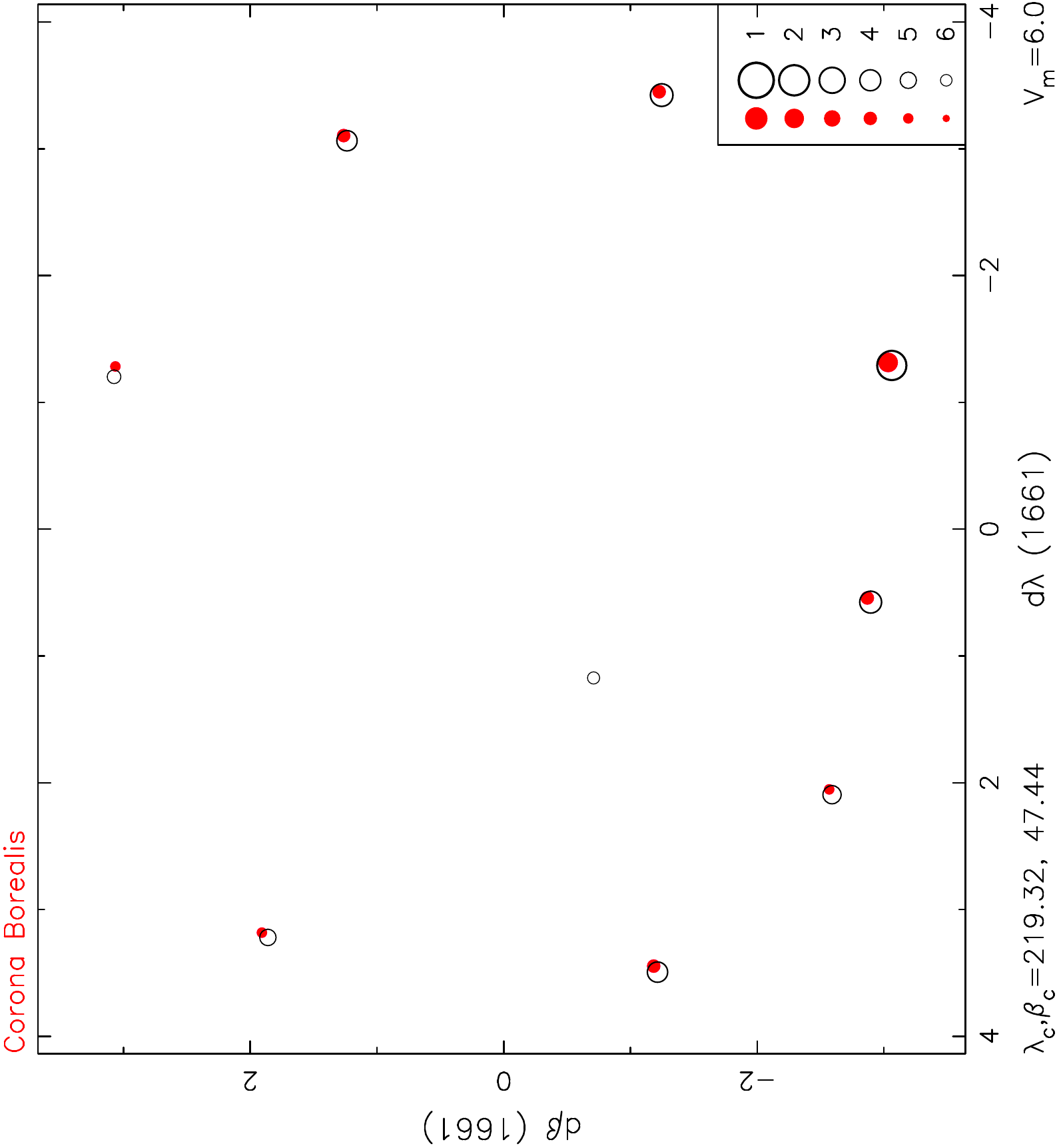}}
\caption{Corono Borealis
 \label{f:corbor}}
\end{figure}

\noindent H\,570, near 3.2,1.9 in Fig.\,\ref{f:corbor}, is identified
by Rybka with $\rho$\,CrB = HIP\,78459 ($V$=5.4, $d$=208\arcmin). Our
identification corresponds to $\iota$\,CrB.

\noindent H\,581. HIP\,61910 corresponds to the combined flux of the
two stars given by Rybka, SAO157447/8.

\noindent H\,593 is P\,Cygni. Hevelius gives the magnitude as 3.0.6 which
we interpret as indicating a range of 3 to 6, in agreement with the
actual variations of this star. Rybka does not identify this entry.

\noindent H\,604. An alternative identification is HIP\,99639, closer but
a magnitude fainter ($V$=4.8, $d$=2\farcm9).

\begin{figure}
\centerline{\includegraphics[angle=270,width=\columnwidth]{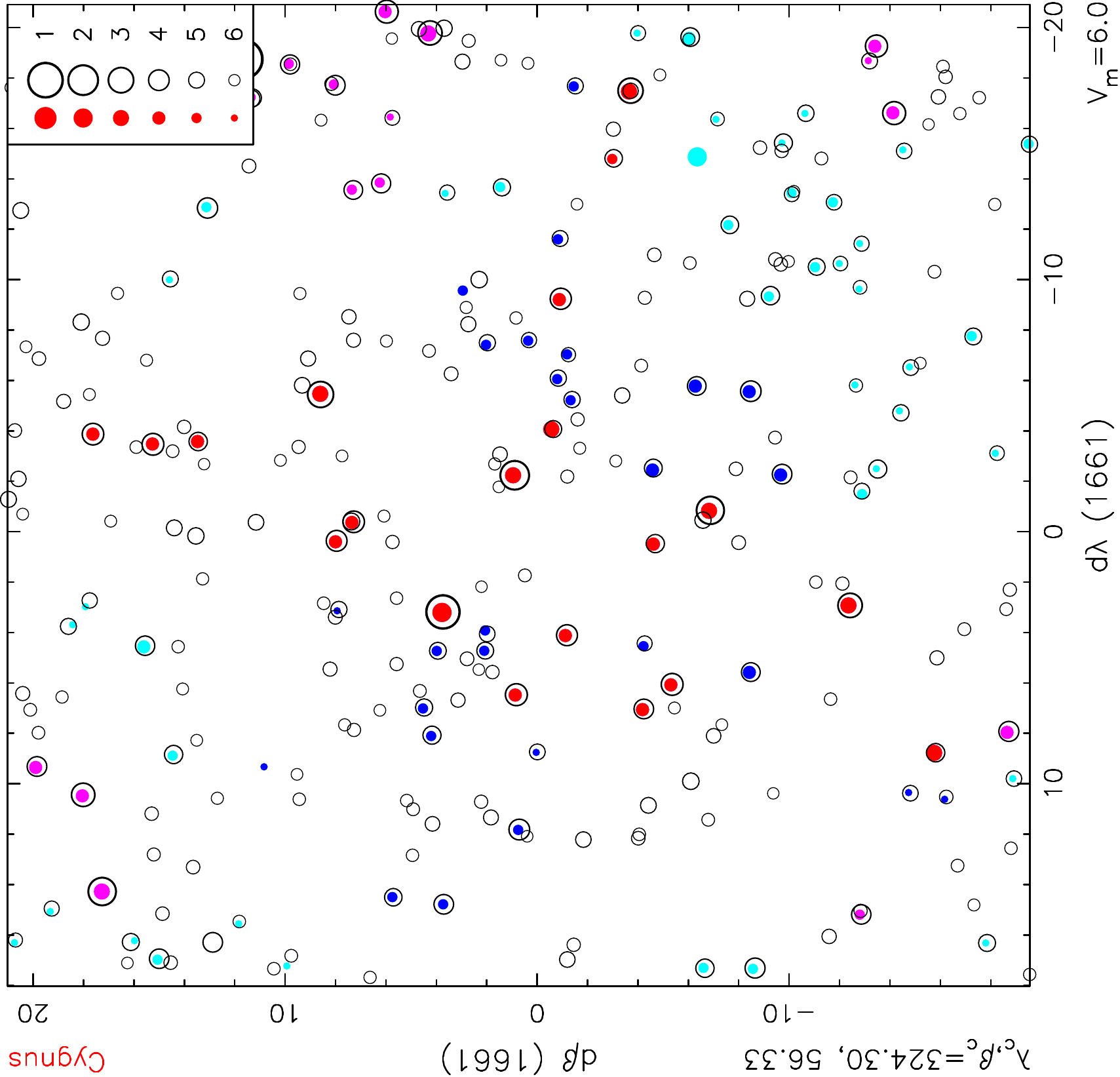}}
\caption{Cygnus
 \label{f:cygnus}}
\end{figure}

\noindent H\,619, near 9.3,10.8 in Fig.\,\ref{f:cygnus}. The nearest 
{\em Hipparcos} star with $V$$\leq$6.0 is HIP\,103956 ($V$=5.9, $d$=81\arcmin;
Rybka identifies with HIP\,103530 $V$=5.6, $d$=4$^\circ$. We consider this
entry unidentified.

\noindent H\,636, near $-$9.6,3.0 in Fig.\,\ref{f:cygnus}. HIP\,97367 is closer
to this entry but much fainter ($V$=5.8, $d$=41\farcm7) than our identification.

\begin{figure}
\centerline{\includegraphics[angle=270,width=\columnwidth]{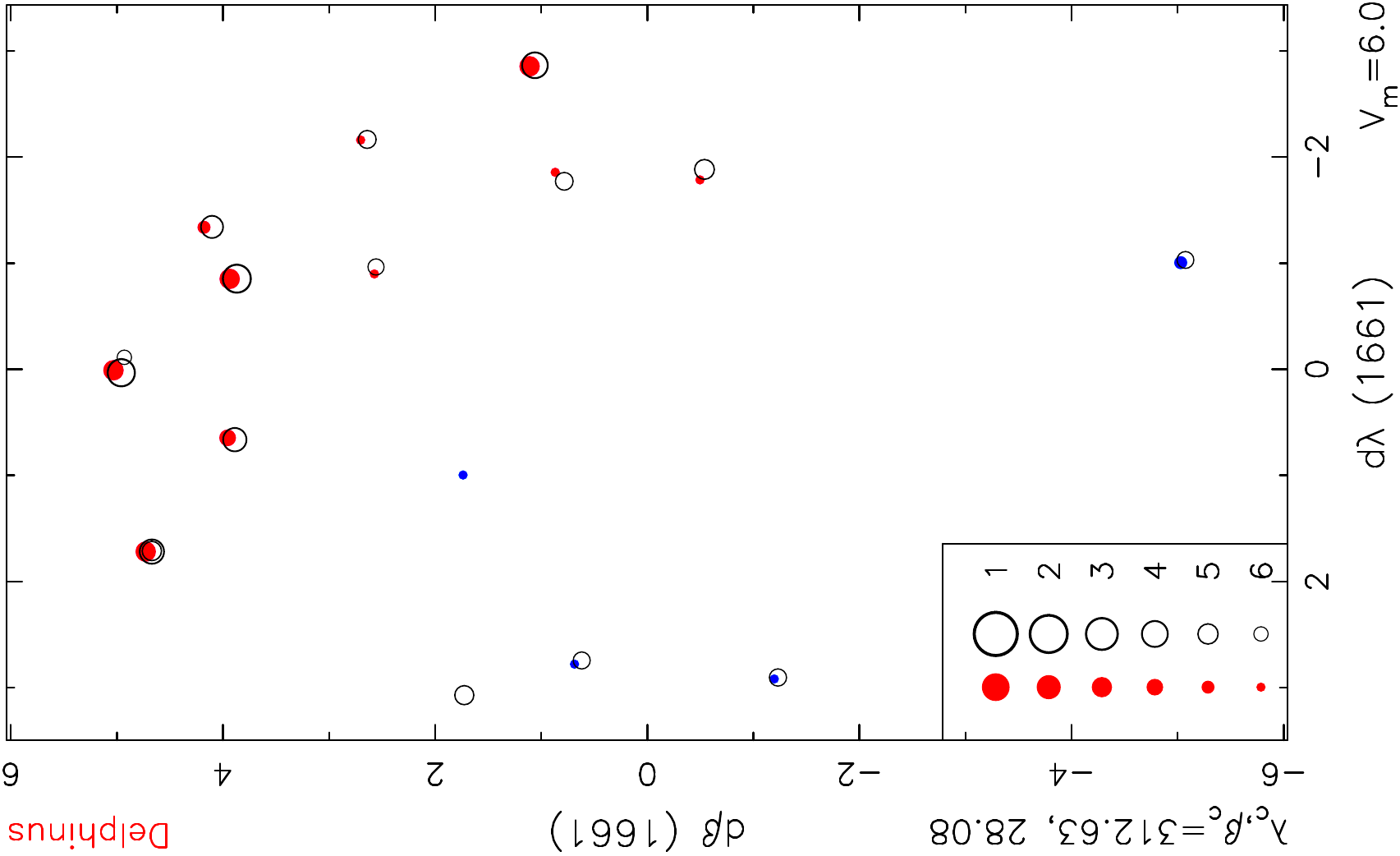}}
\caption{Delphinus
 \label{f:delphinus}}
\end{figure}

\noindent H\,649, near 1.0,1.7 in Fig.\,\ref{f:delphinus}, would be a good match
to HIP\,103294 if a correction of 2$^\circ$ is made to the longitude. The counterpart
given by Rybka is too faint and too far ($V$=6.3, $d\simeq5^\circ$).

\noindent H\,678, H\,679. H\,678 corresponds to the close pair HIP\,86614, HIP\,86620.
The entry K\,94 in the Brahe Star Catalogue corresponding to H\,679, had a position
closer to HIP\,87728, but a description which better matches HIP\,89937. The
position in Hevelius also matches HIP\,89937. See Fig.\,C.5 in Paper\,I

\begin{figure}
\centerline{\includegraphics[angle=270,width=\columnwidth]{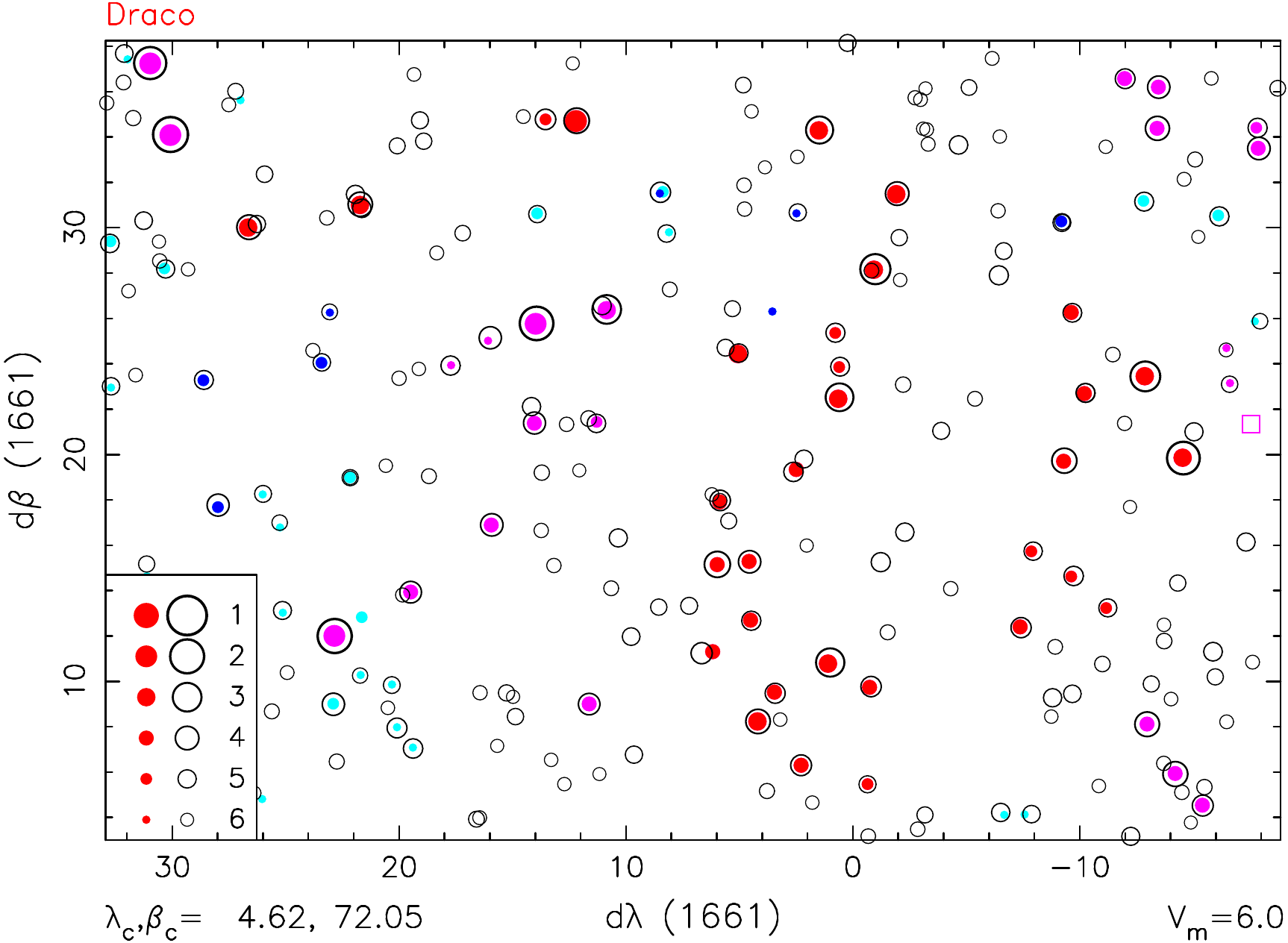}}
\caption{Draco
 \label{f:draco}}
\end{figure}

\noindent H\,690, near 23.1,26.3 in Fig.\,\ref{f:draco} is identified by Rybka
with HIP\,61281 ($d\simeq5\fdg5$).

\noindent H\,698 ($\epsilon$\,Equ) corresponds to the combined flux of
HIP\,103571 and HIP\,103569 (separated by 10\arcsec).

\begin{figure}
\centerline{\includegraphics[angle=270,width=\columnwidth]{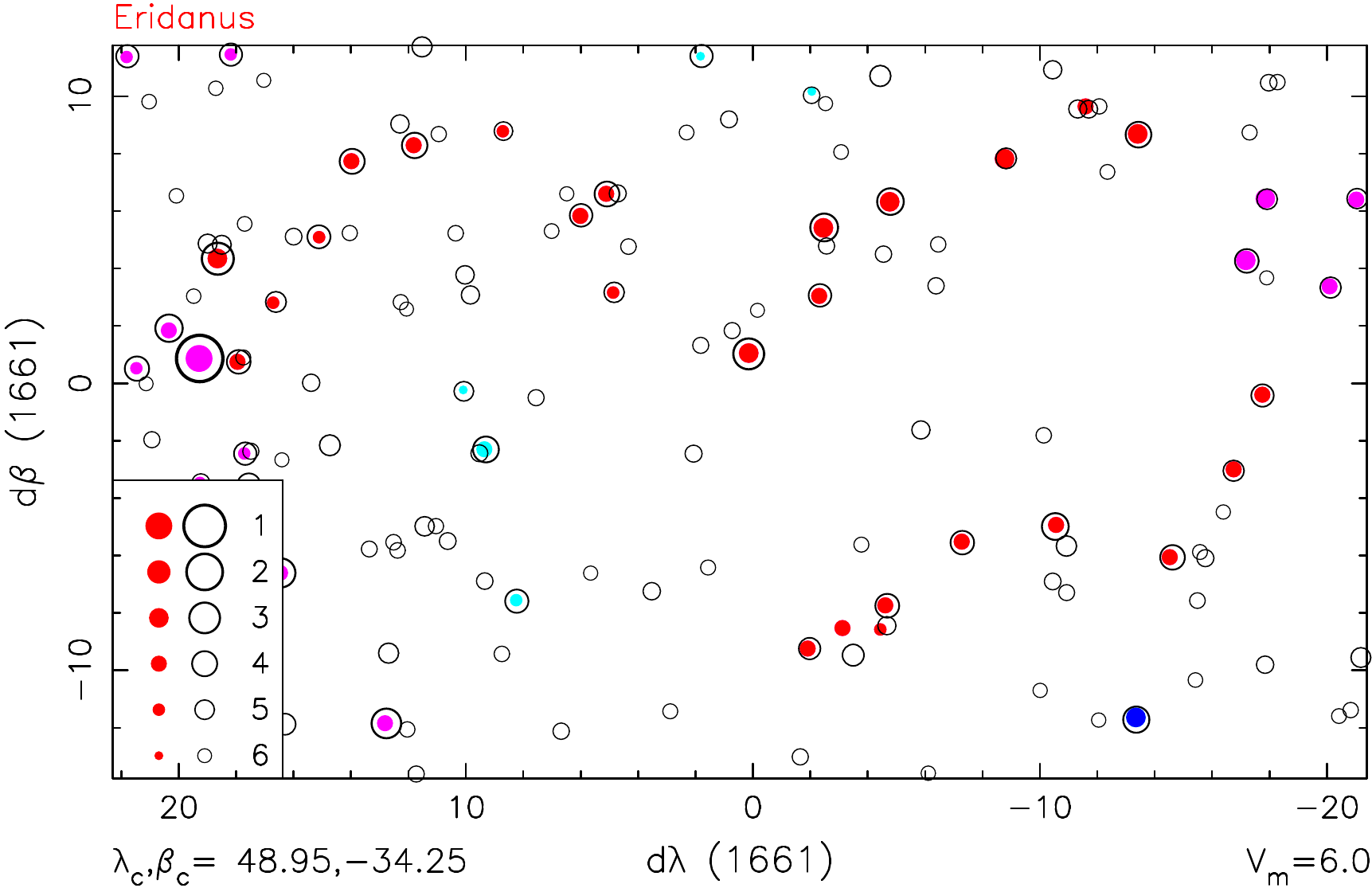}}
\caption{Eridanus
 \label{f:eridanus}}
\end{figure}

\noindent H\,714, near $-$11.6,9.7 in Fig.\,\ref{f:eridanus}, lies
between HIP\,14293 and HIP\,14168, somewhat closer to the latter. The
corresponding entry in Brahe's star catalogue K\,912 is closer to
HIP\,14293. The counterpart given by Rybka, HIP\,16989, is almost
9$^\circ$ off.

\begin{figure}
\centerline{\includegraphics[angle=270,width=\columnwidth]{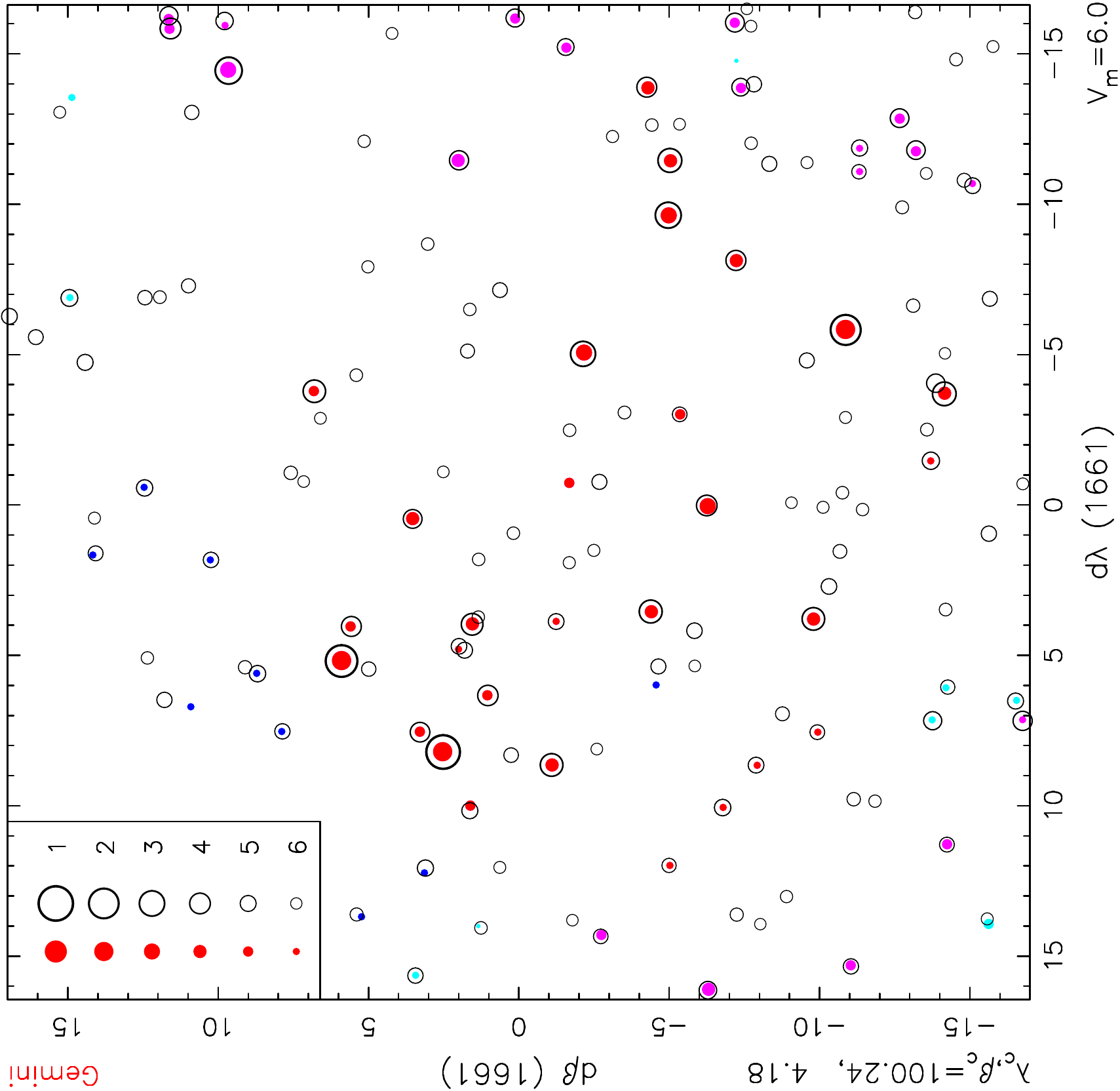}}
\caption{Gemini
 \label{f:gemini}}
\end{figure}

\noindent H\,737, near 4.8,2.0 in Fig.\,\ref{f:gemini}, lies between
HIP\,36429 and HIP\,36393, somewhat closer to the latter. The
corresponding entry in Brahe's star catalogue K\,556 is closer to
HIP\,36429. Rybka mentions both as possible counterparts.

\begin{figure}
\centerline{\includegraphics[angle=270,width=\columnwidth]{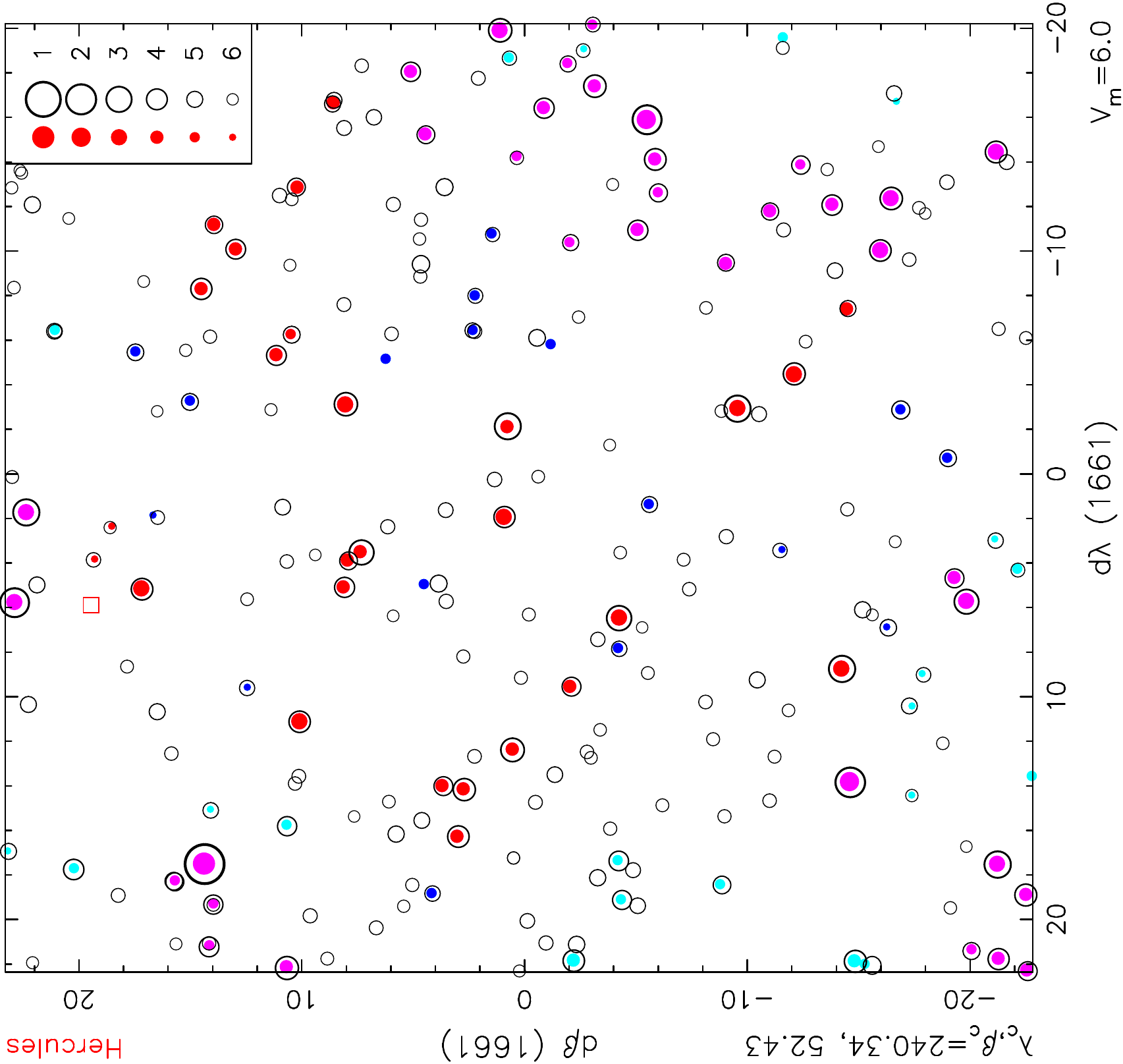}}
\caption{Hercules
 \label{f:hercules}}
\end{figure}

\noindent H\,790, near $-$16.7,8.6 in Fig.\,\ref{f:hercules}, lies
between HIP\,75793 (= $\nu^1$\,Boo) and HIP\,76041 (= $\nu^2$\,Boo),
somewhat closer to the latter. Rybka identifies this entry with
HIP\,75793.

\noindent H\,792, near 3.8,19.3 in Fig.\,\ref{f:hercules}, is
identified by Rybka with HIP\,86173 ($V$=7.3, $d$=4\farcm9).  Our
identification is brighter and closer.

\noindent H\,794, near 5.9,19.5 in Fig.\,\ref{f:hercules}, is identified with
HIP\,87280 by Rybka. Although variable, this star is too faint and too far
($V$=6.8, $d$=26\farcm4) to be a reliable counterpart. We prefer to leave this
star unidentified, as its correpondent entry K\,175 in Brahe's star catalogue
(Paper\,I).

\begin{figure*}
\centerline{\includegraphics[angle=270,width=15.6cm]{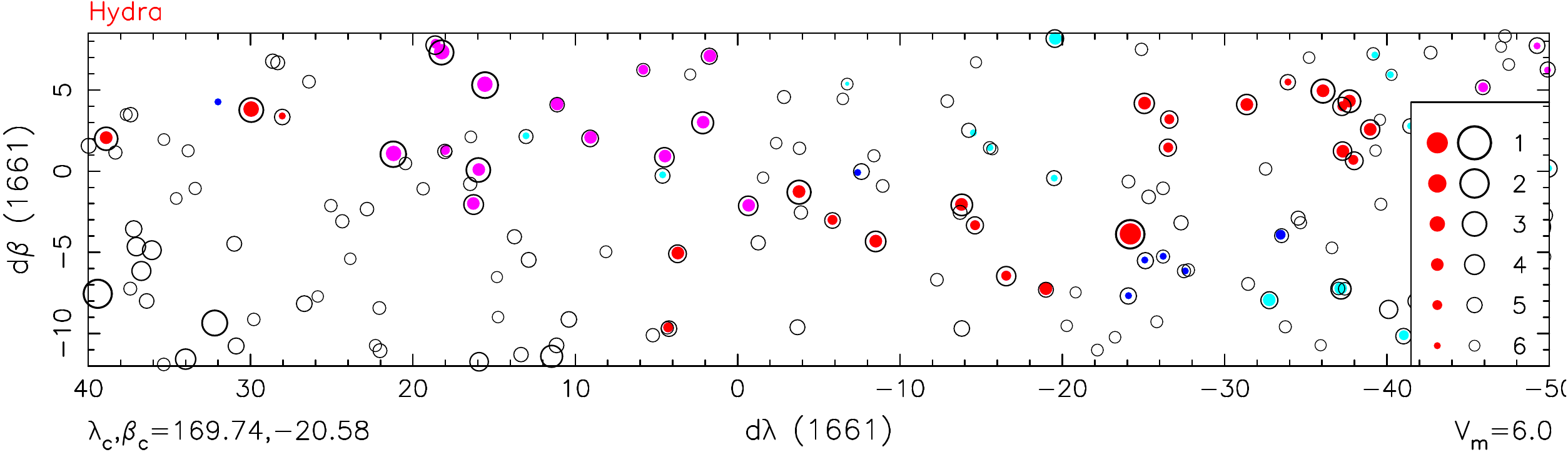}}
\caption{Hydra
 \label{f:hydra}}
\end{figure*}

\noindent H\,831, near 4.3,$-$9.6 in Fig.\,\ref{f:hydra}, is between HIP\,54255,
with which we identify it, and HIP\,54204, slightly further but brighter ($V$=4.9,
$d$=5\farcm6), with which Rybka identifies it.

\noindent H\,837, near $-$27.6,$-$6.1 in Fig.\,\ref{f:hydra}, lies between HIP\,44961,
with which we identify it, and HIP\,44833, with which Rybka identifies it, even
though it is both fainter and further ($V$=5.6, $d$=12\farcm2).

\noindent H\,838, near 32.0,4.3 in Fig.\,\ref{f:hydra} is identified with the Mira
variable R\,Hya by Rybka. Our identification procedure did not find this possibility,
due to the faint average magnitude ($V$=6.4), but the star is highly variable and
we agree with Rybka that this is a plausible identification.

\begin{figure}
\centerline{\includegraphics[angle=270,width=\columnwidth]{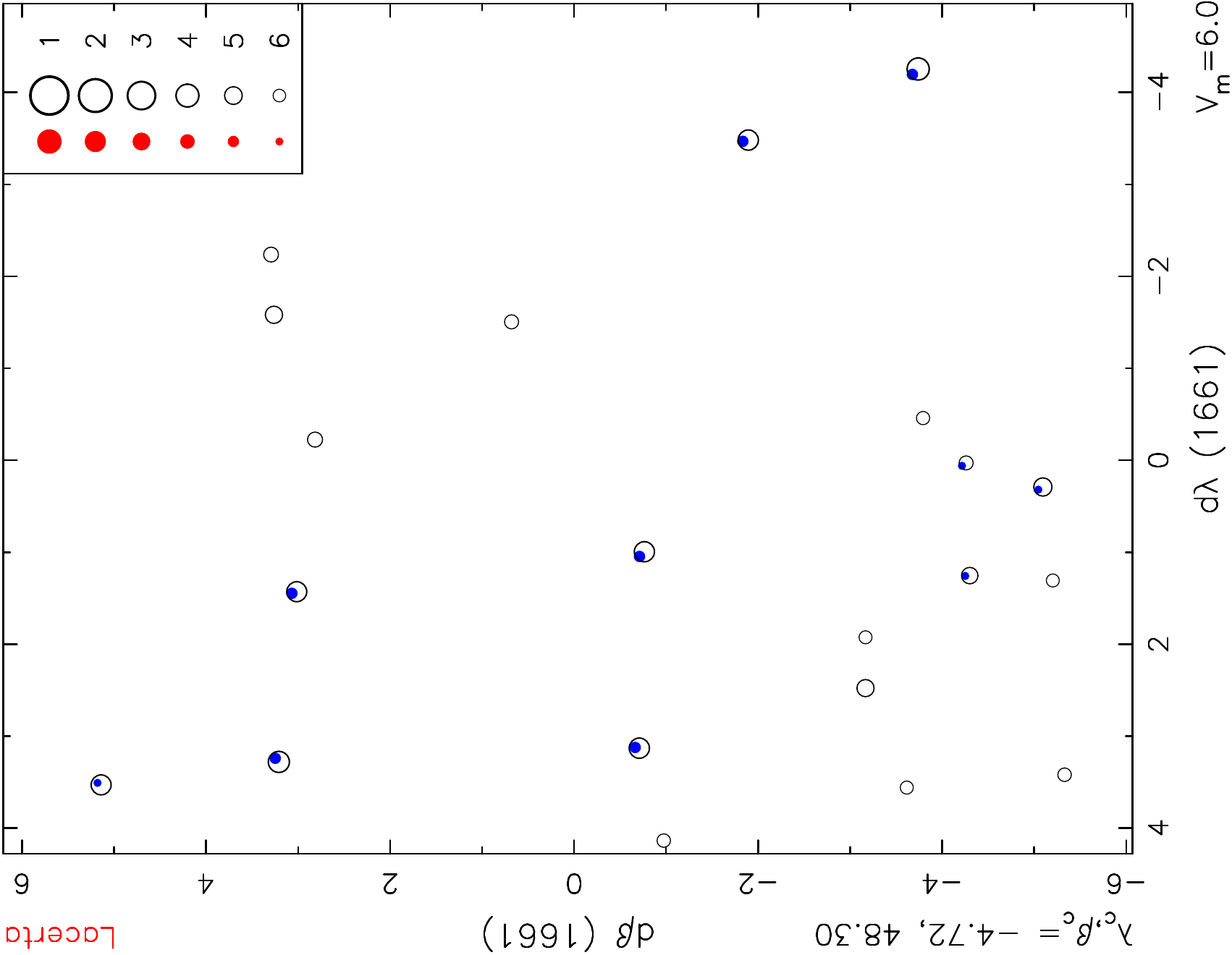}}
\caption{Lacerta
 \label{f:lacerta}}
\end{figure}

\noindent H\,849, near 0.1,$-$4.3 in Fig.\,\ref{f:lacerta}, is the combined
light of HIP\,111544 and HIP\,111546 (separation 22\farcs3)

\begin{figure}
\centerline{\includegraphics[angle=270,width=\columnwidth]{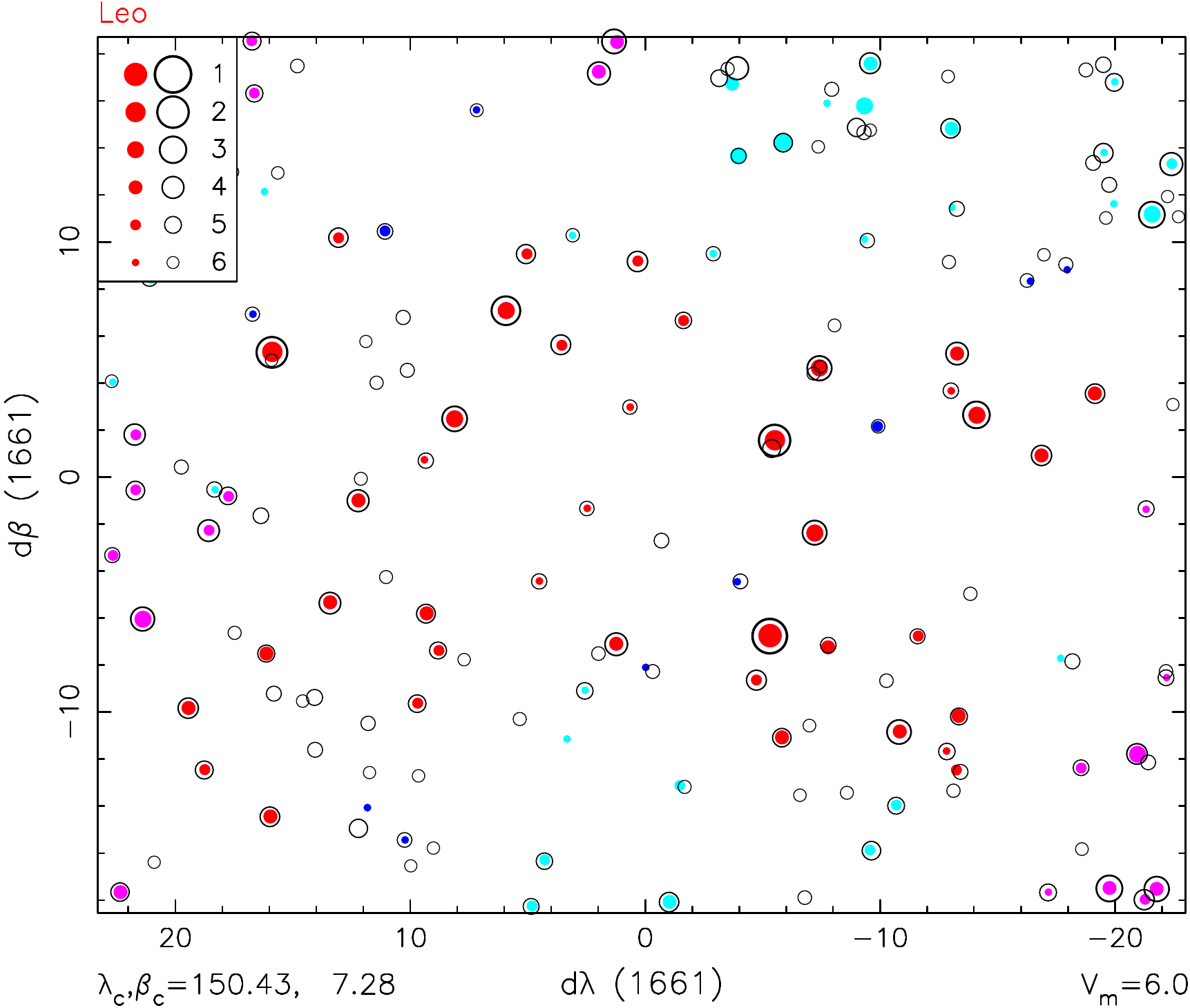}}
\caption{Leo
 \label{f:leo}}
\end{figure}

\noindent H\,887, near $-$1.6,6.7 in Fig.\,\ref{f:leo}, has a position in Hevelius
very different from its position in Brahe, where it lies above HIP\,52422 (near $-$2.9,9.5)

\noindent H\,895, near $-$3.9,$-$4.5 in Fig.\,\ref{f:leo} is identified by Rybka with
HIP\,50755, fainter and far from the correct position ($V$=6.2, $d$=104\arcmin).

\begin{figure}
\centerline{\includegraphics[angle=270,width=\columnwidth]{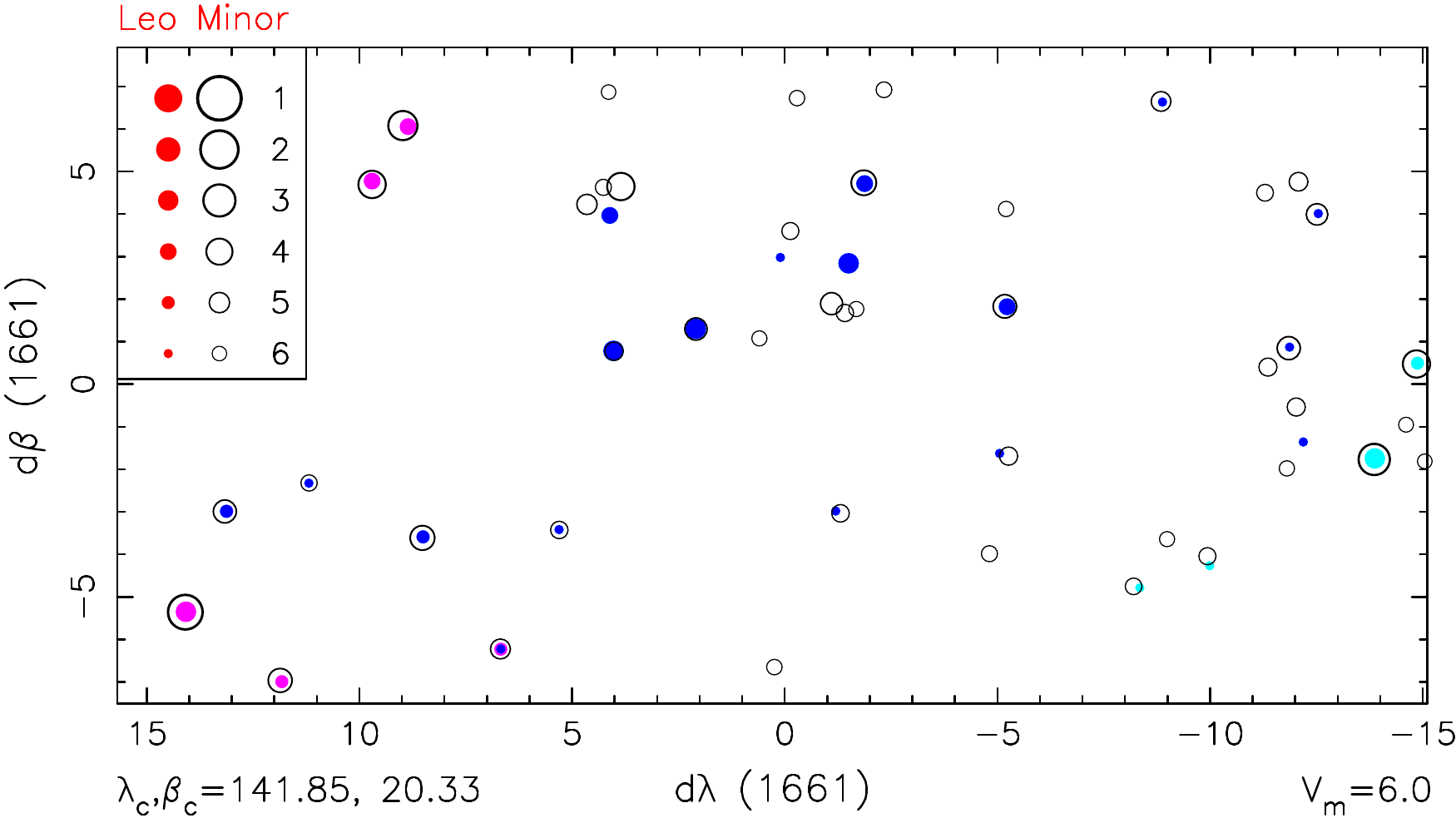}}
\caption{Leo Minor
 \label{f:leominor}}
\end{figure}

\noindent H\,905, near $-$12.2,$-$1.4 in Fig.\,\ref{f:leominor}, lies between
HIP\,46652 ($V$=5.9, $d$=44\farcm1) and HIP\,46735 ($V$=5.4, $d$=50\farcm5).
Rybka identifies H\,905 with the nearer star, we think either star is possible.

\noindent H\,913, near 4.1,4.0 in Fig.\,\ref{f:leominor}, is near three stars,
HIP\,53426 ($V$=5.0), $d$=36\arcmin), HIP\,53377 ($V$=5.7, $d$=40\arcmin) and
HIP\,53229 ($V$=3.8, $d$=44\arcmin). We agree with Rybka in considering
the brightest star to be the more plausible counterpart.

\noindent H\,936 is closer to HIP\,72603 ($\alpha^1$\,Lib, $V$=5.2,
$d$=3\farcm3) than to HIP\,72622 ($\alpha^2$\,Lib), but the latter is much
brighter and therefore the counterpart. H\,956 is the counterpart of $\alpha^1$\,Lib,
and Hevelius notes {\em non nisi Tub. vis.} (not visible without tube, i.e.\ telescope)

\begin{figure}
\centerline{\includegraphics[angle=270,width=\columnwidth]{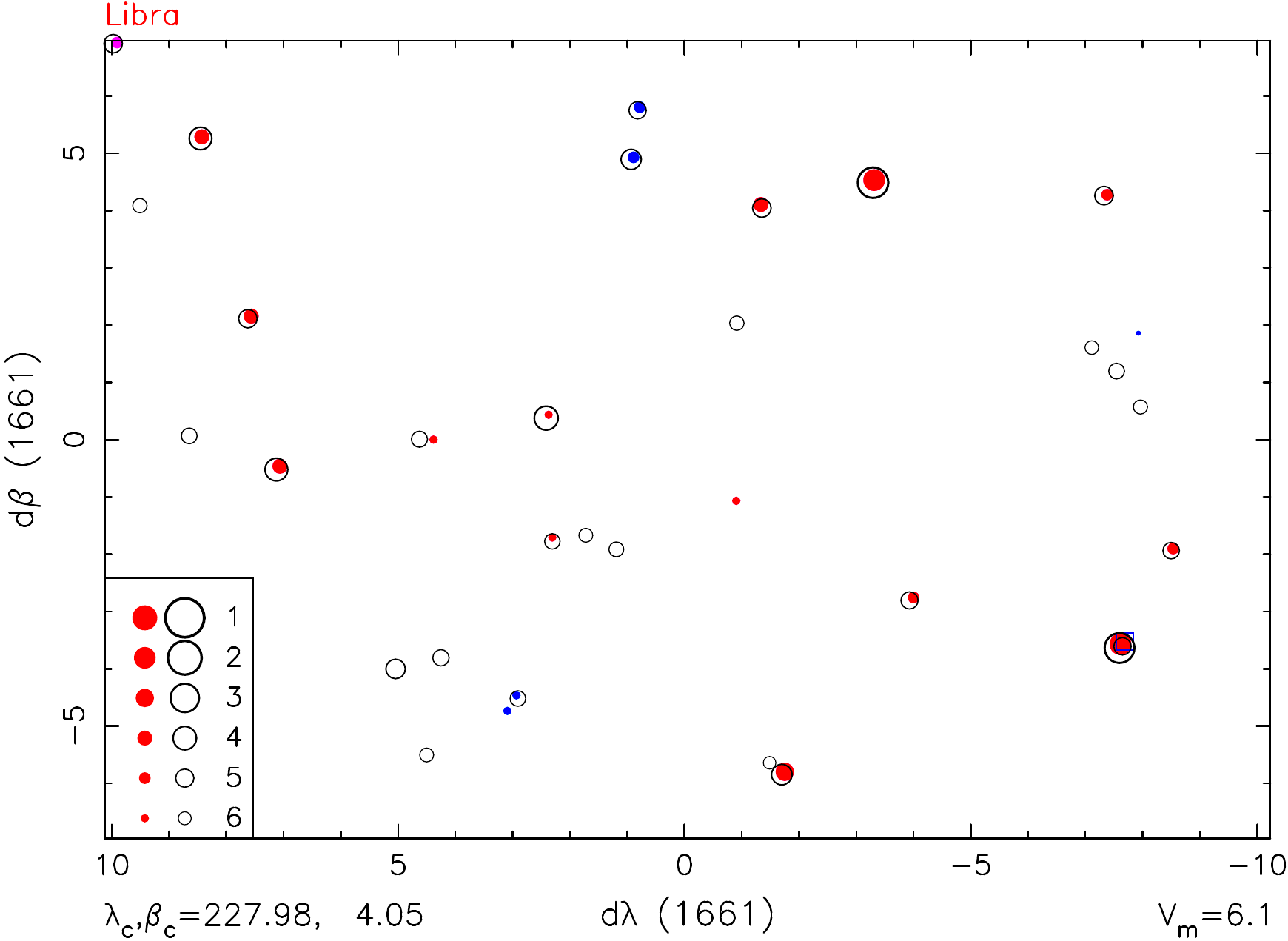}}
\caption{Libra
 \label{f:libra}}
\end{figure}

\noindent H\,954, H\,955, near 3.0,$-$4.5 in Fig.\,\ref{f:libra}, are
identified by Rybka with HIP\,76880 ($d$=125\arcmin) and HIP\,76628
($d$=88\arcmin) respectively, but both are much closer to HIP\,76106.
We identify H\,955 as the closest match to HIP\,76106, and leave H\,954
unidentified.

\noindent H\,956: see H\,936. 

\begin{figure}
\centerline{\includegraphics[angle=270,width=\columnwidth]{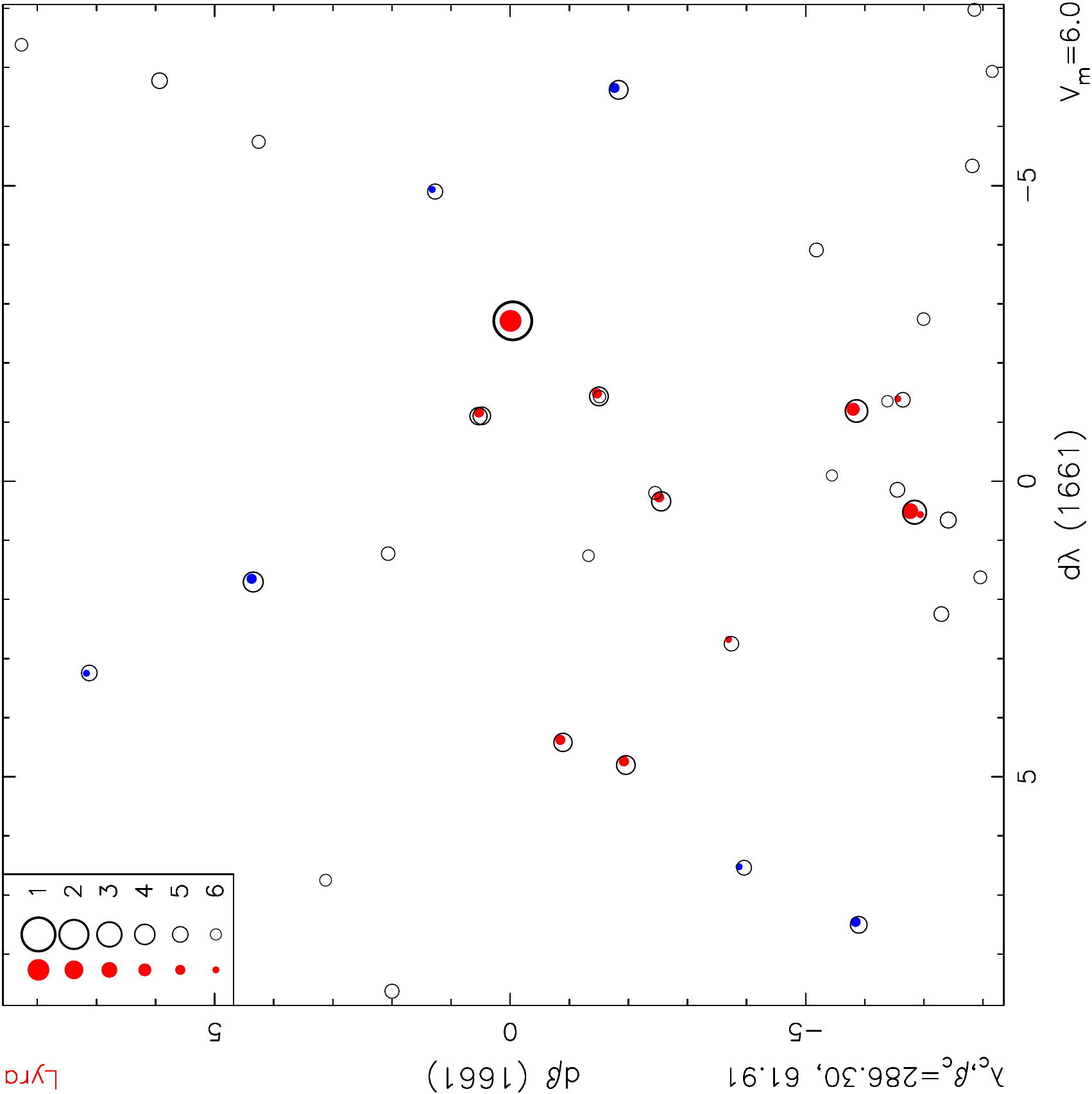}}
\caption{Lyra
\label{f:lyra}}
\end{figure}

\noindent H\,979, near $-$1.2,0.5 in Fig.\,\ref{f:lyra}, is identified by us
with HIP\,91919, but HIP\,91926 is at a comparable distance and marginally
brighter ($V$=4.6, $d$=4\farcm6). Rybka considers H\,979 to be the combined light of
the two Hipparcos stars are separated by 3\farcm5. 

\noindent H\,980, near $-$1.5,$-$1.5 in Fig.\,\ref{f:lyra}, is the
combined light of HIP\,91971 and HIP\,91973 (separated by 44\arcsec).

\noindent H\,981, near 0.3,$-$2.5 in Fig.\,\ref{f:lyra}, lies between HIP\,92791
and the fainter HIP\,92728, closer to the brighter star. Rybka lists both
stars as possible counterparts.

\noindent H\,986, near 0.6,$-$6.9 in Fig.\,\ref{f:lyra}, lies closest to
HIP\,93193, with which H\.978 is identified, and corresponds to the fainter
star to the south, HIP\.93279. In the star catalogue of Brahe its position
is closer to that counterpart.

\noindent H\,1005 (Mon\,13) has no OT number, but Hevelius gives
a Brahe position which after subtraction of his 51\arcmin\ correction
for precession, exactly matches the longitude and latitude of K\,961
(Arg\,11) in \keplere, {\em if} one accepts that Hevelius' value $Z$=4
is an emendation of Brahe's $Z$=6. In his edition of Brahe's catalogue
Rawlins (1993) makes exactly this emendation, which leads to a very
good match of K\,961 with HIP\,37447 ($\alpha$\,Mon).

\noindent H\,1027 is identified by us with HIP\,26268 ($d$=2\farcm7) and by Rybka
with HIP\,26237 ($d$=6\farcm4). The corresponding entry in Brahe's star catalogue,
K\,870, is closer to HIP\,26237: see Fig.\,C.48 in Paper\,I.

\begin{figure}
\centerline{\includegraphics[angle=270,width=\columnwidth]{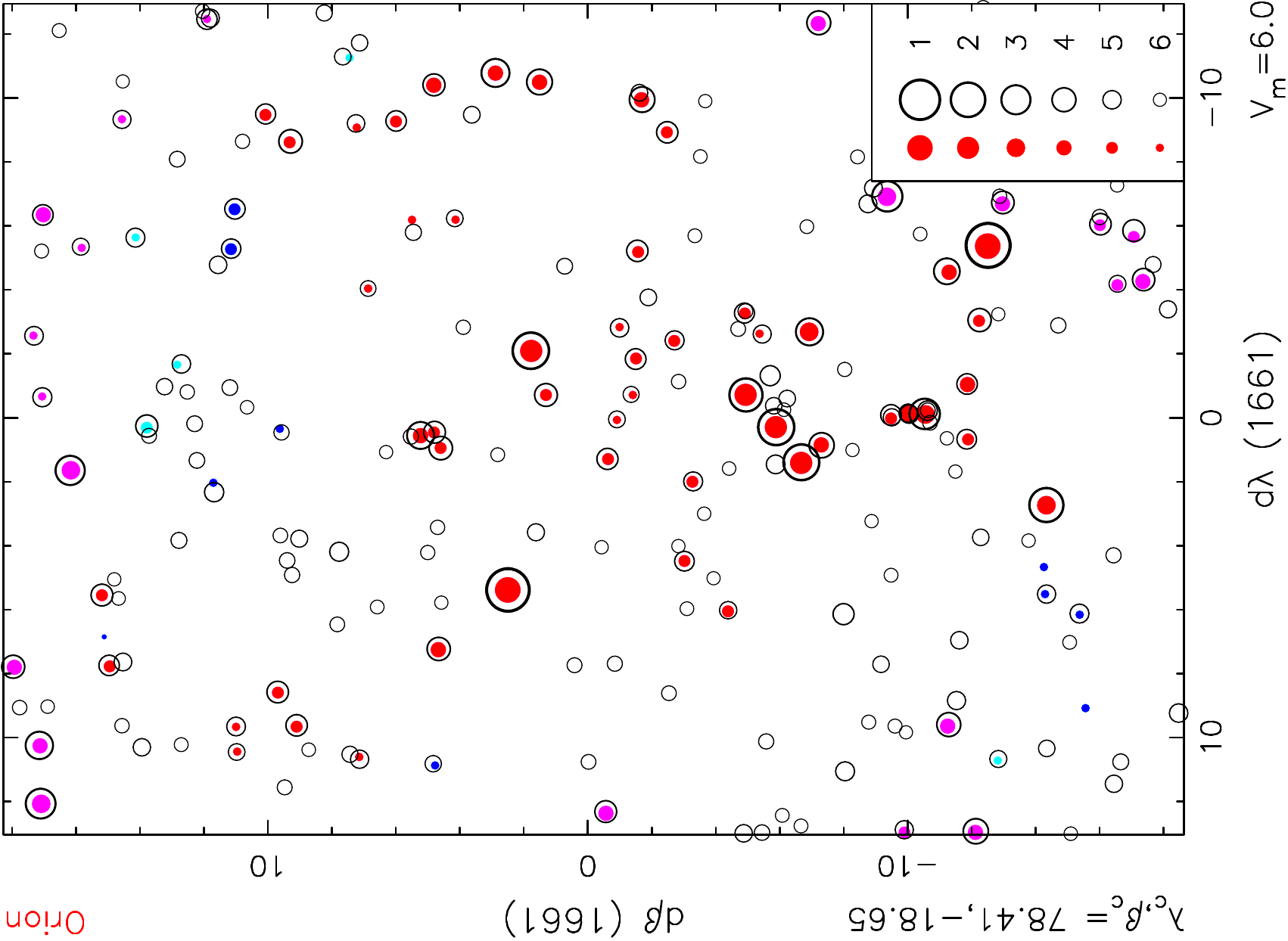}}
\caption{Orion
\label{f:orion}}
\end{figure}

\noindent H\,1063-5. Hevelius lists H\,1064 as two magnitudes fainter
than H\,1063 and H\,1065. This, and the close match in the positions of 
H\,1063 and H\,1965, near 5.6,15.2 and 7.8,15.0 in Fig.\,\ref{f:orion},
with HIP\,27913 (=$\chi^1$\,Ori) and HIP\,28716 (=$\chi^2$\,Ori), respectively,
excludes the interpretation of Rybka, who identifies H\,1064 with $\chi^2$\,Ori
and H\,1065 with HIP\,29650 (near 10.3,13.9)

\noindent H\,1073, near 4.7,$-$14.3 in Fig.\,\ref{f:orion}, is closer to HIP\,28325
which however is identified with H\,1075. We agree with Rybka that this leaves
HIP\,27713, near 3.8,$-$13.8 as a plausible identification for H\,1073.

\begin{figure}
\centerline{\includegraphics[angle=270,width=\columnwidth]{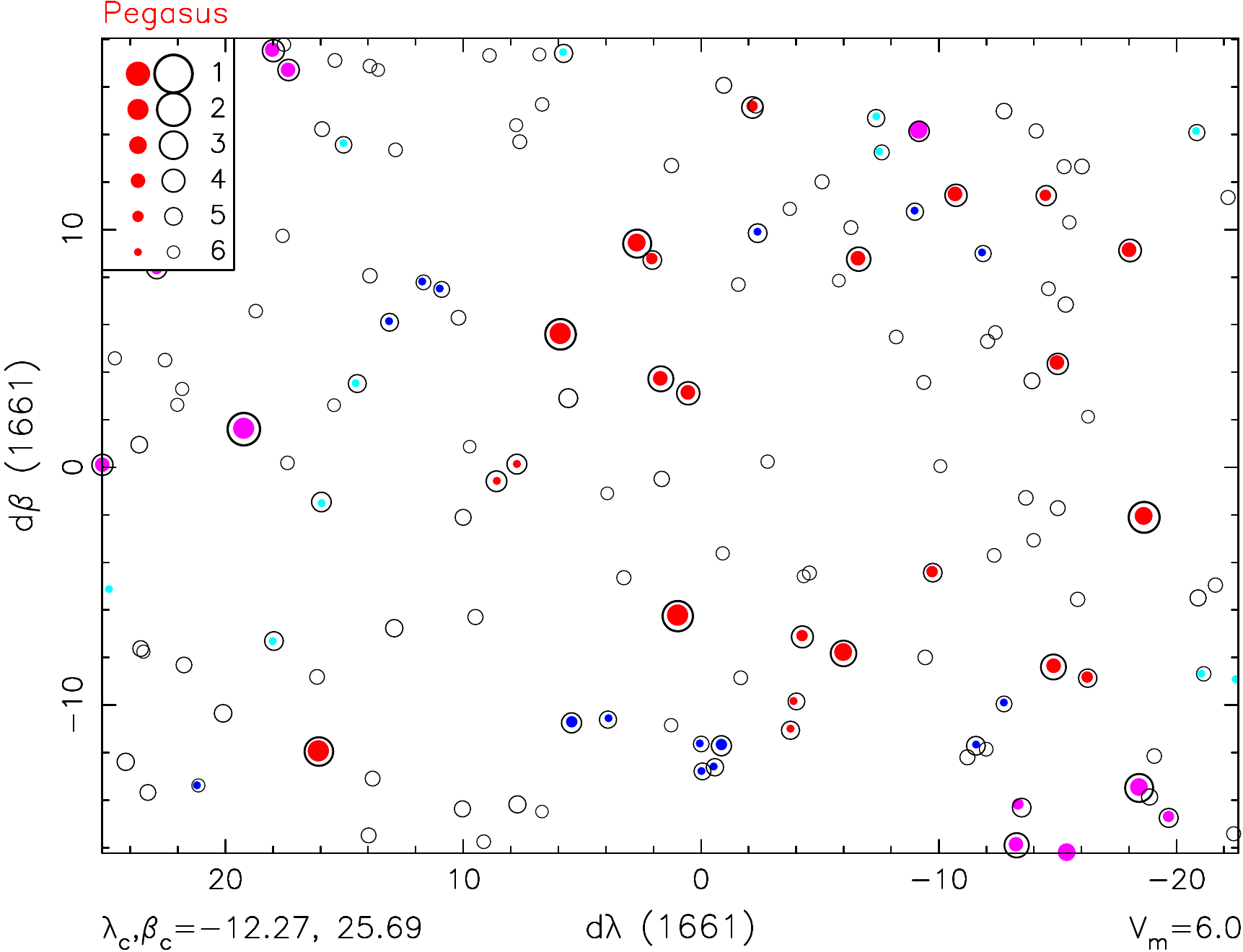}}
\caption{Pegasus
\label{f:pegasus}}
\end{figure}

\noindent H\,1099, near 8.6,$-$0.6 in Fig.\,\ref{f:pegasus}, has a
{\em very} different position in \hevelius\ from it entry K\,441 in
\keplere, which we identified with HIP\,112997 (near $-$0.9,$-$3.6).

\begin{figure}
\centerline{\includegraphics[angle=270,width=\columnwidth]{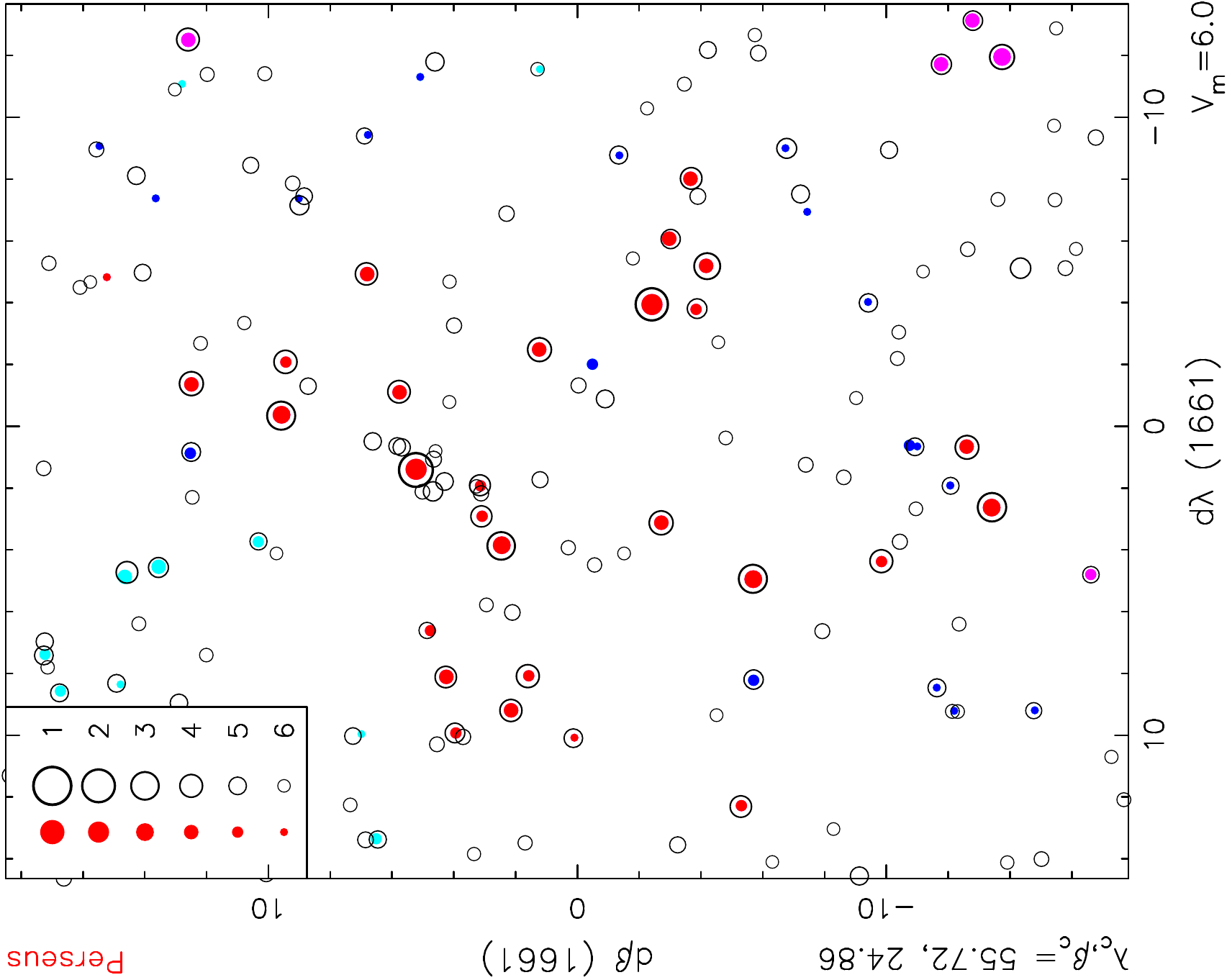}}
\caption{Perseus
\label{f:perseus}}
\end{figure}

\noindent H\,1127, near $-$4.8,15.2 in Fig.\,\ref{f:perseus}, is identified by
us with HIP\,10729 just north of it. The corresponding entry in \keplere\ K\,268 has
a position closer to HIP\,11060 (near $-$5.0,14.1) and is identified accordingly.
Rybka identifies H\,1127 with HIP\,11060.

\noindent H\,1148, near $-$7.4,9.0 in Fig.\,\ref{f:perseus}, lies
closest to HIP\,11220 ($V$=5.2, $d$=11\farcm5), but we agree with
Rybka that HIP\,11313 is the more likely counterpart

\noindent H\,1151, near 0.7,$-$11.0 in Fig.\,\ref{f:perseus}, is identified with the
same star as H\,1145, and thus a repeat entry

\noindent H\,1159, near 9.2,$-$12.2 in Fig.\,\ref{f:perseus}, lies between
HIP\,20579 and HIP\,20591, slightly closer to the former, with which we identify
it. Rybka mentions both as possible counterparts.

\begin{figure}
\centerline{\includegraphics[angle=270,width=\columnwidth]{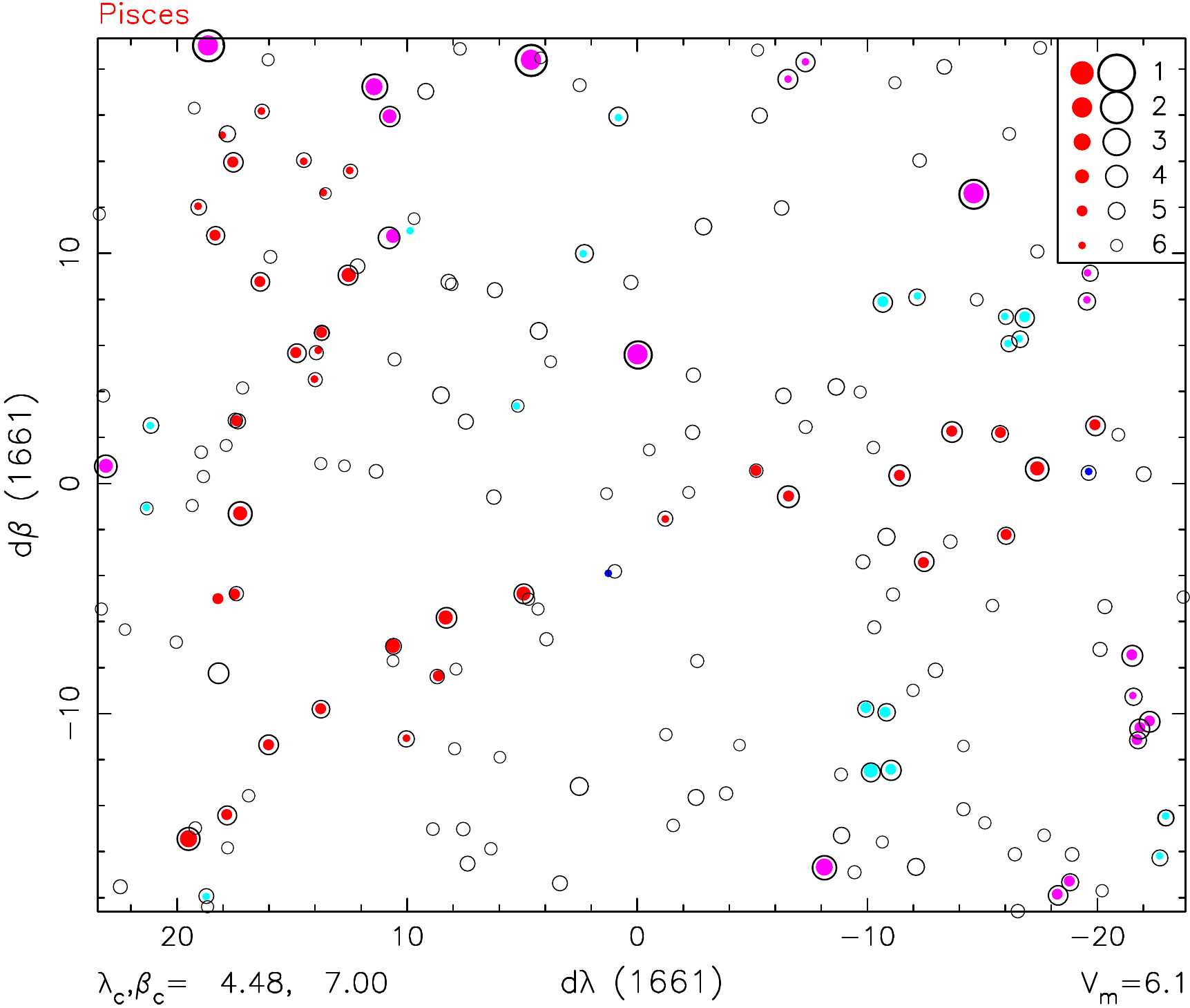}}
\caption{Pisces
\label{f:pisces}}
\end{figure}

\noindent H\,1185, near 17.4,2.7 in Fig.\,\ref{f:pisces}, lies between HIP\,6732,
which we take as the counterpart, and the slightly brighter and slightly more distant
HIP\,6706 ($V$=5.4, $d$=4\farcm2).

\noindent H\,1188, near 18.2,$-$5.0 in Fig.\,\ref{f:pisces}, lies
closest to HIP\,7535 as did its counterpart in \kepler; in
\keplere\ we emended its latitude sign from B to A which gives an
excellent match with HIP\,8198 (near 18.2,$-$8.3). Remarkably,
Hevelius also gives a northern latitude.

\noindent H\,1191, near 13.7,12.6 in Fig.\,\ref{f:pisces}, is identified with
a $V$=6.1 star, as was its counterpart K\,808 in \keplere.

\begin{figure}
\centerline{\includegraphics[angle=270,width=\columnwidth]{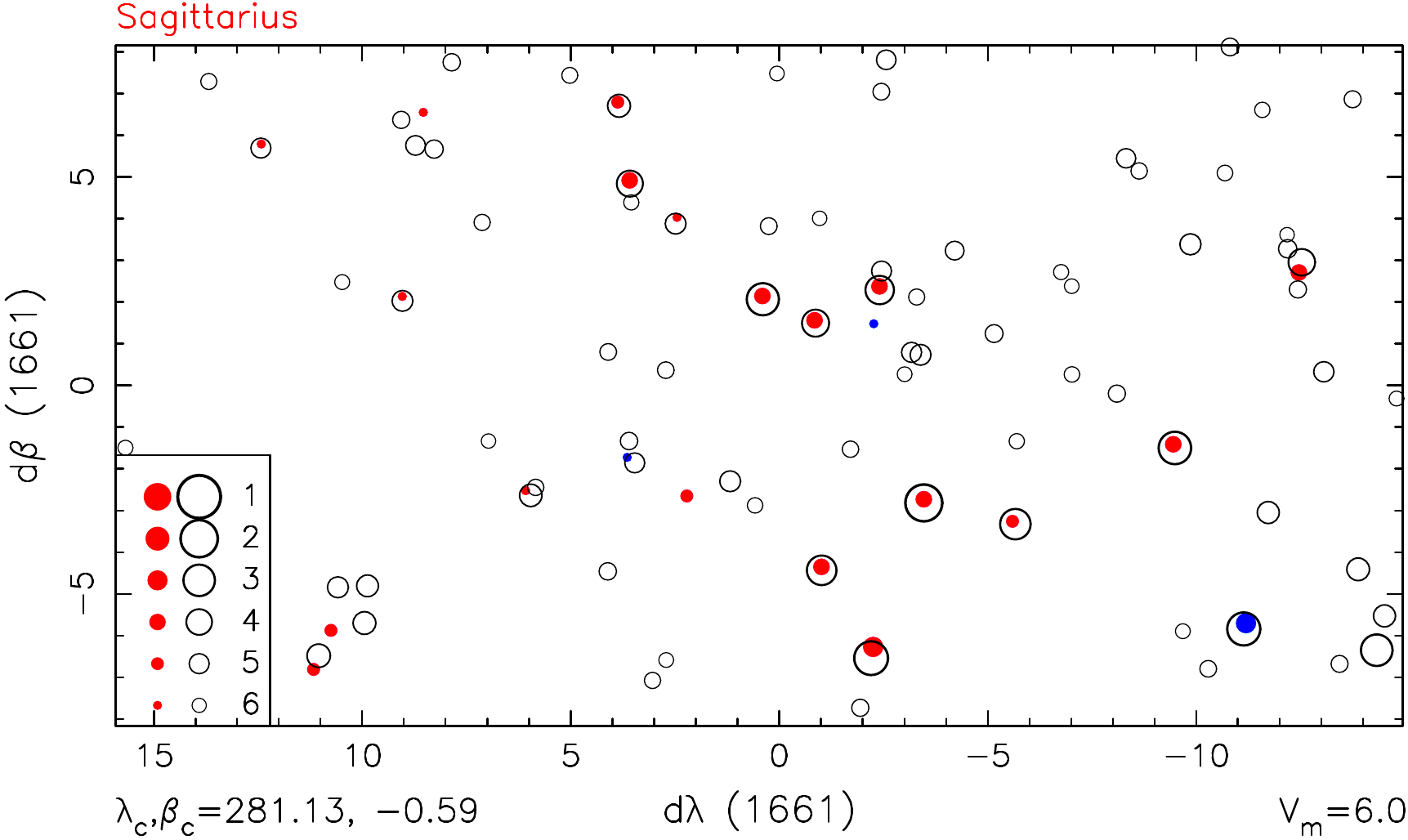}}
\caption{Sagittarius
\label{f:sagittarius}}
\end{figure}

\noindent H\,1226, near 8.5,6.6 in Fig.\,\ref{f:sagittarius}, is identified by
us with HIP\,97063. Rybka identifies H\,1226 with HIP\,96950 (near 8.7,5.8) which
is our identification of the corresponding entry K\,709 in \keplere, which had
a position closer to HIP\,96950.

\noindent H\,1229, near 10.8,$-$5.9 in Fig.\,\ref{f:sagittarius}, is closest to
HIP\,98688, which however is the counterpart of H\,1232. Rybka identifies H\,1229
with HIP\,98066 (near 9.9,$-$4.8, $V$=4.7) but HIP\,98162 (near 10.0,$-$5.7) is
closer and slightly brighter.

\noindent H\,1234, near $-$2.3,1.5 in Fig.\,\ref{f:sagittarius}, is
identified by Rybka with the pair $\nu^1$-$\nu^2$\,Sgr (HIP\,92761 and
HIP\,92845, near $-$3.3,0.8). We think this is plausible, but not
certain.  $\nu^1$-$\nu^2$\,Sgr are separated by 13\farcm8;
HIP\,92845 is closest to H\,1234.

\begin{figure}
\centerline{\includegraphics[angle=270,width=\columnwidth]{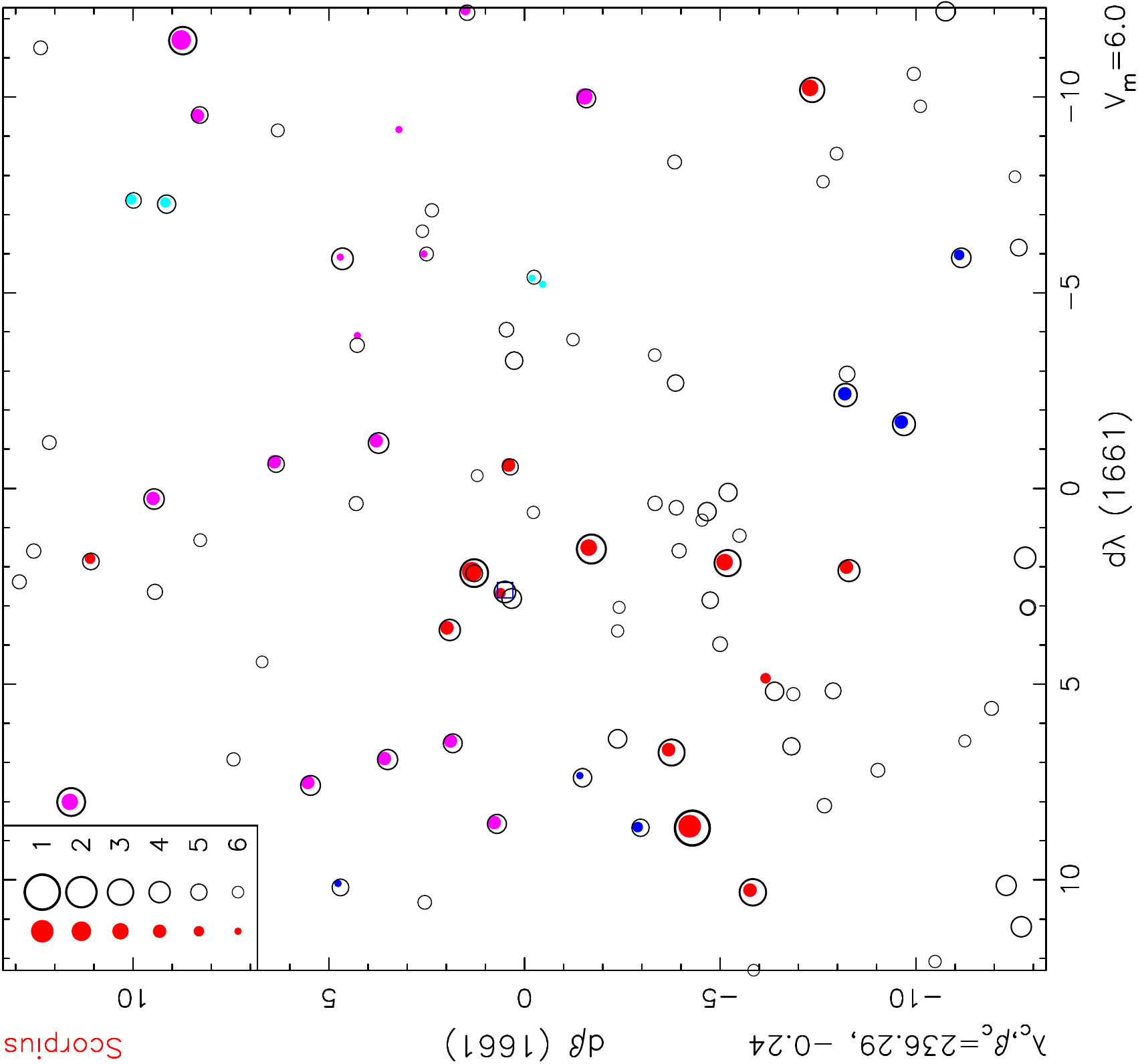}}
\caption{Scorpius
\label{f:scorpius}}
\end{figure}

\noindent H\,1248, near 7.3,$-$1.4 in Fig.\,\ref{f:scorpius}, is identified by
Rybka with HIP\,80079 (near 6.4,$-$2.4), rather further from H\,1248 than our
identification.

\noindent H\,1250 is located very close to H\,1245 by Hevelius: {\em
  Prope Supr. Front.  Duar. Austr. Infer. adhaeret alteri, non nisi
  Tubo visibilis: ex occultatione stabilita}. H\,1245 is Hip\,78933
($\omega^1$\,Sco), and it would be tempting to identify H\,1250 with
HIP\,78990 ($\omega^2$\,Sco). As Rybka remarks, two stars separated by
14\farcm6 do not fit this description.

\noindent H\,1280 and H\,1281 have positions rather different from the
corresponding entries in \keplere, K\,354 and K\,355. In
\keplere\ these stars have very uncertain identifications (see
Fig.\,C.20 of Paper\,I), but the more accurate positions in Hevelius
lead to unambiguous identifications.

\noindent H\,1293 corresponds to the combined light of HIP\,81634 and HIP\,81641,
separated by 1\farcm1 and thus not separable by the naked eye.

\begin{figure}
\centerline{\includegraphics[angle=270,width=\columnwidth]{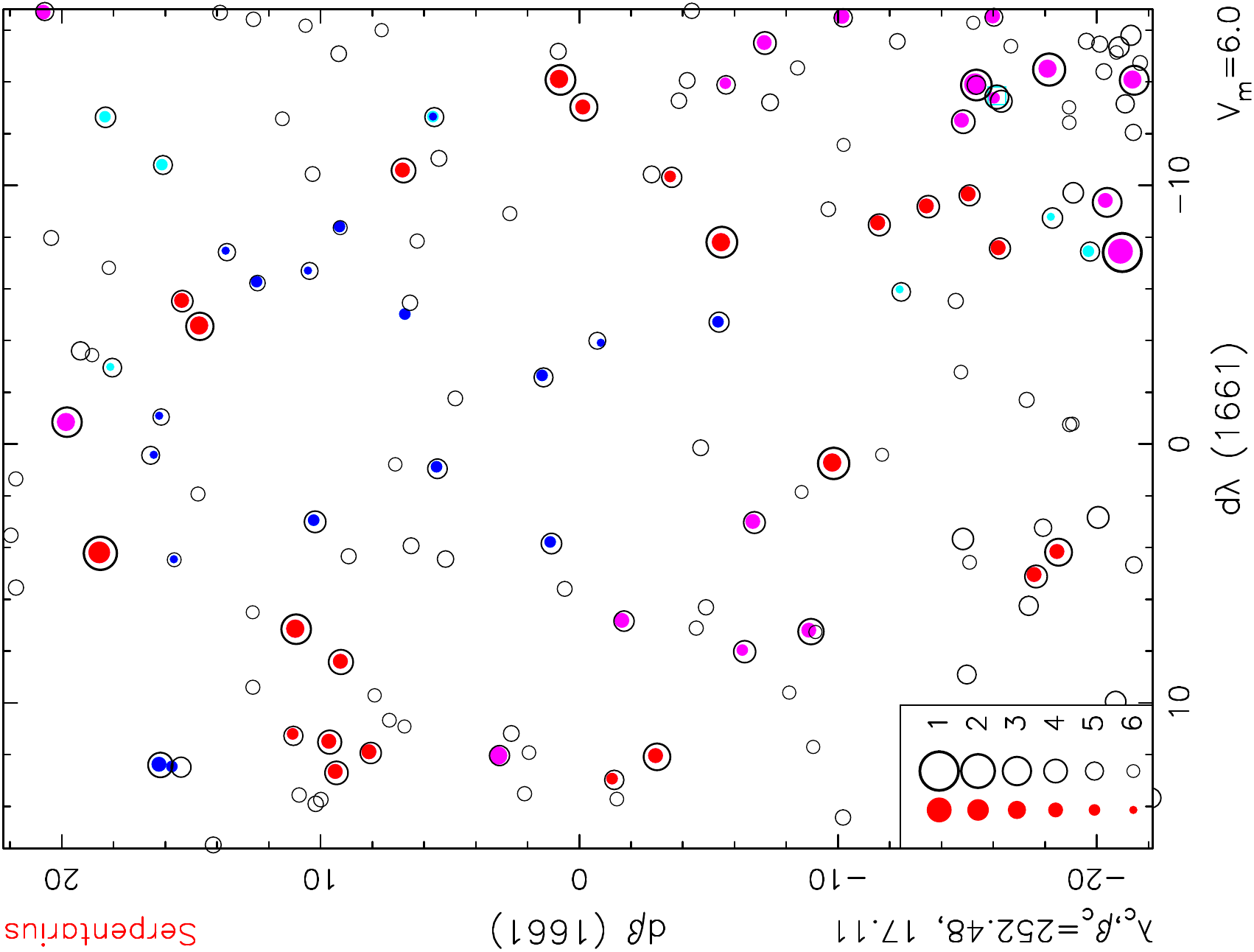}}
\caption{Serpentarius
\label{f:serpentarius}}
\end{figure}

\noindent H\,1296, near 0.4,16.5 in Fig.\,\ref{f:serpentarius}, is identified
by Rybka with HIP\,84626, near 3.2,$-$17.9, clearly not correct.

\noindent H\,1337 is identified by Rybka with HIP\,51592, indeed closer than
our suggested identification, but also much fainter ($V$=7.1, $d$=34\farcm2)

\noindent H\,1346 lies right between HIP\,20885 ($\theta^1$\,Tau) and HIP\,20894
($\theta^2$\,Tau), which stars are separated by 5\farcm6, and is probably the
combined light of the two.

\begin{figure}
\centerline{\includegraphics[angle=270,width=\columnwidth]{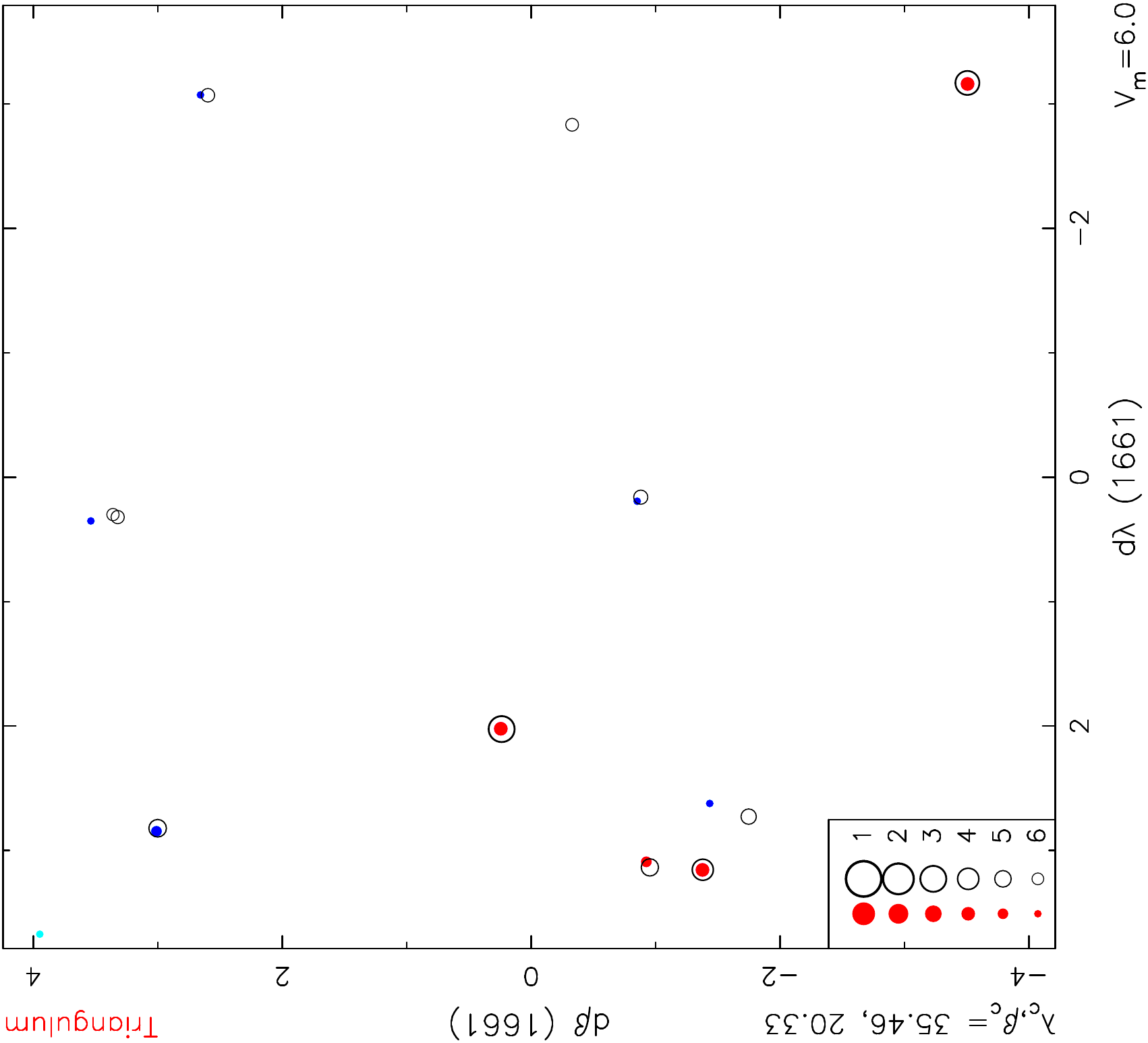}}
\caption{Triangulum Maius
\label{f:triangulum}}
\end{figure}

\noindent H\,1397, near 0.4,3.5 in Fig.\,\ref{f:triangulum}, lies north of HIP\,9001
and HIP\,9021. We identify with the nearer HIP\,9001, Rybka with the slightly brighter
HIP\,9021 ($V$=5.7), but since the separation between these two stars is just
3\farcm3, H\,1397 may correspond to the combined light of them.

\begin{figure}
\centerline{\includegraphics[angle=270,width=\columnwidth]{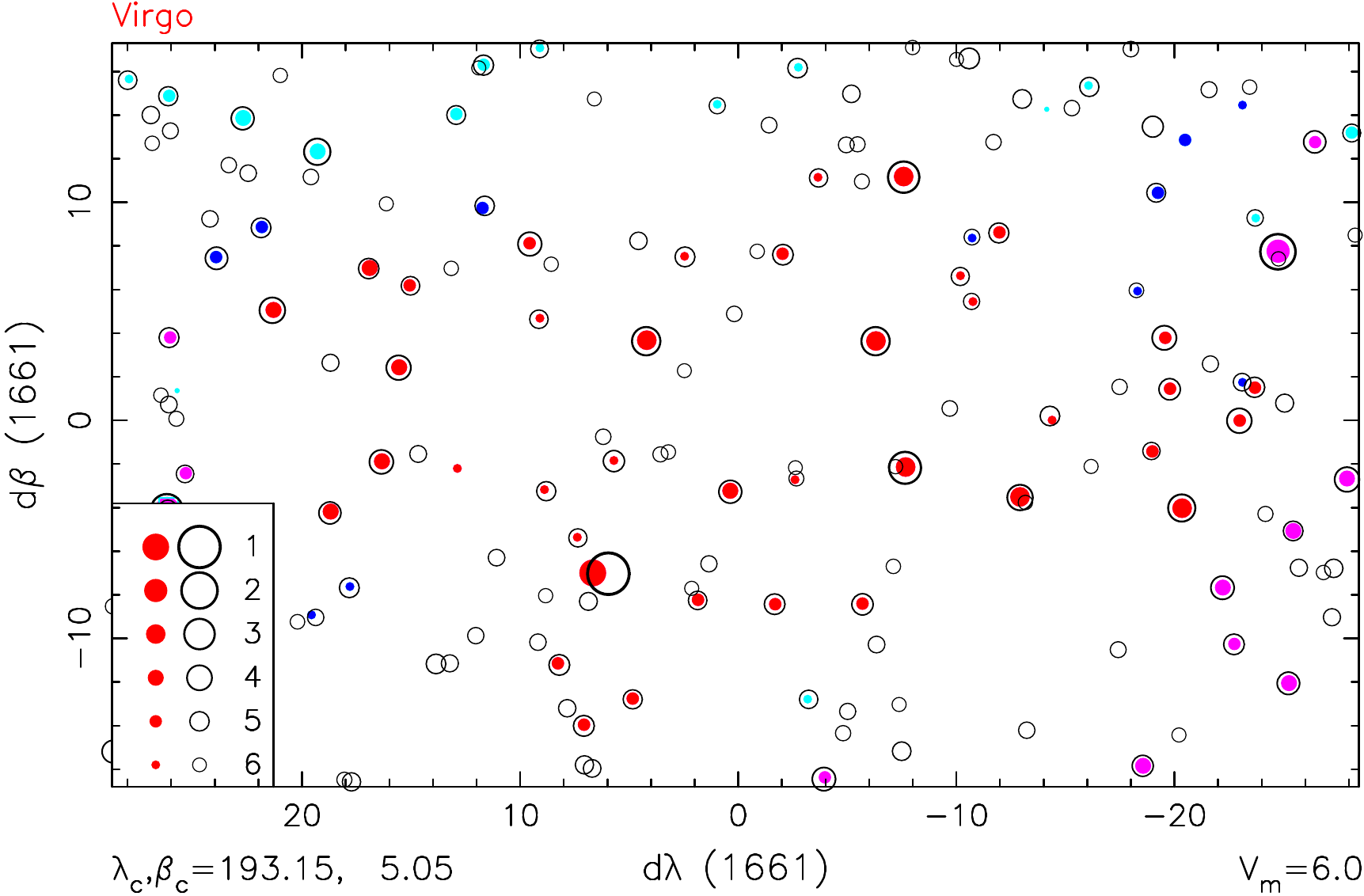}}
\caption{Virgo
\label{f:virgo}}
\end{figure}

\noindent H\,1442, near $-$23.1,14.5 in Fig.\,\ref{f:virgo}, lies closest to HIP\,58858
($V$=5.9, $d$=58\farcm4), but we agree with Rybka that HIP\,59501 (near $-$21.6,15.2)
is the more plausibe identification, with a similar offset as H\,1450 (near $-$20.5,12.9)
to its identification HIP\,60202.

\noindent H\,1451, near 17.8,$-$7.6 in Fig.\,\ref{f:virgo}, is identified by Rybka
with 40\,Vir = $\psi$\,Vir, which however is the identification of H\,1437.

\begin{figure}
\centerline{\includegraphics[angle=270,width=\columnwidth]{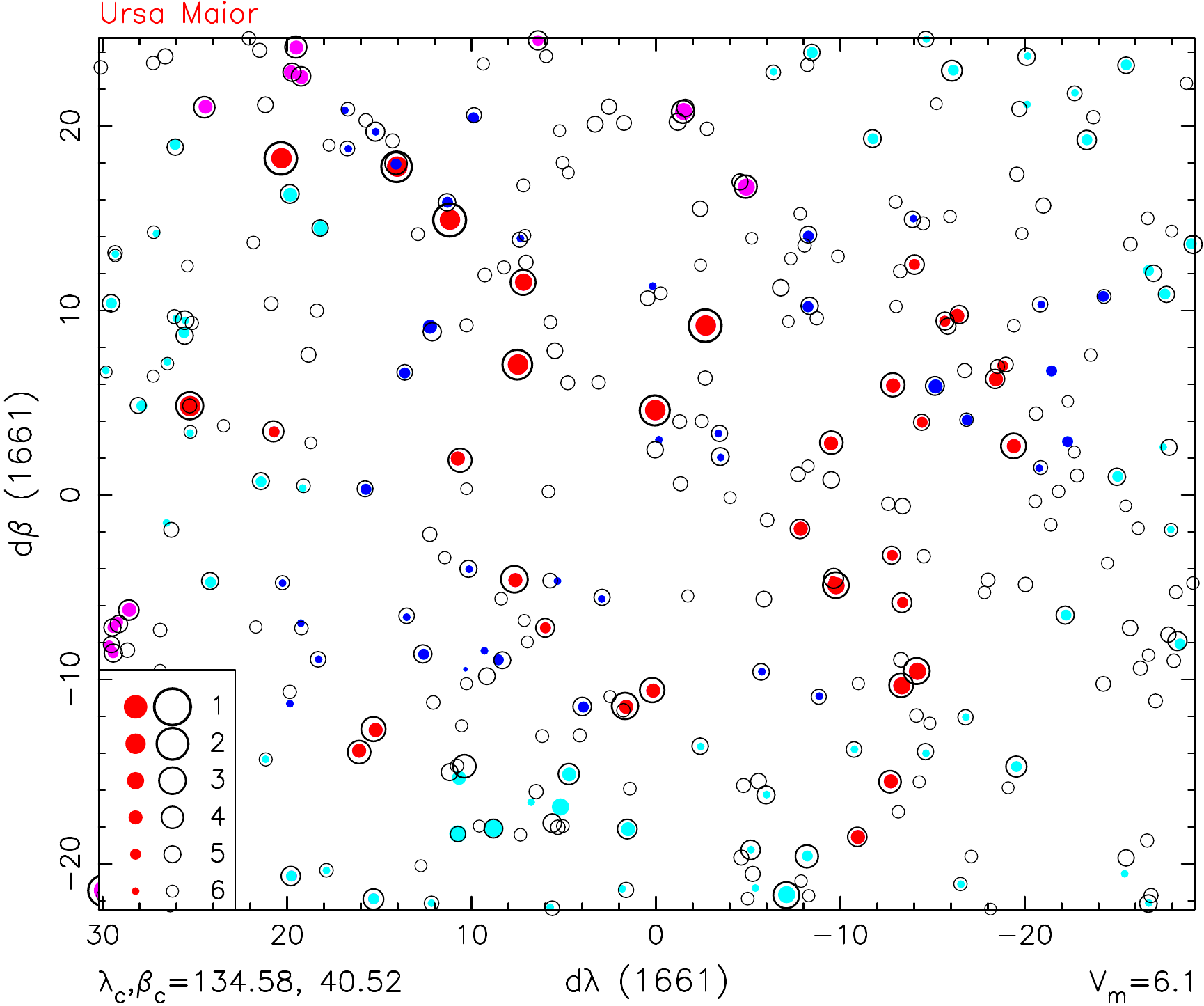}}
\caption{Ursa Maior
\label{f:ursamaior}}
\end{figure}

\noindent H\,1460, near 25.3,4.8 in Fig.\,\ref{f:ursamaior}, is the combined light
of HIP\,63121 and HIP\,63125, separated by 19\arcsec.

\noindent H\,1499, near $-$21.5,6.7 in Fig.\,\ref{f:ursamaior}, is unidentified.

\noindent H\,1505 and H\,1519, respectively near 10.3,$-$9.5 and
9.3,$-$8.5 in Fig.\,\ref{f:ursamaior}, are identified by Rybka with
HIP\,54136 and HIP\,53838, respectively near 10.3,$-$10.2 and 9.2,$-$9.8.
This is an attractive solution, but the offsets are large and we consider
it uncertain.

\noindent H\,1521, near 0.2,11.3 in Fig.\,\ref{f:ursamaior} lies close
to both HIP\,56035, which we suggest as the identification, and
HIP\,56290, slighly brighter but further ($V$=5.5, $d$=43\farcm0),
suggested by Rybka as the identification.

\begin{figure}
\centerline{\includegraphics[angle=270,width=\columnwidth]{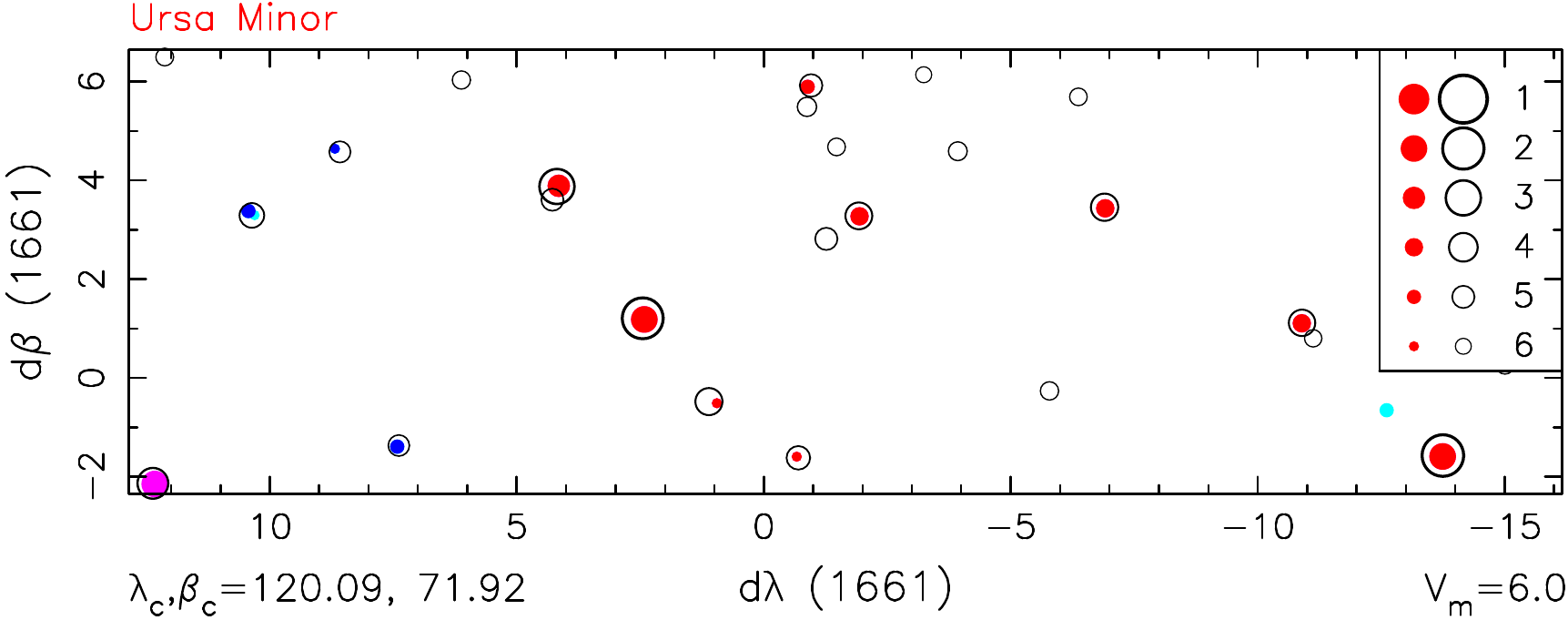}}
\caption{Ursa Minor
\label{f:ursaminor}}
\end{figure}

\noindent H\,1536, near 10.4,3.4 in Fig.\,\ref{f:ursaminor}, is
identified by Rybka with HIP\,74605, near 8.6,4.6, which is a close match
for H\,1537, and identified by us accordingly

\noindent H\,1537, near 8.7,4.6  in Fig.\,\ref{f:ursaminor}, is
identified by Rybka with HIP\,74272 ($V$=6.2, $d$=38\farcm8).

\begin{figure}
\centerline{\includegraphics[angle=270,width=\columnwidth]{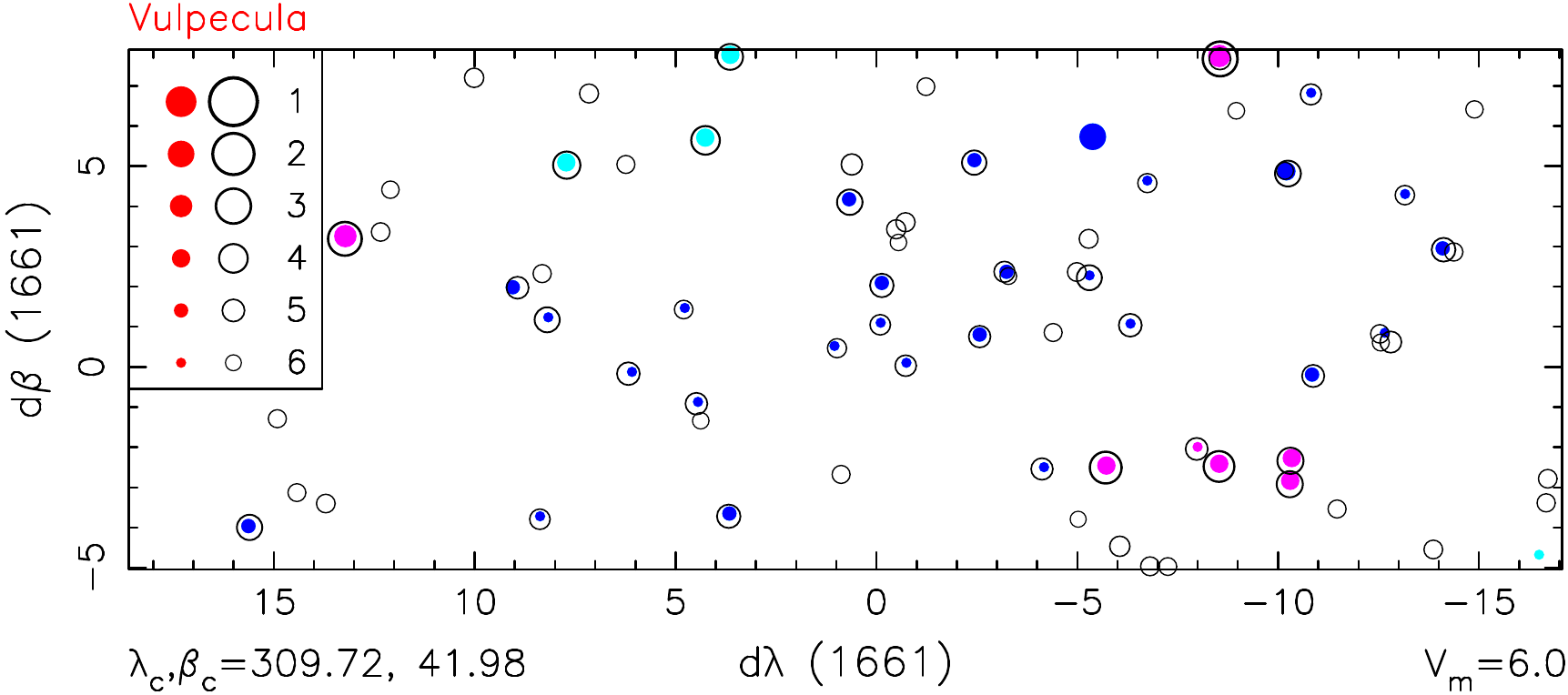}}
\caption{Vulpecula
\label{f:vulpecula}}
\end{figure}

\noindent H\,1540, near $-$5.4,5.7 in Fig.\,\ref{f:vulpecula}, in Nova
Vul 1670, whose modern counterpart CK Vul was discovered at $m_R$=20.7
in 1983 (Shara et al.\ 1985).  \nocite{shara85} We use the position
given in that paper to compute the offset with respect to the position
given by Hevelius, since CK\,Vul is too faint for the Hipparcos
catalogue.

\noindent H\,1561, near $-$12.7,0.9 in Fig.\,\ref{f:vulpecula}, is
close to three stars. We identify it with the nearest. The brightest of
the three, HIP\,95498 ($V$=5.1, $d$=16\farcm6) is too far, the
identification suggested by Rybka, HIP95582 ($V$=5.8, $d$=15\farcm9),
is both too far and too faint.

\section{Figures \label{s:figures}}

To illustrate and clarify our identifications we provide a Figure for
each constellation.  In these figures the stars listed with the
constellation in \hevelius\ are shown red when Hevelius gives them an
OT value ($>0$) indicating their presence in \kepler\ or in {\em
  Secunda classis}, and blue otherwise (OT=0).  Other stars listed in
\hevelius\ are light-red (OT$>$0) and light-blue (OT=0), respectively.
To minimize deformation of the constellations, we determine the center
of the constellation $\lambda_c,\beta_c$ from the extremes in
$\lambda$ and $\beta$, compute the rotation matrix which moves this
center to $(x,y,z)=(1,0,0)$, and then apply this rotation to each of
the stellar positions $\lambda_i,\beta_i$.  (For exact details see
Paper\,I.)  The resulting $y,z$ values correspond roughly to
differences in longitude and latitude, exact at the center
$\lambda_c,\beta_c$ and increasingly deformed away from the center. We
plot the rotated positions of the stars in \hevelius\ as
$d\lambda\equiv y$ and $d\beta\equiv z$ with filled circles.  The same
rotation matrix is applied to all stars down to a magnitude limit
$V_\mathrm{m}$ (usually $V_m=6.0$) from the {\em Hipparcos Catalogue}
and those in the field of view are plotted as open circles.  The
symbol sizes are determined from the magnitudes as indicated in the
legenda. The used values for $\lambda_\mathrm{c}$, $\beta_\mathrm{c}$
and $V_\mathrm{m}$ are indicated with each Figure.

Where necessary we show enlarged detail Figures; for easy comparison with
the Figures showing the whole constellation, these detail Figures use the
same rotation center (and thus rotation matrix).

\clearpage

\begin{figure}
\centerline{\includegraphics[angle=270,width=0.8\columnwidth]{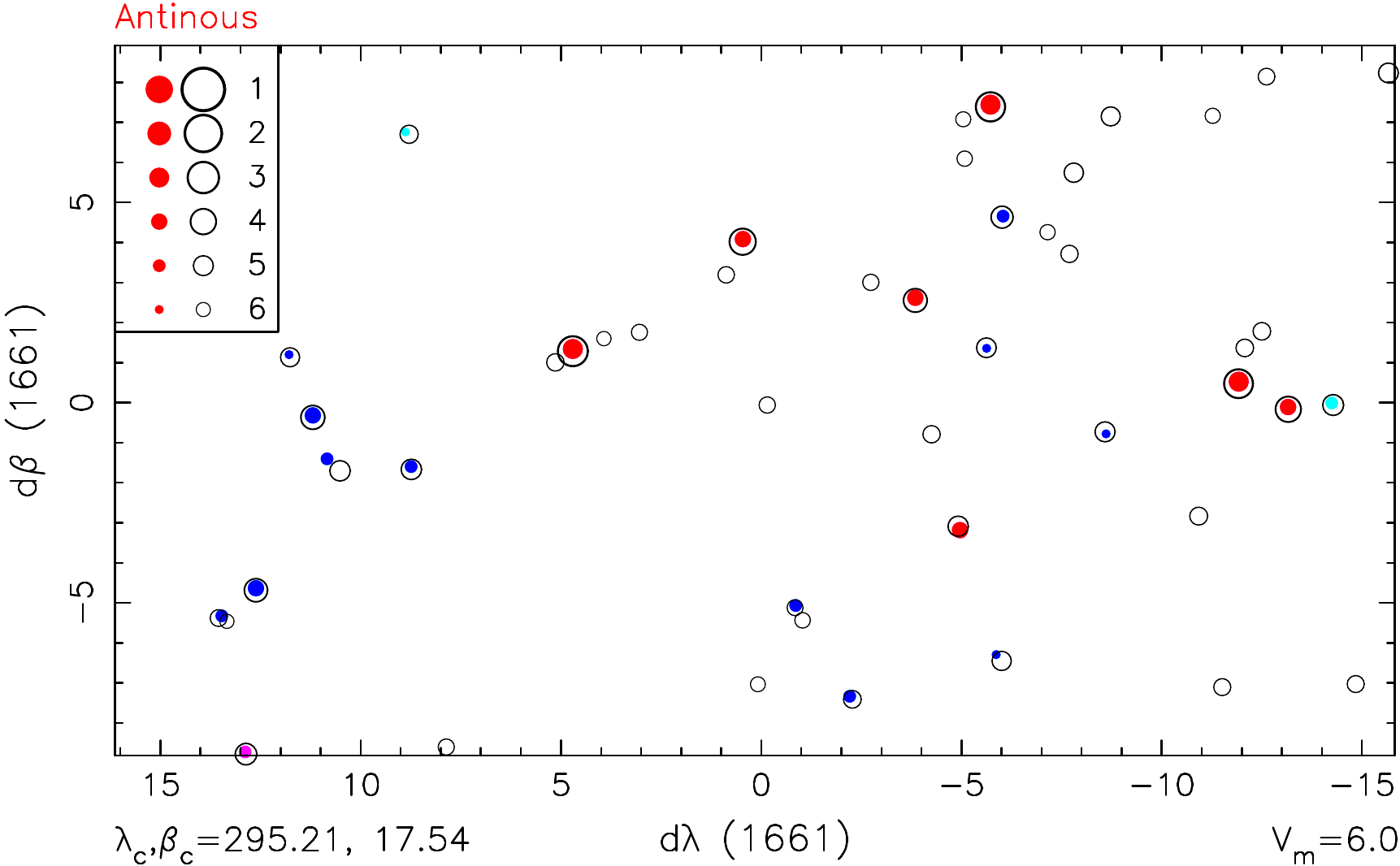}}
\caption{Antinous
 \label{f:antinous}}
\end{figure}

\begin{figure}
\centerline{\includegraphics[angle=270,width=\columnwidth]{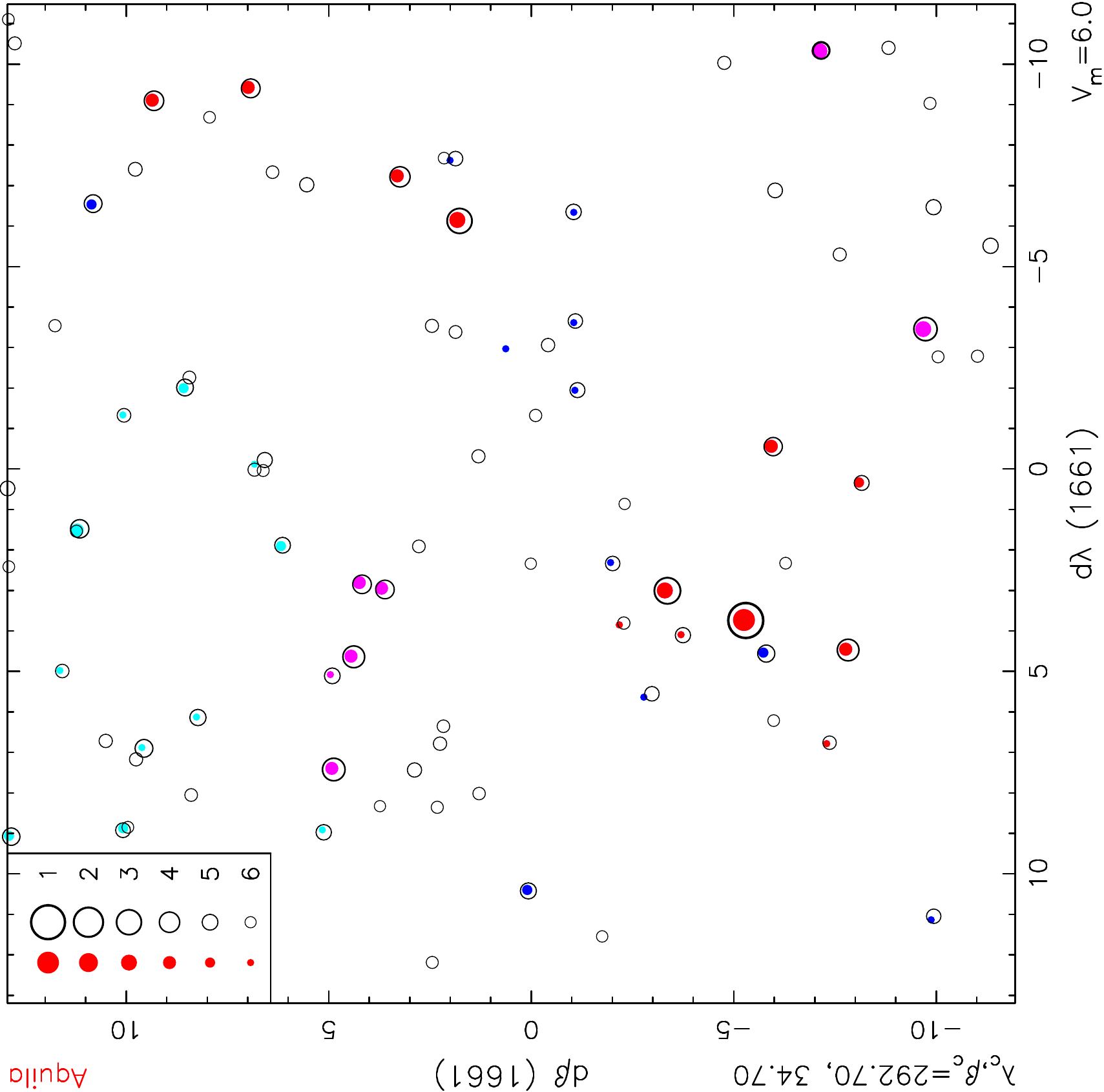}}
\caption{Aquila
 \label{f:aquila}}
\end{figure}

\begin{figure}
\centerline{\includegraphics[angle=270,width=\columnwidth]{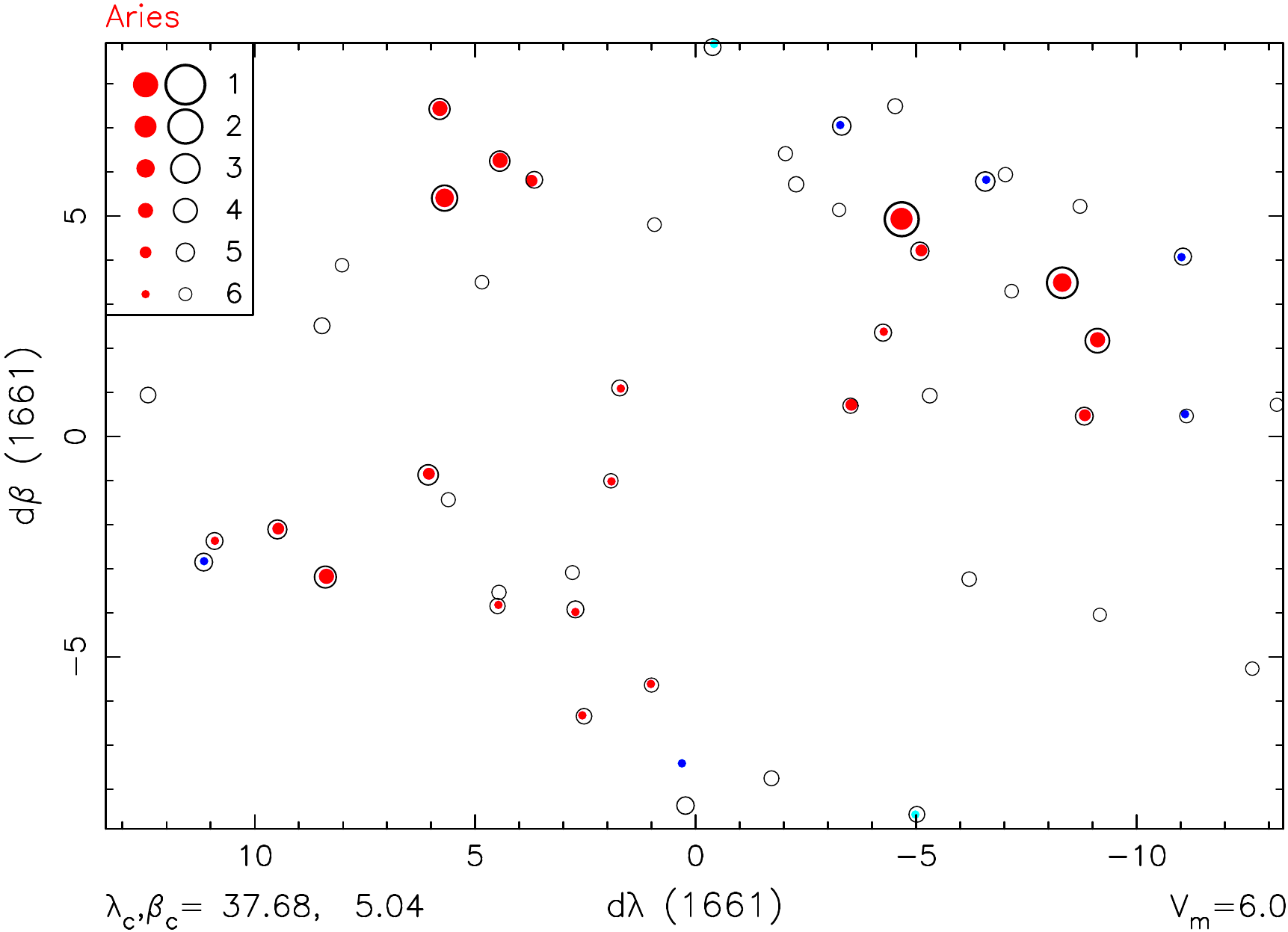}}
\caption{Aries
 \label{f:aries}}
\end{figure}

\clearpage

\begin{figure}
\centerline{\includegraphics[angle=270,width=\columnwidth]{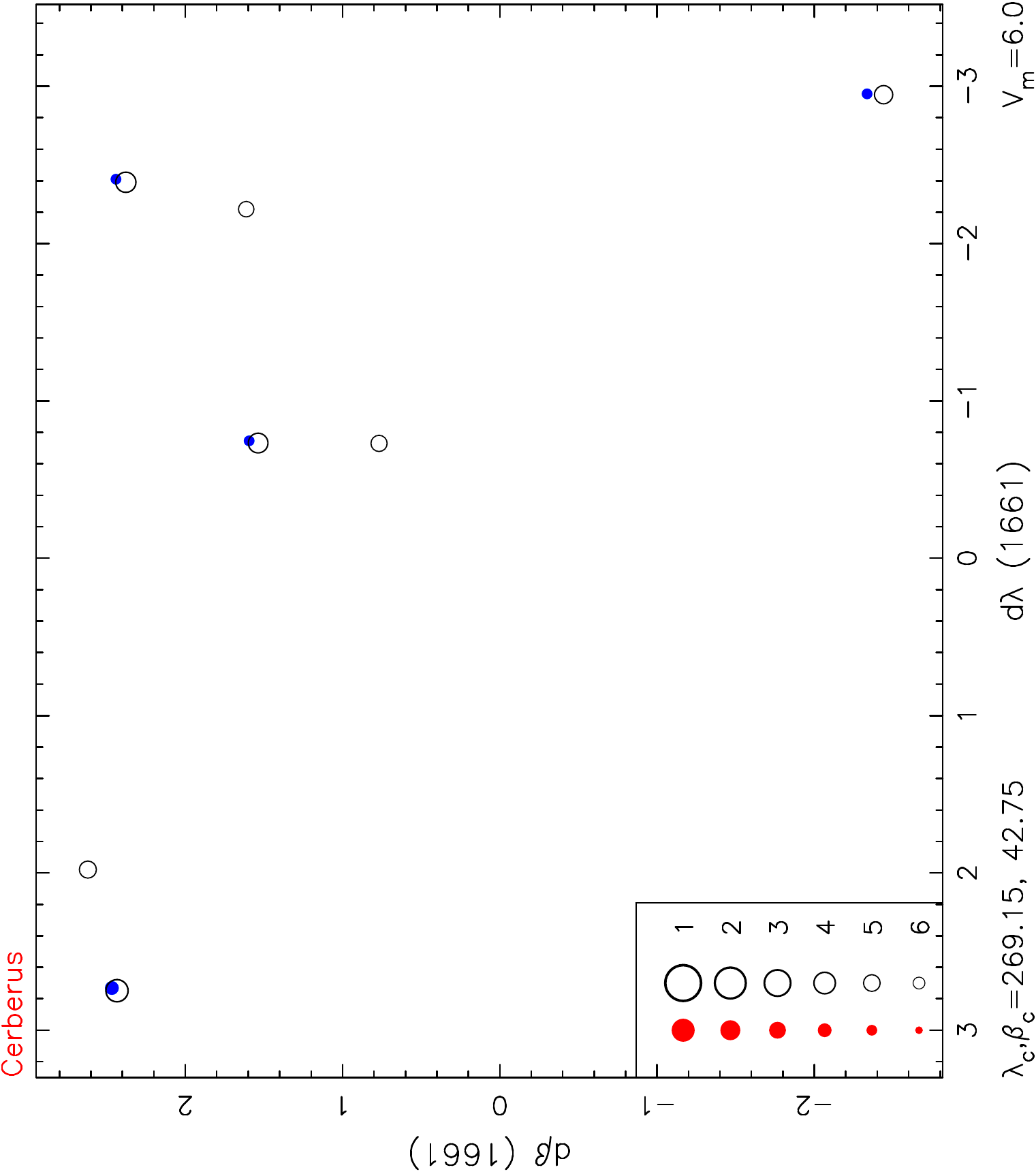}}
\caption{Cerberus
 \label{f:cerberus}}
\end{figure}

\begin{figure}
\centerline{\includegraphics[angle=270,width=0.7\columnwidth]{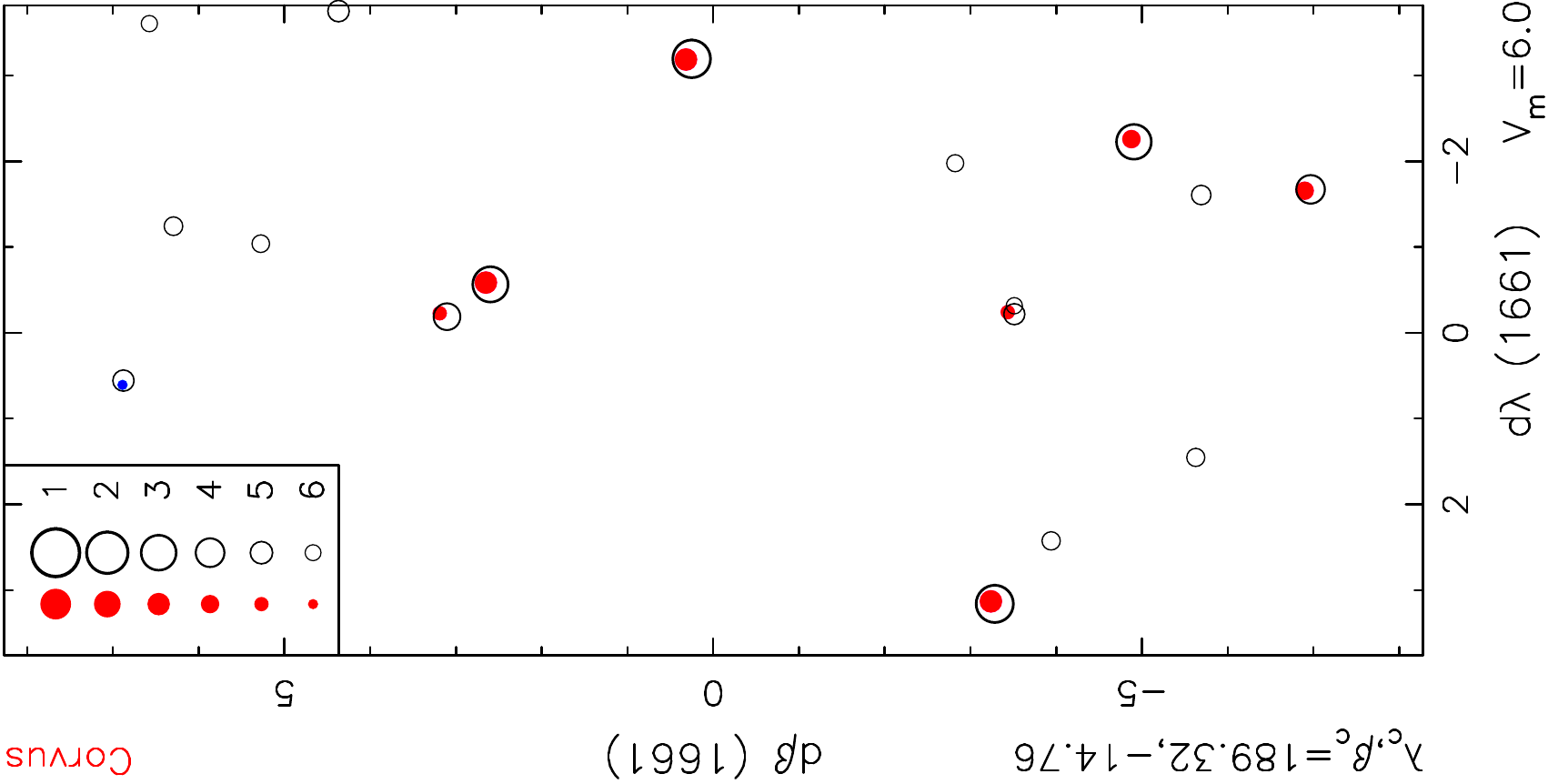}}
\caption{Corvus
 \label{f:corvus}}
\end{figure}

\begin{figure}
\centerline{\includegraphics[angle=270,width=\columnwidth]{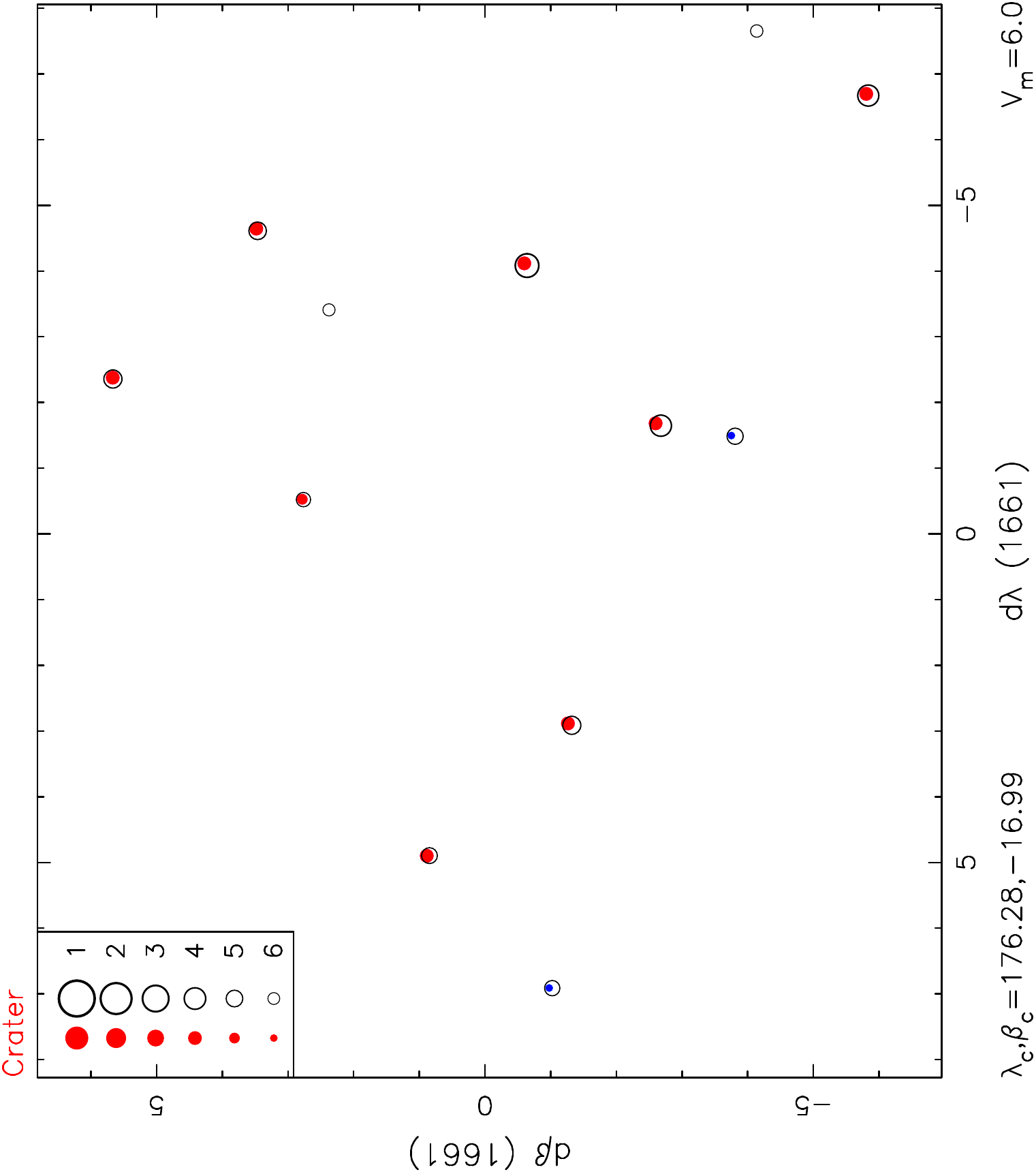}}
\caption{Crater
 \label{f:crater}}
\end{figure}

\begin{figure}
\centerline{\includegraphics[angle=270,width=0.7\columnwidth]{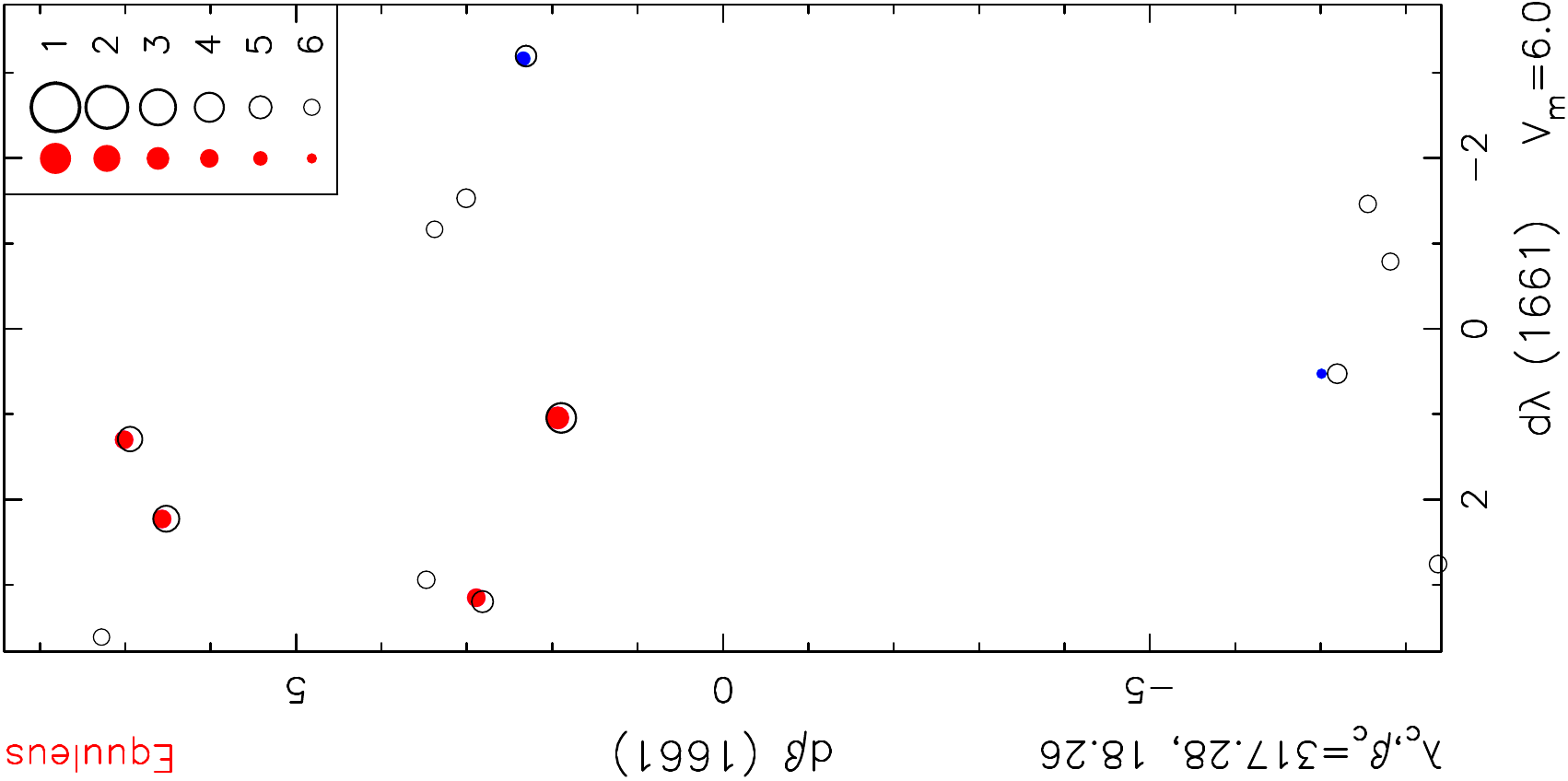}}
\caption{Equuleus
 \label{f:equuleus}}
\end{figure}

\clearpage

\begin{figure}
\centerline{\includegraphics[angle=270,width=\columnwidth]{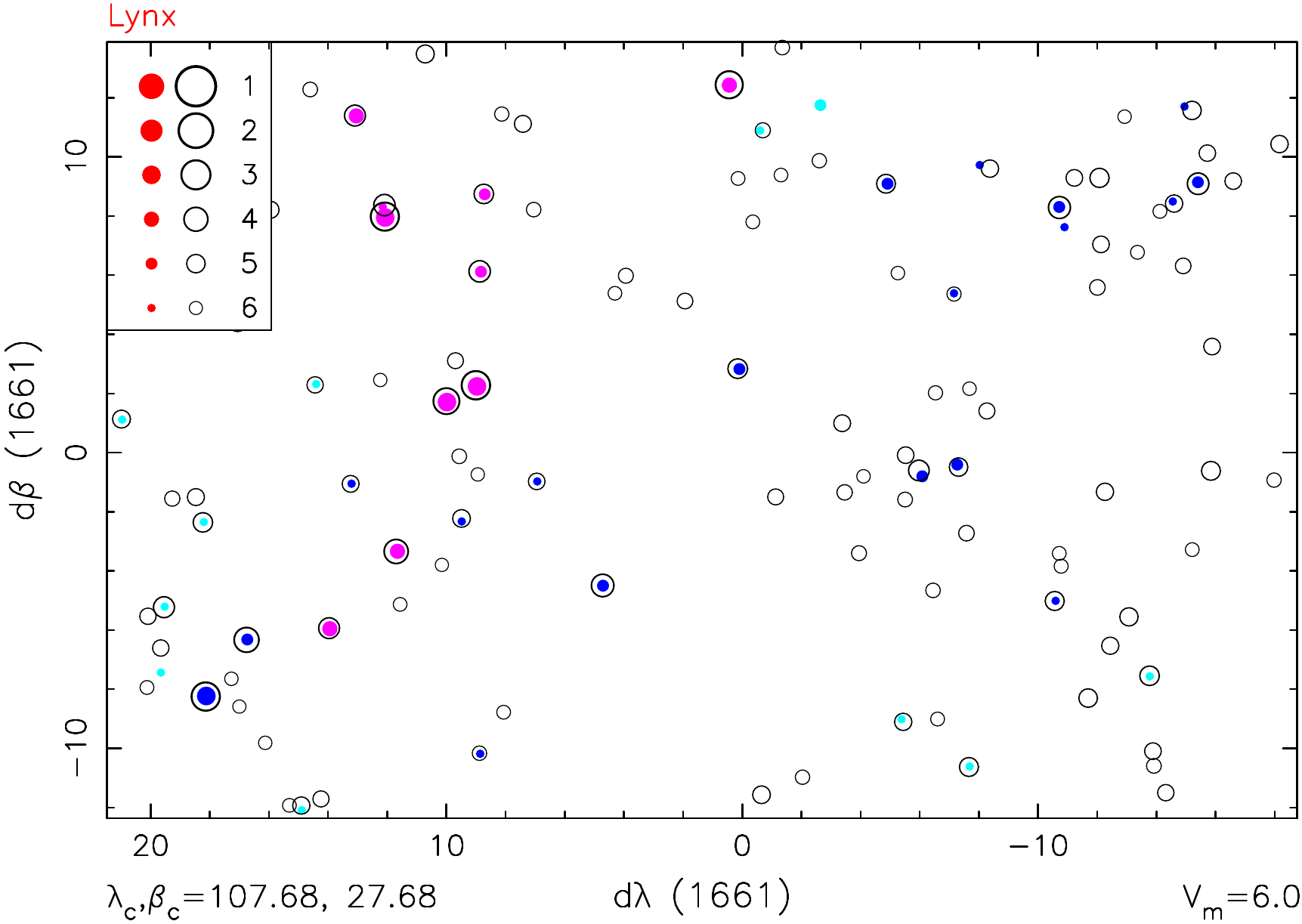}}
\caption{Lynx
\label{f:lynx}}
\end{figure}

\begin{figure}
\centerline{\includegraphics[angle=270,width=\columnwidth]{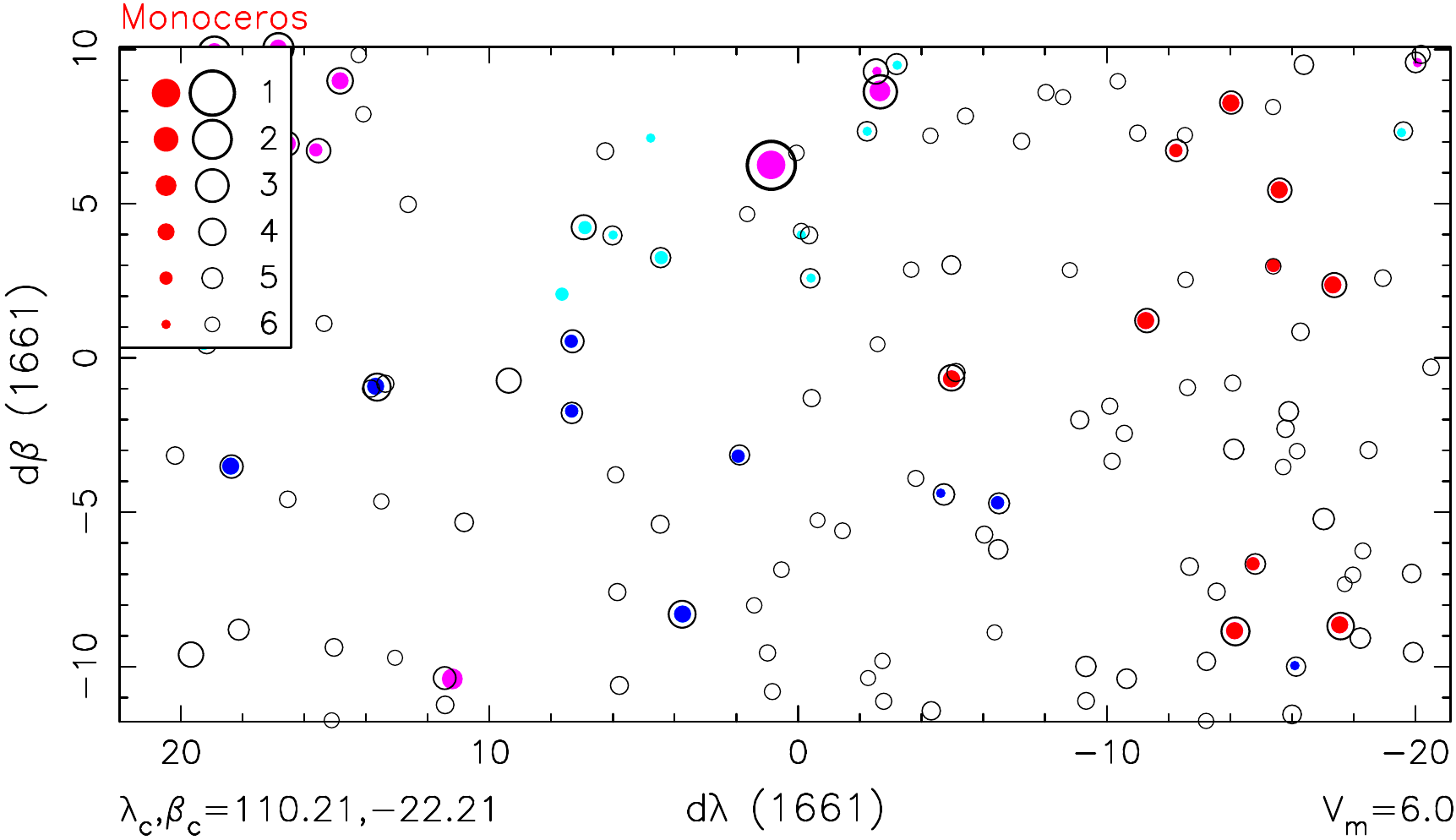}}
\caption{Monoceros
\label{f:monoceros}}
\end{figure}

\begin{figure}
\centerline{\includegraphics[angle=270,width=0.5\columnwidth]{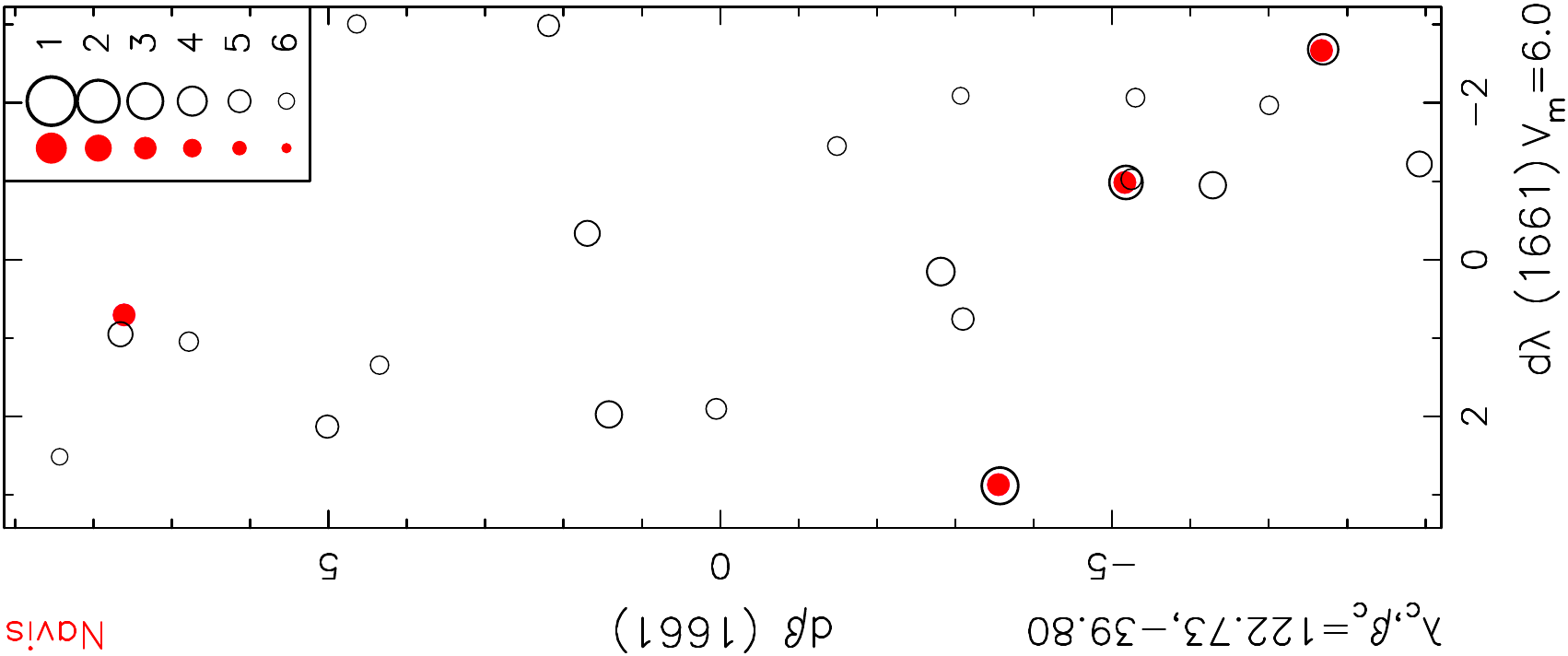}}
\caption{Navis
\label{f:navis}}
\end{figure}

\clearpage

\begin{figure}
\centerline{\includegraphics[angle=270,width=\columnwidth]{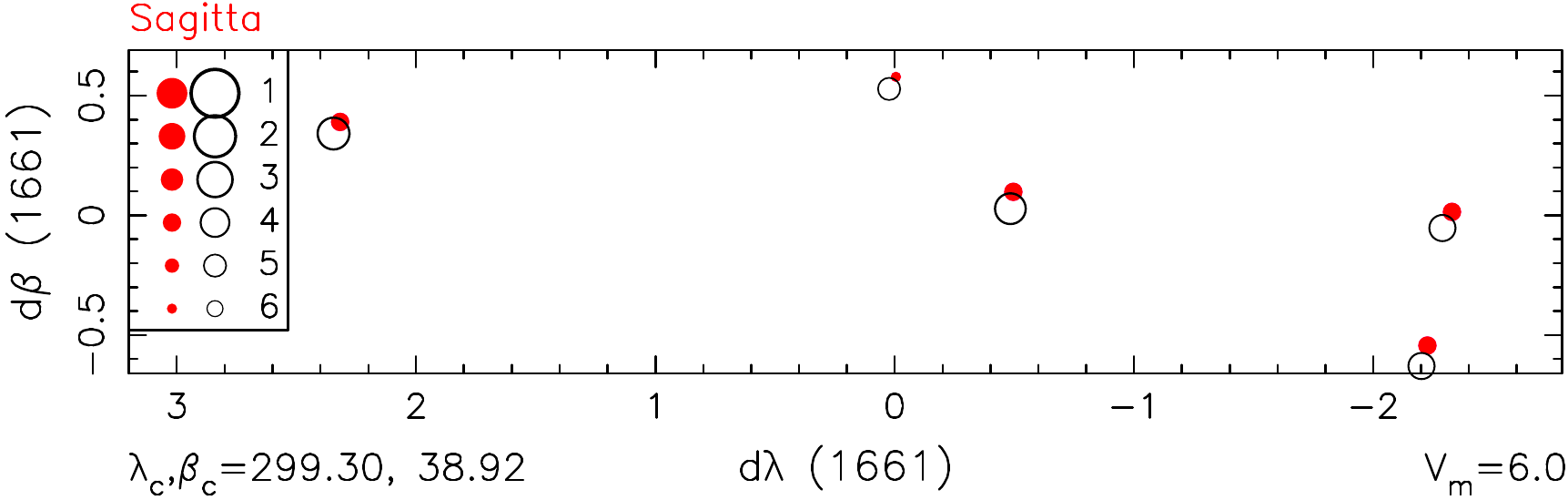}}
\caption{Sagitta
\label{f:sagitta}}
\end{figure}

\begin{figure}
\centerline{\includegraphics[angle=270,width=\columnwidth]{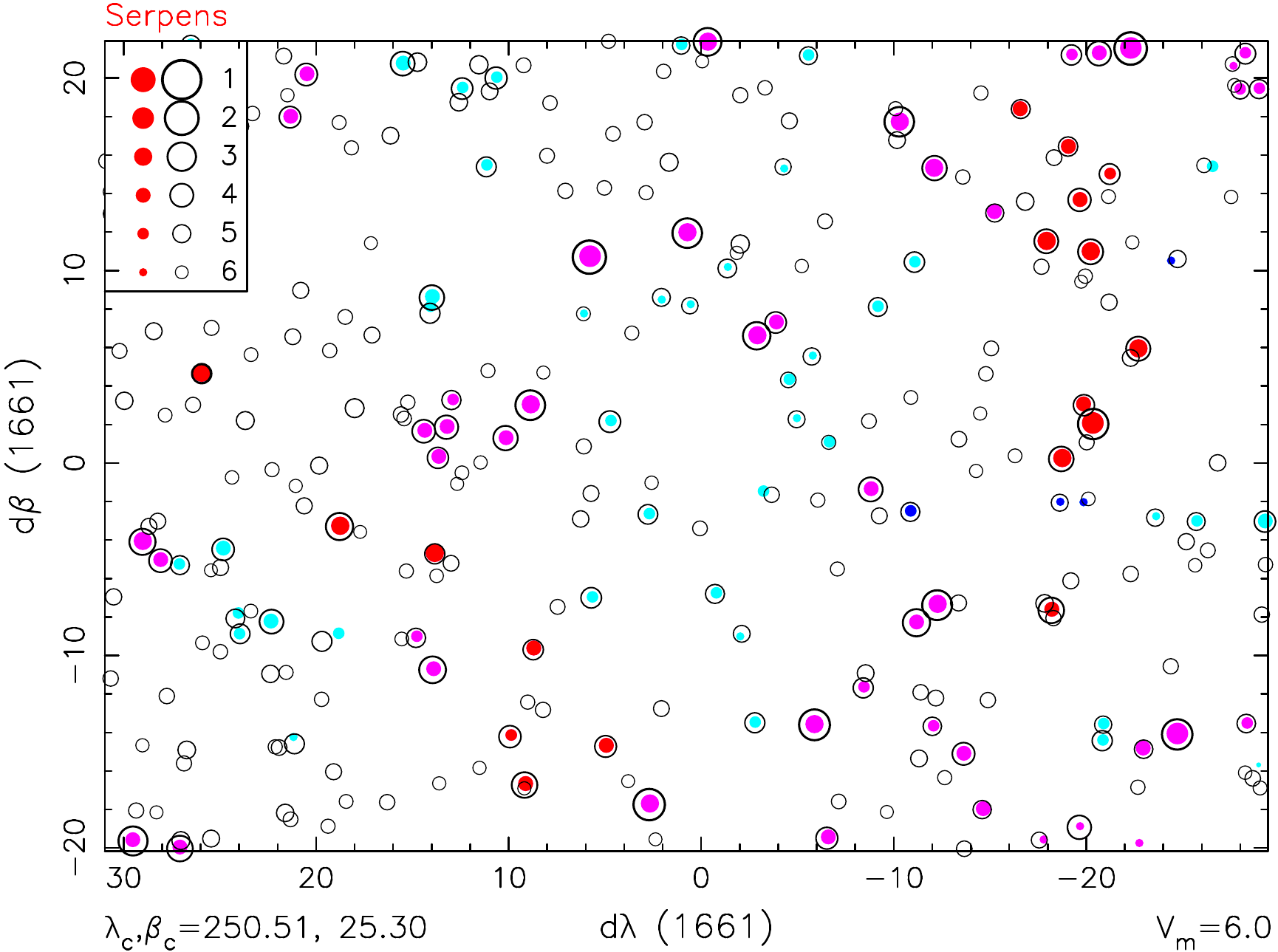}}
\caption{Serpens
\label{f:serpens}}
\end{figure}

\clearpage

\begin{figure}
\centerline{\includegraphics[angle=270,width=\columnwidth]{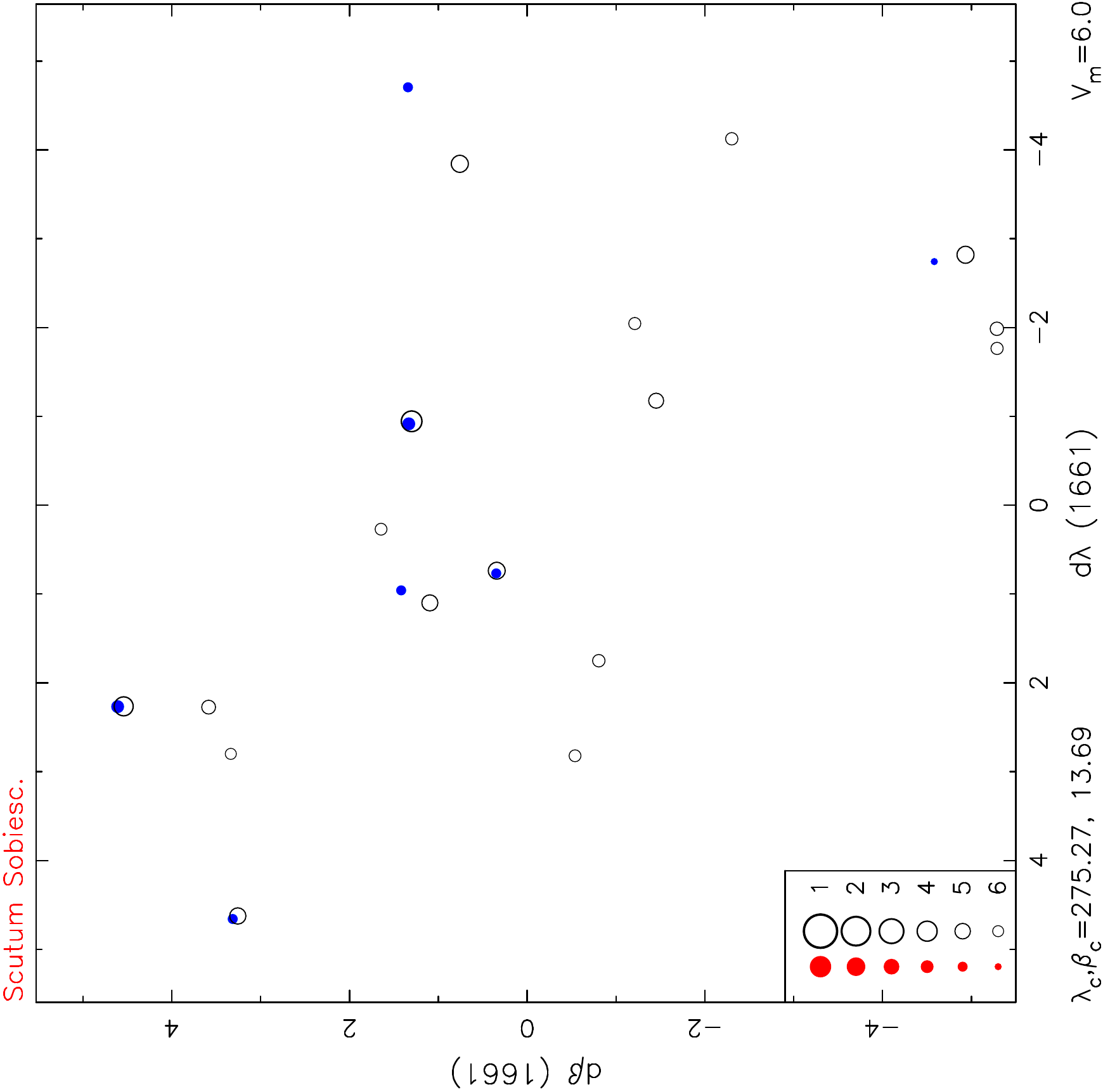}}
\caption{Scutum Sobiescanum
\label{f:scutum}}
\end{figure}

\begin{figure}
\centerline{\includegraphics[angle=270,width=\columnwidth]{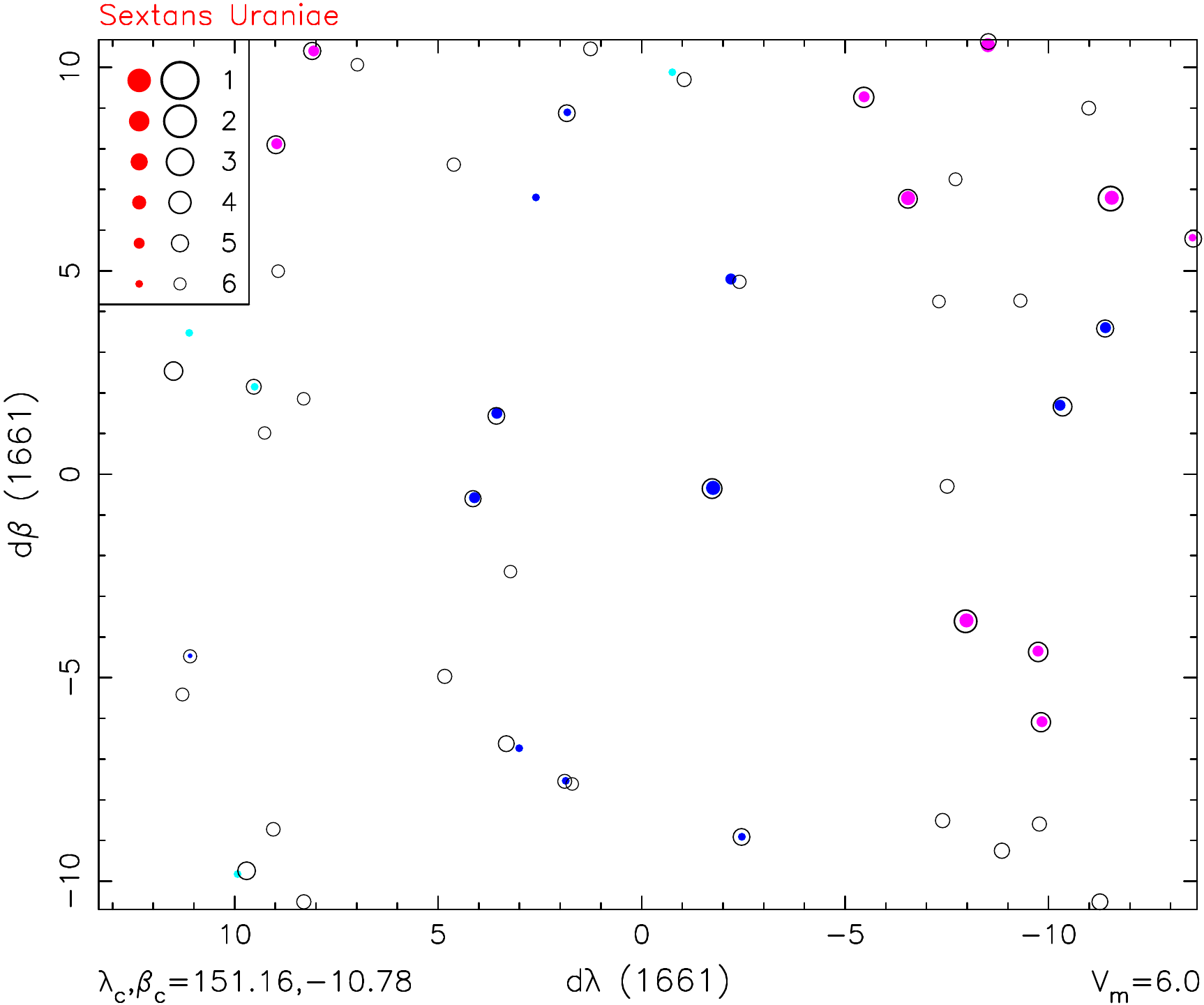}}
\caption{Sextans Uraniae
\label{f:sextans}}
\end{figure}

\begin{figure}
\centerline{\includegraphics[angle=270,width=\columnwidth]{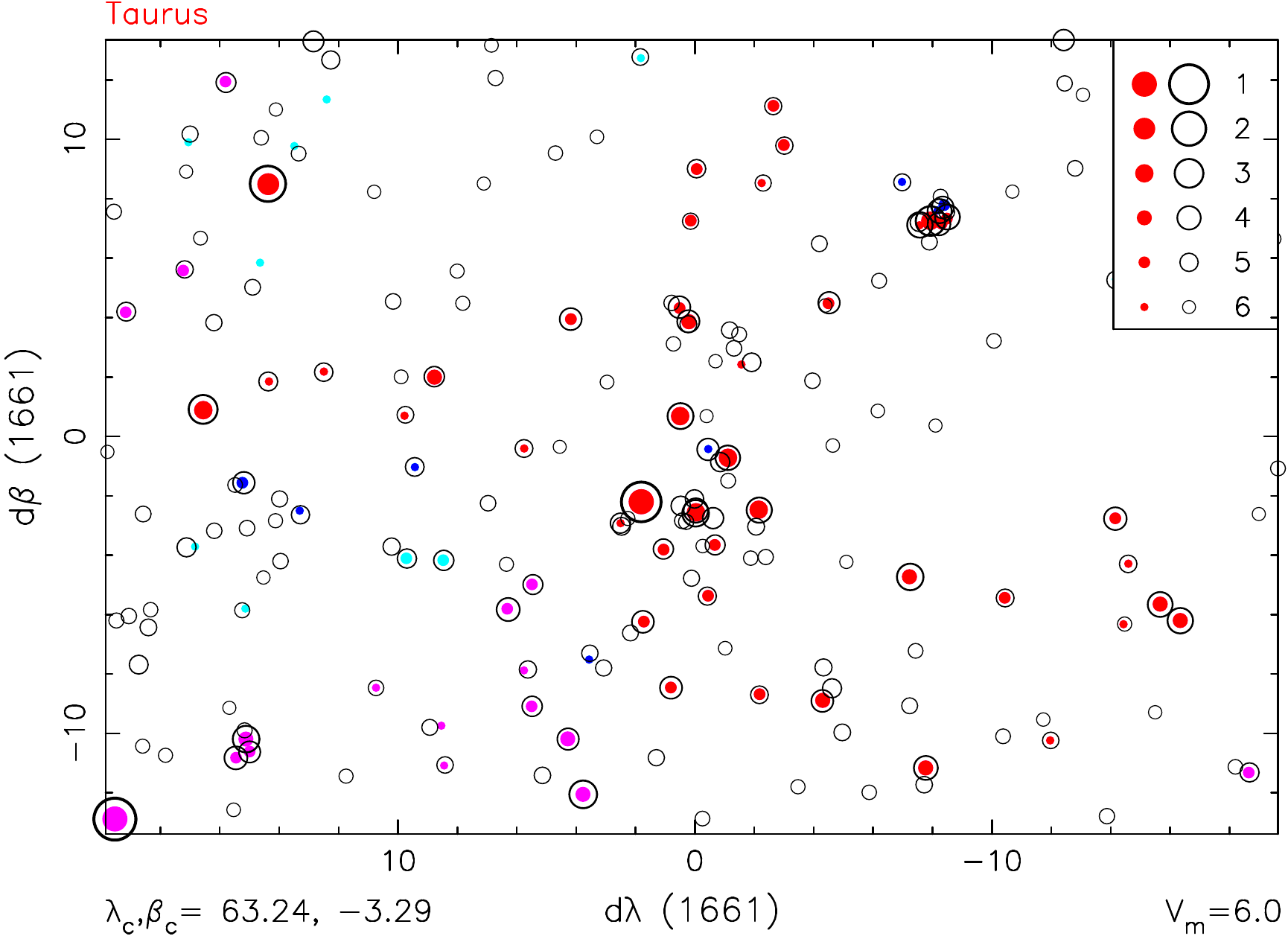}}
\caption{Taurus; for Pleiades see Fig.\,\ref{f:pleiades}.
\label{f:taurus}}
\end{figure}

\begin{figure}
\centerline{\includegraphics[angle=270,width=\columnwidth]{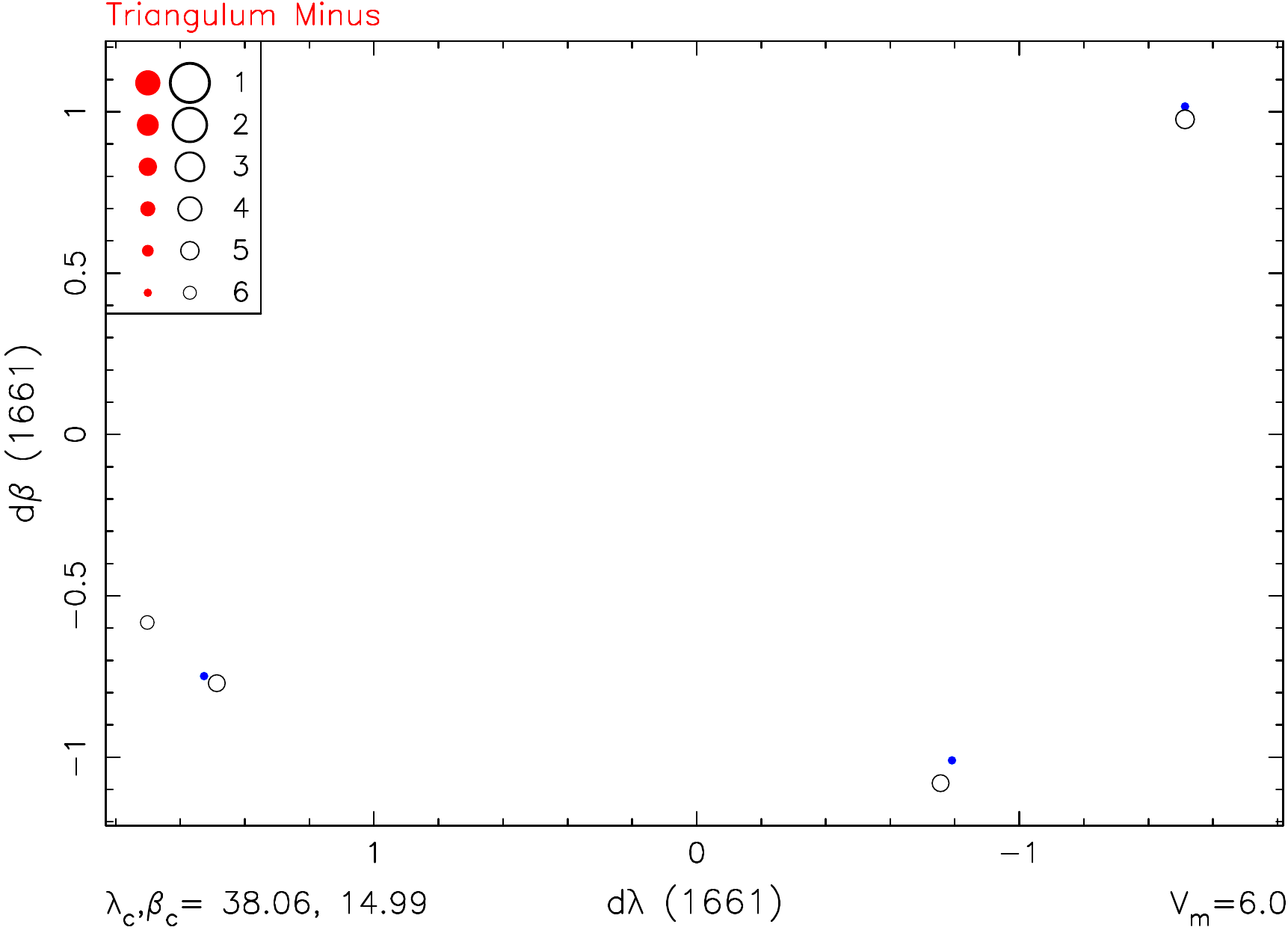}}
\caption{Triangulum Minus
\label{f:triminus}}
\end{figure}

\end{appendix}

\end{document}